\documentclass{article}

\usepackage[textheight=.76\paperheight, letterpaper]{geometry} 
\usepackage[bf,small,raggedright]{titlesec}

 \csname
@addtoreset\endcsname{equation}{section}

\usepackage[T1]{fontenc} 
\usepackage[utf8]{inputenc} 
\usepackage{microtype} 
\usepackage{lmodern} 
\usepackage{booktabs} 
\usepackage{mdframed} 
\usepackage{graphicx}
\usepackage{todonotes} 
\usepackage{nicefrac}

\usepackage{upgreek}
\usepackage{dsfont}

\usepackage{cite}

\usepackage{hyperref}
\usepackage{color}
\usepackage{amsmath}
\usepackage{slashed}
\usepackage{cancel}
\definecolor{dark-red}{rgb}{0.4,0.15,0.15}
\definecolor{dark-blue}{rgb}{0.15,0.15,0.4}
\definecolor{medium-blue}{rgb}{0,0,0.5}
\hypersetup{colorlinks, linkcolor={dark-red}, citecolor={dark-blue}, urlcolor={medium-blue}}

\usepackage{amsmath,amsfonts,amssymb}
\usepackage{mleftright} 
\usepackage{tensor} 
\usepackage{braket} 
\usepackage{mathrsfs} 
\usepackage{cleveref}

\usepackage{pict2e}

\makeatletter
\DeclareRobustCommand{\loplus}{\mathbin{\mathpalette\dog@lsemi{+}}}
\DeclareRobustCommand{\lotimes}{\mathbin{\mathpalette\dog@lsemi{\times}}}
\DeclareRobustCommand{\roplus}{\mathbin{\mathpalette\dog@rsemi{+}}}
\DeclareRobustCommand{\rotimes}{\mathbin{\mathpalette\dog@rsemi{\times}}}

\newcommand{\dog@rsemi}[2]{\dog@semi{#1}{#2}{-90,90}}
\newcommand{\dog@lsemi}[2]{\dog@semi{#1}{#2}{270,90}}
\newcommand{\dog@semi}[3]{%
  \begingroup
  \sbox\z@{$\m@th#1#2$}%
  \setlength{\unitlength}{\dimexpr\ht\z@+\dp\z@\relax}%
  \makebox[\wd\z@]{\raisebox{-\dp\z@}{%
    \begin{picture}(1,1)
    \linethickness{\variable@rule{#1}}
    \roundcap
    \put(0.5,0.5){\makebox(0,0){\raisebox{\dp\z@}{$\m@th#1#2$}}}
    \put(0.5,0.5){\arc[#3]{0.5}}
    \end{picture}%
  }}%
  \endgroup
}
\newcommand{\variable@rule}[1]{%
  \fontdimen8  
  \ifx#1\displaystyle\textfont3\else
    \ifx#1\textstyle\textfont3\else
      \ifx#1\scriptstyle\scriptfont3\else
        \scriptscriptfont3\relax
  \fi\fi\fi
}
\makeatother

\newcommand{\D}{\text{d}}
\newcommand{\beq}{\begin{equation}}
\newcommand{\eeq}{\end{equation}}
\newcommand{\beqn}{\begin{eqnarray}}
\newcommand{\eeqn}{\end{eqnarray}}
\newcommand{\pa}{\partial}
\newcommand{\nn}{\nonumber}

\newcommand{\GN}{{\cal G}} 

\newcommand{\cB}{{\cal B}}

\newcommand{\cE}{\mathscr{E}}
\newcommand{\cF}{{\cal F}}

\newcommand{\cM}{{\cal M}}

\newcommand{\e}{\epsilon} 
\newcommand{\ve}{\varepsilon}

\newcommand{\m}{\mu}
\newcommand{\n}{\nu}

\title{Holographic Lorentz and Carroll Frames}

\author{Andrea Campoleoni${}^{\, a,}$\footnote{Research Associate of the Fund for Scientific Research -- FNRS, Belgium.}\ , Luca Ciambelli${}^{\, b}$, Arnaud Delfante${}^{\, a,}$\footnote{FRIA grantee of the Fund for Scientific Research -- FNRS, Belgium.}\ ,\\[3pt]
Charles Marteau${}^{\, c}$, P. Marios Petropoulos${}^{\, d}$, Romain Ruzziconi${}^{\, e}$}
\date{}

\begin{document}

\maketitle

\vspace{-15pt}

\begin{center}

${}^a${\small
Service de Physique de l'Univers, Champs et Gravitation,\\
Universit{\'e} de Mons -- UMONS,
20 place du Parc, 7000 Mons, Belgium}
\vspace{5pt}

${}^b${\small Physique Math\'ematique des Interactions Fondamentales \& International Solvay Institutes,\\ Universit\'e Libre de Bruxelles, Campus Plaine -- CP 231, 1050 Bruxelles, Belgium}
\vspace{5pt}

${}^c${\small Department of Physics and Astronomy,\\
University of British Columbia,
Vancouver, BC V6T 1Z1, Canada}
\vspace{5pt}

${}^d${\small Centre de Physique Th\'eorique, 
        Ecole Polytechnique, CNRS\footnote{\emph{Centre National de la Recherche Scientifique}, Unit\'e Mixte de Recherche UMR 7644.}\\
        Institut Polytechnique de Paris,
        91128 Palaiseau Cedex, France    }
\vspace{5pt}

${}^e${\small 
Institute for Theoretical Physics,\\ TU Wien,
Wiedner Hauptstrasse 8, 1040 Vienna, Austria}
\vspace{5pt}

\end{center}

\vspace{20pt}

\begin{abstract}
Relaxing the Bondi gauge, the solution space of three-dimensional gravity in the metric formulation has been shown to contain an additional free function that promotes the boundary metric to a Lorentz or Carroll frame, in asymptotically AdS or flat spacetimes. 
We pursue this analysis and show that the solution space also admits a finite symplectic structure, obtained taking advantage of the built-in ambiguities.  
The smoothness of the flat limit of the AdS symplectic structure selects a prescription in which the holographic anomaly
appears in the boundary Lorentz symmetry, that rotates the frame. This anomaly turns out to be cohomologically equivalent to the standard holographic Weyl anomaly and survives in the flat limit, thus predicting the existence of quantum anomalies in conformal Carrollian field theories. 
We also revisit these results in
the Chern--Simons formulation, where the prescription for the symplectic structure admitting a smooth flat limit follows from the variational principle, and we compute the charge algebra in the boundary conformal gauge.
\end{abstract}

\vfill
\hfill RR050.072022
\newpage

\tableofcontents

\section{Introduction}
 
The study of the classical phase space of three-dimensional gravity in asymptotically anti-de Sitter  spacetimes has been a prolific research area since the eighties, with a revamped interest in the early days of the AdS/CFT correspondence.
Featuring the classical phase space essentially amounts to impose boundary conditions on the metric, i.e., to select the allowed metric fluctuations at infinity, as first discussed in the seminal paper by Brown and Henneaux \cite{Brown:1986nw}. This choice does not require to fix any particular gauge for the metric, but it is often convenient to opt for a given gauge and then discuss the behavior at infinity  of a reduced set of metric components only. A posteriori, the results of Brown and Henneaux can indeed be recovered by fixing the Fefferman--Graham gauge \cite{FG1, Fefferman:2007rka} and choosing the boundary metric to be Minkowski: this leads to the well-known asymptotic symmetries given by two copies of the Virasoro algebra with central charge $c=\nicefrac{3\ell}{2\GN}$, where $\GN$ denotes Newton's constant and $\ell$ the AdS$_3$ radius. The condition on the boundary metric was relaxed in \cite{Troessaert:2013fma} to allow for fluctuations of its conformal factor that preserve the vanishing of the corresponding curvature, a condition that ensures a well-posed variational problem. This relaxation leads to an augmented asymptotic symmetry algebra that contains two additional affine $\mathfrak{u}(1)$ algebras.
In \cite{Alessio:2020ioh}, it was then proposed to allow for all possible values of the conformal factor while still working in the Fefferman--Graham gauge. In that case the variational problem is ill-posed, which unveils the presence of the holographic Weyl anomaly. Nevertheless, one can still compute surface charges that are integrable and that belong to a representation of the conformal group together with Weyl rescalings.

Alternative boundary conditions in three dimensions have also been considered in \cite{Compere:2013bya, Perez:2016vqo, Afshar:2016kjj, Ojeda:2019xih}.\footnote{See also \cite{Donnay:2015abr, Grumiller:2019fmp, Adami:2020ugu, Adami:2020amw, Adami:2021nnf} for a recent account on the phase space for finite-distance boundaries.}
The most general phase space for which the variational problem is well-posed was identified in \cite{Grumiller:2016pqb} by relaxing the Fefferman--Graham gauge. However, as we shall discuss shortly, a well-posed variational principle renders the Weyl anomaly invisible in the residual diffeomorphisms, contrary to the analyses in \cite{Papadimitriou:2005ii, Ciambelli:2019bzz, Alessio:2020ioh, Fiorucci:2020xto, Ruzziconi:2020wrb, Geiller:2021vpg}. In this work, we propose to abandon the Fefferman--Graham gauge and explore an alternative setup, which  has also the advantage to allow for a smooth flat-space limit, thus making contact with the current efforts to identify a holographic description of asymptotically flat spacetimes. More precisely, in \cite{Campoleoni:2018ltl, Ciambelli:2020eba, Ciambelli:2020ftk}, another partial gauge fixing ---~originally motivated by fluid/gravity correspondence \cite{Ciambelli:2018wre} and not associated with a well-posed variational principle~--- was considered and its solution space was shown to generalize that of the Bondi gauge by trading the boundary metric for a Cartan frame. 
Here we go beyond the study of the solution space presented in \cite{Ciambelli:2020eba, Ciambelli:2020ftk} and complete the derivation of its associated symplectic structure only explored in some specific examples in \cite{Campoleoni:2018ltl}. This effort will be shown to promote some of the additional free functions in the bulk metric to new boundary degrees of freedom for both asymptotically AdS and asymptotically flat spacetimes.

In order to illustrate the key features of our partial gauge fixing, let us recall the corresponding relevant aspects of the Fefferman--Graham gauge.
In that gauge, the solution space is characterized by a boundary metric $g_{\mu\nu}$ and a boundary energy--momentum tensor $T_{\mu\nu}$, and the variation of the on-shell action (properly renormalized) is 
\begin{equation}
    \delta S=\frac{1}{2} \int_{\partial \cM} \!\D^{D}x \sqrt{-g}\,T^{\mu\nu}\delta g_{\mu\nu}
    \label{VarAction}
\end{equation}
in any dimension \cite{deHaro:2000vlm}, see also \cite{Skenderis:2002wp, Papadimitriou:2005ii}.
The bulk metric actually induces a conformal class of metrics on the boundary as firstly shown in \cite{FG1, Imbimbo:1999bj, Schwimmer:2000cu, Rooman:2000ei, Rooman:2000zi, Fefferman:2007rka} and recently reconsidered, e.g., in \cite{Ciambelli:2019bzz, Anastasiou:2020zwc, Jia:2021hgy}: 
the boundary metric that characterizes the solution space should be understood as a representative of this conformal class.
Moving from one representative to another is not innocuous: when the spacetime dimension  is odd, there exists a non-trivial surface charge associated with bulk diffeomorphisms which are mapped to Weyl transformations on the boundary \cite{Alessio:2020ioh, Fiorucci:2020xto, Geiller:2021vpg}. This implies that a boundary Weyl transformation results in a physically distinct solution. The associated surface charge is not conserved, but this does not invalidate the option to use it in order to label distinguishable physical configurations.
    
Until now all considerations were purely gravitational, but one can reinterpret them in a holographic language. 
Evaluating \eqref{VarAction} on a bulk diffeomorphism inducing on the  boundary a Weyl rescaling generated by $\sigma$ plus a diffeomorphism generated by $\xi^\mu$ gives, in three bulk dimensions,
\begin{equation}\label{weylANintro}
    \delta_{(\xi,\sigma)} S\sim c \int_{\partial \cM} \!\D^{2}x \sqrt{-g} \ \sigma\,R \, .
\end{equation}
We recognize the usual boundary Weyl (also known as conformal) anomaly sourced by the Brown--Henneaux central charge \cite{Henningson:1998gx}.
Moreover, the non-conservation of Weyl charges mentioned earlier can be interpreted, at least when $D=2$, as an anomalous Ward identity for the boundary Weyl symmetry \cite{Alessio:2020ioh}.

In the Fefferman--Graham gauge, the bulk metric comprises a space-like radial coordinate, along which one approaches the conformal boundary. Another class of gauges that has been widely explored are those in which the boundary is approached along a null direction. The best-known example is the Bondi gauge, originally introduced to study asymptotically flat spacetimes at null infinity in the presence of gravitational radiation \cite{Bondi:1962px, Sachs:1962wk, Sachs:1962zza}.
This gauge also exists in asymptotically AdS spacetimes and, crucially, it admits a well-defined flat limit as opposed to the Fefferman--Graham gauge \cite{Barnich:2012aw, Barnich:2013sxa, Poole:2018koa, Compere:2019bua, Ciambelli:2018wre, Compere:2020lrt,  Geiller:2022vto}.
In three-dimensional gravity, the option to take a flat limit was already successfully exploited to analyze some features of a putative flat-space holographic duality \cite{Barnich:2012aw, Barnich:2012xq, Bagchi:2012xr,Bagchi:2012cy, Detournay:2014fva, Bagchi:2014iea, Bagchi:2015wna, Hartong:2015usd, Basu:2017aqn, Campoleoni:2018ltl, Ciambelli:2020eba, Ciambelli:2020ftk, Geiller:2021vpg}.
In the following we shall focus on this class of gauges.

In \cite{Poole:2018koa, Compere:2019bua, Ciambelli:2020eba, Ciambelli:2020ftk}, it was shown that the solution spaces of the Fefferman--Graham and Bondi gauges are in one-to-one correspondence. This is achieved by exhibiting a rather involved diffeomorphism that relates the two. The dictionary requires a choice of boundary frame that is used to project the two independent data of the Fefferman--Graham gauge, i.e. the metric and energy--momentum tensor, in order to relate them to Bondi data. This points to a natural way to relax the Bondi gauge, which is to restore this broken frame covariance. This is done by  partially fixing the gauge in the bulk and, in particular, by letting the metric component that mixes the null radial and the spatial directions be non-zero. For these reasons, we  refer here to this choice as covariant Bondi gauge, although it was originally named derivative expansion in \cite{Ciambelli:2018wre, Campoleoni:2018ltl}, due to its interpretation within the fluid/gravity correspondence (c.f.~\cite{Bhattacharyya:2007vjd, Haack:2008cp, Hubeny:2011hd, Ciambelli:2018xat, Petkou:2022bmz}).  
The covariant Bondi gauge thus introduces a dependence on the boundary Cartan frame as well as new residual symmetries corresponding to its (hyperbolic) rotations. The group of residual bulk diffeomorphisms becomes a product of boundary diffeomorphisms, Weyl rescalings and Lorentz boosts \cite{Ciambelli:2020eba, Ciambelli:2020ftk}.

A Cartan frame could certainly be introduced in the Fefferman--Graham gauge as an alternative to the boundary metric. For instance, in three bulk spacetime dimensions, one could express the boundary metric as
\begin{equation}
    g_{\mu\nu}=\ell^2\left(-u_\mu u_\nu+\ast u_\mu \ast\! u_\nu\right),
\end{equation}
with $u_\mu$ and $\ast u_\nu$ the components of the two Cartan forms.  
However, the variation of the action  \eqref{VarAction} would remain sensitive to the energy--momentum tensor and to the metric variation, but not separately to  $\delta u_\m$ and $\delta \!\ast\!u_\m$. 
Therefore, no asymptotic charge could be associated with the hyperbolic rotation of the frame: the symmetry is de facto pure-gauge. This is not the case in our covariant Bondi gauge, where the bulk metric depends explicitly on the boundary Cartan frame.  In order to establish which residual symmetries are physical, one needs to study the symplectic structure that gravity induces on the asymptotic data. In the process, one can utilize the built-in ambiguities \cite{Lee:1990nz, Iyer:1994ys, Wald:1999wa} to render that structure finite \cite{Papadimitriou:2005ii, Compere:2008us, Compere:2018ylh, Compere:2020lrt, Fiorucci:2020xto, Ruzziconi:2020wrb, Freidel:2021fxf,  Geiller:2021vpg, Chandrasekaran:2021vyu}; different choices may lead to different finite presymplectic potentials and so to different surface charges.\footnote{Another worth exploring pattern is to consider the extended phase spaces recently proposed in \cite{Ciambelli:2021vnn, Ciambelli:2021nmv, Freidel:2021dxw}, which are similarly affected by ambiguities \cite{Speranza:2022lxr}.}
This also determines which symmetries are anomalous as we shall appreciate in this work.

A detailed analysis of the presymplectic potential associated with the covariant Bondi gauge shows that there exists a prescription which leads to the usual Fefferman--Graham symplectic potential $\Theta^{(r)} \sim T^{\mu\nu}\delta  g_{\mu\nu}$.
As explained above, this implies that the Lorentz component of the residual symmetries is not a  symmetry of the boundary theory, and thus cannot acquire an anomalous contribution. This is not an issue in itself, nevertheless such a choice has disadvantages when it comes to considering the flat limit, since the presymplectic potential diverges for infinite $\ell$.  

Remarkably, there exists another prescription giving a presymplectic potential that remains finite in the flat-space limit. With this choice, rotations of the Cartan frame are associated with non-trivial surface charges and the variation of the on-shell action is sensitive to them:
\begin{equation}
\delta S = \int_{\partial \cM} \!\D^{2}x\sqrt{-g}\left( J^\mu \, \delta u_\mu+J_\ast^\mu \, \delta\!\ast\!u_\mu \right). \label{variation_Lorentz}
\end{equation}
Here the couple of currents $(J^\mu, J^\mu_\ast)$ plays a role analogous to the energy--momentum tensor in the Fefferman--Graham gauge: the variation of the action has the familiar holographic interpretation (vev)$\,\times\,\delta$(source).  
As we shall see, these currents combine in an energy--momentum tensor which is traceless and skew-symmetric, thus providing compelling evidence for the presence of an anomaly in the boundary Lorentz symmetry.
This is confirmed by evaluating \eqref{variation_Lorentz} on a bulk diffeomorphism inducing on the boundary a Weyl rescaling and a diffeomorphism, generated by $\sigma$ and $\xi^\mu$ as in \eqref{VarAction}, together with a two-dimensional Lorentz boost generated by $\eta$:
\begin{equation}\label{lorentzAN}
    \delta_{(\xi,\sigma,\eta)} S\sim c \int_{\partial \cM} \!\D^{2}x \, \sqrt{-g} \, \eta \, F \, .
\end{equation}
Notice that the infinitesimal Lorentz boost $\eta$ does not preserve the on-shell action, thus disclosing a holographic anomaly, while the other symmetries of the theory do not appear and therefore are not anomalous. The holographic anomaly is sourced by the field strength of the Weyl connection, as introduced in \cite{Ciambelli:2019bzz}. 

The appearance of the anomaly in the Lorentz rather than Weyl symmetry calls for a deeper understanding of the anomaly structure in the holographic boundary theory. 
It is possible to classify the anomalies and central charges that a two-dimensional theory possessing  Weyl--Lorentz symmetry can admit.\footnote{A forthcoming publication \cite{preprint_Luca} will be mainly devoted to classify Weyl--Lorentz anomalies using BRST techniques (see, e.g., \cite{Henneaux:1992ig} and references therein). These techniques have already been used to classify Weyl anomalies in the second-order formalism in \cite{Boulanger:2007ab, Boulanger:2007st}.} As we shall briefly discuss in appendix~\ref{app:BRST}, the outcome is that there are three possible central charges in a two-dimensional Weyl--Lorentz theory, and the two expressions \eqref{weylANintro} and \eqref{lorentzAN} are shown to be cohomologically equivalent.\footnote{This is the reason why one should more properly refer to the holographic anomaly as a mixed Weyl--Lorentz anomaly, but, to distinguish these two different representatives in the same cohomology, we will refer to \eqref{weylANintro} as Weyl anomaly (or anomaly in the Weyl sector), and to \eqref{lorentzAN} as Lorentz anomaly (or anomaly in the Lorentz sector).} This proves that the ambiguities in the covariant phase-space formalism amount to displace the central charge from one sector of the theory to another. It is  noteworthy that this simple three-dimensional setup  highlights these rather generic  features of the boundary field theory.

It is also remarkable that only one choice of ambiguities guarantees a finite flat limit. In such a limit, the Weyl--Lorentz structure reduces to a Weyl--Carroll structure, related to those that have been introduced in \cite{Duval:2014uva, Duval:2014lpa, Ciambelli:2019lap} generalizing the Carrollian structures of \cite{Henneaux:1979vn, Duval:1990hj, Duval:2014uoa, Ashtekar:2014zsa} (see also \cite{Herfray:2021qmp, Bergshoeff:2022eog}). This prescription makes the anomaly explicit in the Lorentz sector, and, in the flat limit, in the Carroll-boost sector. 
We indeed show in section~\ref{sec:Flat} that in the flat limit the phase-space mirrors its AdS counterpart in terms of surface charges and symmetries, the main difference being that the boundary geometry is now encoded in a Carroll frame and Lorentz boosts are replaced by ultra-relativistic Carroll boosts. A classification of anomalies in $\mathfrak{bms}_3$-invariant field theories\footnote{A concrete example of field theories with conformal Carroll symmetry is given by the scalar field theories that have been recently considered in  \cite{Bagchi:2019xfx, Gupta:2020dtl, Bagchi:2022eav, Rivera-Betancour:2022lkc, Baiguera:2022lsw}. Other relevant works on Carrollian field theories include, e.g., \cite{Bagchi:2019clu, Henneaux:2021yzg, deBoer:2021jej}.} is at present missing, albeit some examples have been discussed in \cite{Bagchi:2021gai}. These differ however from our result, which is thus a new holographic prediction, calling for further  investigation.

The surface charges turn out to be non-integrable as well as non-conserved, for both prescriptions for the symplectic potential. Integrability is however restored by tuning  the boundary metric into conformal gauge.
When implementing this choice within the prescription admitting a flat limit, Weyl transformations turn out to be pure-gauge while Lorentz boosts are generated by integrable but non-conserved charges. The non-conservation is another signal either of the Lorentz anomaly for finite $\ell$ or of the Carroll-boost anomaly at infinite $\ell$, similarly to what happens in the Fefferman--Graham gauge for the Weyl anomaly \cite{Alessio:2020ioh}. Obtaining integrable charges allows for a detailed characterization of the algebras of asymptotic symmetries. We study this in detail in section~\ref{sec:first-order} by employing the Chern--Simons formulation of three-dimensional gravity \cite{Achucarro:1986uwr, Witten:1988hc}. Aside from the usual technical simplifications brought by this setup, one can find a simple Chern--Simons connection that gives rise to the set of surface charges associated with the symplectic structure admitting a flat limit. Furthermore, in this formalism we establish the boundary terms to be added to the bulk action, such that the variational principle correctly reproduces our symplectic structure.

\section{Covariant Bondi gauge in AdS and holographic frames}\label{sec:DE}

In this section, we first review the definition of the covariant Bondi gauge and identify the residual diffeomorphisms preserving it, following \cite{Campoleoni:2018ltl, Ciambelli:2020eba, Ciambelli:2020ftk}. 
We then compute the symplectic form on the boundary phase space. The latter is sensitive to the choice of boundary counterterms in the action or, equivalently, presents a well-known ambiguity when defined starting from a bulk Lagrangian. We present two prescriptions for fixing these ambiguities leading to a finite result: the first reproduces the usual symplectic form derived in the Fefferman--Graham gauge, while the second promotes rotations of the boundary Cartan frame into asymptotic symmetries of the theory. Only the second option admits a flat limit, which will be discussed in section~\ref{sec:Flat}, and in this case the holographic Weyl anomaly is displaced onto the Lorentz sector of the asymptotic symmetries. For a generic boundary Cartan frame, the surface charges are neither integrable nor conserved, irrespective of the  prescription for the presymplectic potential. Still, we conclude this section showing that integrability can be attained by restricting the boundary metric to the conformal gauge.

\subsection{Relaxing the Bondi gauge}

\paragraph{Solution space}
The covariant Bondi gauge, firstly introduced in three dimensions in \cite{Campoleoni:2018ltl}, is defined by the line element  
\begin{equation} \label{DE}
\D s_{\text{AdS}}^2= \frac{2}{k^2}\, \text{u}  \left(\D r+ r \, \text{A} \right)+r^2 g_{\mu\nu} \, \D x^\mu \D x^\nu+\frac{8\pi \GN}{k^4}\, \text{u} \left(\varepsilon\, \text{u}+\chi \ast\!\text{u} \right) ,
\end{equation}
where $r$ is a null radial coordinate and $\{x^\mu, \mu=0,1\}=\{k u,\phi\}$ are adimensional coordinates charting  the boundary. We have also introduced the inverse AdS radius $k=\ell^{-1}$, so that the flat limit is reached when $k\to 0$, while $\GN$ denotes Newton's constant. The conformal boundary is located at $r\to \infty$. Compared to the usual Bondi gauge \cite{Bondi:1962px, Sachs:1962wk, Sachs:1962zza}, this partial gauge fixing allows for a non-vanishing $\D r\D\phi$ component. 

The line element is parameterized in terms of the following boundary quantities: the metric $g_{\m\n}$, the couple of one-forms $(\text{u}=u_\mu \D x^\mu,\ast\text{u}=\ast u_\mu \D x^\mu)$, the Weyl connection $\text{A} = A_\m \D x^\m$ and the  scalars $\varepsilon$ and $\chi$, such that the $r$ dependence is explicit in \eqref{DE}. However, not all boundary data are independent: $\ast\text{u}$ is  the Hodge dual\footnote{Our conventions are: $\cE_{\mu\nu} = \sqrt{-g}\, \varepsilon_{\mu\nu}$ with $\varepsilon_{01} = +1$ and $\cE^{\mu\rho} \cE_{\rho\nu} = \delta^\mu_\nu$. The Hodge dual of a one-form is then defined as $\ast v_\mu = \cE_{\mu\nu} v^\nu$. Notice that the convention for the Hodge dual differs from \cite{Campoleoni:2018ltl, Ciambelli:2020eba}, but agrees with \cite{Ciambelli:2020ftk}.} of $\text{u}$ ($\ast u_\mu=\cE_{\mu\nu}u^\nu$ and $ u_\mu=\cE_{\mu\nu}\!\ast\!u^\nu$). The boundary being two dimensional, one can equivalently consider these one-forms as independent and use them to define a Cartan frame:
\begin{equation}
g_{\mu\nu}=\frac{1}{k^2} \left(-u_\mu u_\nu+\ast u_\mu \!\ast\!u_\nu \right) .
\label{Bmetric}
\end{equation}
This Cartan frame is not orthonormal since, by definition, $u^\mu$ and $\ast u^\mu$ are orthogonal but normalized to $-k^2$ and $k^2$ instead of $-1$ and $1$, respectively.
In \cite{Campoleoni:2018ltl, Ciambelli:2020ftk}, the background metric and the normalized vector $u^\m$ were mainly chosen as independent boundary data, and the vector congruence $u^\m$ was interpreted as the velocity of a two-dimensional fluid hosted on a curved background metric $g_{\m\n}$ with local energy density $\ve$ and heat density $\chi$. Here, following \cite{Ciambelli:2020eba}, we mainly consider $\text{u}$ and $\ast\text{u}$ as the two independent one-forms of the boundary dyad, while $\ve$ and $\chi$ specify the  components of the boundary energy--momentum tensor.

The bulk metric also involves the Weyl connection $\text{A}$, introduced in the fluid/gravity correspondence in \cite{Loganayagam:2008is} (see also \cite{Ciambelli:2019bzz}). It depends on the one-forms $(\text{u},\ast\text{u})$ as follows,
\begin{equation}
\text{A}=\frac{1}{k^2} \left(\Theta^\ast \!\ast\! \text{u}-\Theta\, \text{u} \right) ,
\label{WeylConnection}
\end{equation}
where we introduced the expansion $\Theta=\nabla_{\!\mu} u^\mu$ and its dual $\Theta^\ast=\nabla_{\!\mu}\! \ast\! u^\mu$. Equivalently, these quantities can be defined from the exterior derivative of the Cartan frame,
\begin{equation}
\text{d}\ast\!\text{u} =\frac{\Theta}{k^2}\ast \!  \text{u} \wedge\text{u} \, ,
\qquad
\text{d}\text{u} =\frac{\Theta^\ast}{k^2}\ast \!  \text{u} \wedge \text{u} \, . \label{def-expansions}
\end{equation}
The one-form $\text{A}$ is dubbed Weyl connection because if one performs a Weyl rescaling on the boundary frame ---~$(\text{u},\ast\text{u})\to\mathcal{B}^{-1}(\text{u},\ast\text{u})$, where ${\cal B}$ is a nowhere vanishing boundary function~--- it transforms as a connection,
\begin{equation} \label{A_transf}
    \text{A}\to \text{A}-\text{d}\ln\mathcal{B} \, .
\end{equation}
Requiring the bulk line element \eqref{DE} be invariant under boundary Weyl transformations induced by $r \to \cB \, r$, one obtains that $(\ve,\chi) \to \cB^2(\ve,\chi)$. In the covariant Bondi gauge, differently from the  Bondi or Fefferman--Graham gauges, but similarly to the Weyl--Fefferman--Graham gauge \cite{Ciambelli:2019bzz}, boundary Weyl transformations are thus induced by simple bulk diffeomorphisms.\footnote{The covariant Bondi gauge discussed here is a three-dimensional tayloring of the so-called derivative expansion, introduced in fluid/gravity correspondence \cite{Bhattacharyya:2007vjd, Haack:2008cp, Hubeny:2011hd, Ciambelli:2018xat}, where Weyl covariance was a constructive motive.} The curvature of the Weyl connection is
\begin{equation}
    F_{\mu\nu} = \partial_\mu A_\nu - \partial_\nu A_\mu = \frac{1}{k^2} \left(\partial_\mu \Theta^\ast \ast\! u_\nu - \partial_\nu \Theta^\ast \ast\! u_\mu - \partial_\mu \Theta\, u_\nu + \partial_\nu \Theta \, u_\mu \right), \label{WeylCurvature}
\end{equation}
while its Hodge dual reads
\begin{equation}
    F = \frac{1}{2}\, \cE^{\mu\nu} F_{\mu\nu} = \frac{1}{k^2} \left(u^\mu \partial_\mu \Theta^\ast - \ast u^\mu \partial_\mu \Theta \right) . \label{WeylF}
\end{equation}

The metric \eqref{DE}  solves Einstein's equations only if the six independent boundary functions in $(\text{u},\ast\text{u})$ and $\ve$, $\chi$ satisfy suitable differential equations. A relevant quantity for the description of the equations of motion is the Brown--York energy--momentum tensor \cite{Brown:1992br, deHaro:2000vlm}. It was shown in \cite{Campoleoni:2018ltl} that for the line-element \eqref{DE} it reads
\begin{equation}
     T_{\mu\nu} = \frac{1}{2k} \left(\widetilde{T}_{\mu\nu} + \widehat{T}_{\mu\nu} \right) ,
\end{equation}
where
\begin{subequations}
\begin{align}
\widetilde{\text{T}} & =\frac{\varepsilon}{k^2}\, (\text{u}^2+\ast\text{u}^2) + \frac{\chi}{k^2} \, \left(\text{u}\ast\!\text{u}+\ast\text{u}\,\text{u} \right)+ \frac{R}{8\pi \mathcal{G}k^2}\ast\!\text{u}^2 \, , \\[5pt]
\widehat{\text{T}} & = \frac{1}{8\pi \GN k^4} \left(u^\mu\partial_\mu\Theta+ \ast u^\mu \partial_\mu\Theta^\ast-\frac{k^2}{2}R\right) 
(\text{u}^2+\ast\text{u}^2) - \frac{1}{4\pi \GN k^4} \ast\! u^\mu\partial_\mu\Theta\, \left(\text{u}\ast\!\text{u}+\ast\text{u}\,\text{u} \right) ,
\end{align}
\end{subequations}
and where $R$ stands for the Ricci scalar of the boundary metric. 
Both contributions are tensors decomposed in the basis $(\text{u},\ast\text{u})$. The first  clarifies the role of $\varepsilon$ and $\chi$ as the energy density and flow on the boundary.\footnote{The boundary is two-dimensional, therefore the transverse momentum flow is carried by a scalar.}
Moreover, the viscous stress tensor (also a scalar in two dimensions) involves  the energy density and the boundary scalar curvature. The second contribution includes only the frame and derivatives thereof and induces corrections to the energy density and energy flow. 

The resulting Brown--York energy--momentum tensor is symmetric and Einstein's equations imply 
\begin{equation} \label{holographic-Ward_1}
    \nabla_{\!\mu} T^{\mu\nu} = 0 \, , \qquad
    T^\mu{}_\mu=\frac{R}{16\pi \mathcal{G}k} \, .
\end{equation}
These equations can be spelled in terms of $\varepsilon$, $\chi$, and the Cartan frame as
\begin{subequations}\label{bella1}
\begin{align}
u^\mu \left(\partial_\mu+2A_\mu\right) \, \ve &= -\ast\! u^\mu \left(\partial_\mu+2A_\mu\right) \left(\chi-\frac{F}{4\pi \mathcal{G}}\right) , \\
 u^\mu \left(\partial_\mu+2A_\mu\right) \, \chi &=- \ast\! u^\mu \left(\partial_\mu+2A_\mu \right) \ve \, .
\end{align}
\end{subequations}
Recalling that $u^\mu$ and $\ast u^\mu$ are timelike and spacelike, the first equation shows that the local energy density variation is dictated by the gradient of the energy flow and the Weyl curvature. The second equation tells us that the local energy flow variation is controlled by the gradient of energy density.  We notice that all derivatives are Weyl-covariant, and so are both  equations (see \cite{Campoleoni:2018ltl} for details).

\paragraph{Symmetries}
 
Diffeomorphisms preserving  the line element \eqref{DE} were already studied in \cite{Ciambelli:2020eba, Ciambelli:2020ftk}. They are generated by asymptotic Killing vectors depending on four parameters: two are gathered into the generator of boundary diffeomorphisms $\xi^\mu(x)$, one, $\sigma(x)$, induces Weyl rescalings of the boundary geometry, and the last one, $\eta(x)$, parameterizes local Lorentz boosts of the boundary frame, which is the main novelty offered by the covariant Bondi gauge.
Asymptotic Killing vectors are expressed  in closed form as follows:
\begin{equation}
    v=\left(\xi^\mu-\frac{1}{k^2 r}\, \eta \ast\! u ^\mu\right) \partial_\mu+\left(r \, \sigma+\frac{1}{k^2} \left(\ast u ^\nu \, \partial_\nu \eta+\Theta^\ast \eta\right) + \frac{4\pi\mathcal{G}}{k^2r}\, \chi\,\eta\right) \partial_r \, .
    \label{AKVGeneral}
\end{equation}
We observe that $\eta$ appears only at subleading orders in $r$, in terms depending explicitly on the solution  --- they involve $\chi$ and $\ast\text{u}$.

A bulk diffeomorphism generated by the vector \eqref{AKVGeneral} is reflected on the boundary Cartan frame in a simple manner:
\begin{subequations} \label{transfo dyad}
\begin{align}
    \delta_{({\xi},\sigma, \eta)} \textrm{u} & = \mathcal{L}_{\xi} \textrm{u} + \sigma\, \textrm{u} + \eta \ast\! \textrm{u} \, , \label{transfo-dyad_1} \\[5pt] 
    \delta_{(\xi, \sigma, \eta)}\! \ast \!\textrm{u} & = \mathcal{L}_{\xi} \ast\!\textrm{u} + \sigma \ast \! \textrm{u} + \eta\, \textrm{u} \, .
    \label{transfo-dyad_2}
\end{align}
\end{subequations}
As anticipated, $\xi$ generates  a boundary diffeomorphism, while the $\eta$-transformation is the infinitesimal version of the two-dimensional local Lorentz boost
\begin{equation}
    \begin{pmatrix}
    \text{u}' \\
    \ast\text{u}'
    \end{pmatrix}
    =
    \begin{pmatrix}
    \cosh{\eta} & \sinh{\eta}\\
    \sinh{\eta} & \cosh{\eta}
    \end{pmatrix}
    \begin{pmatrix}
    \text{u} \\
    \ast\text{u}
    \end{pmatrix}.
    \label{LorentzTransfo}
\end{equation}
Finally, from \eqref{transfo dyad} one obtains the variation of the boundary metric \eqref{Bmetric} as
\begin{equation}
    \delta_{(\xi, \sigma, \eta)} g_{\mu\nu}=\mathcal{L}_{\xi}g_{\mu\nu}+2 \, \sigma \, g_{\mu\nu} \, ,
    \label{transfo metric}
\end{equation}
showing that $\sigma$ acts as a Weyl rescaling.

Using the transformations of $\varepsilon$ and $\chi$ available in \cite{Ciambelli:2020eba, Ciambelli:2020ftk}, we note that the Brown--York energy--momentum tensor transforms as
\begin{equation}
    \delta_{(\xi, \sigma, \eta)} T_{\mu\nu}=\mathcal{L}_{\xi}T_{\mu\nu}+\frac{1}{16\pi\mathcal{G}k}\left(\mathcal{L}_{\partial\sigma} g_{\mu\nu}-(g^{\rho\lambda} \,\mathcal{L}_{\partial\sigma} g_{\rho\lambda}) \, g_{\mu\nu}\right),
    \label{transfo stress}
\end{equation}
where we have used the compact notation $\partial \sigma \equiv g^{\mu\nu} \partial_\nu \sigma \partial_\mu$. This quantity transforms like a tensor under diffeomorphisms, non-linearly under a Weyl transformation, and it is insensitive to Lorentz boosts.

It should be emphasized  that when working with the dyad instead of a boundary metric the Weyl and Lorentz transformations acquire a certain resemblance, see eq.~\eqref{transfo dyad}. This similarity becomes even more striking  when computing the algebra obtained using the following modified Lie bracket of two asymptotic Killing vectors \cite{Schwimmer:2008yh, Barnich:2010eb}:
\begin{equation} \label{modified-Lie}
    [v_1,v_2]_{\text{M}} \equiv [v_1,v_2]-\delta_{v_1}v_2+\delta_{v_2}v_1 \, ,
\end{equation}
where $[v_1,v_2]$ is the Lie bracket.
This modified bracket takes into account the dependence of the vectors on fields that transform themselves under the symmetry. We obtain the algebra
\begin{equation}
        \left[\left(\xi_1, \sigma_1, \eta_1\right),\left(\xi_2, \sigma_2, \eta_2\right)\right]_{\text{M}} = \left(\xi_{12},\sigma_{12},\eta_{12} \right),
\end{equation}
with
\begin{equation}
        \xi_{12}=[\xi_1,\xi_2] \, , \qquad
        \sigma_{12}=\xi_1^\mu \, \partial_\mu \sigma_2-\xi_2^\mu \, \partial_\mu \sigma_1 \, , \qquad
        \eta_{12}=\xi_1^\mu \, \partial_\mu \eta_2-\xi_2^\mu \, \partial_\mu \eta_1 \, . \label{symmetry algebra commutators}
\end{equation}
We observe that Weyl and Lorentz transformations form two Abelian sub-algebras while diffeomorphisms act on them producing the semi-direct  structure
\begin{equation} \label{symmetry-group}
    \mathrm{Diff}\loplus \left(\mathrm{Weyl}\oplus \mathfrak{so}(1,1)\right).
\end{equation}

Before moving to the study of the symplectic structure, we would like to further comment on the similarity between the Weyl and Lorentz sectors. Both symmetries are local and supplemented with gauge connections built out of the Cartan frame. The Weyl connection is $\text{A}$, given in eq.~\eqref{WeylConnection}, which transforms under Weyl infinitesimal rescalings ($\mathcal{B}\simeq 1-\sigma$) as
\begin{equation}
    \text{A}\to \text{A}+\text{d}\sigma \, .
\end{equation}
The Lorentz (or spin) connection is encoded in the Hodge dual of the Weyl connection, i.e.\ $\upomega^{10} = \ast \text{A}$ in the orthonormal Cartan frame $\uptheta^0 = \frac{\text{u}}{k}$ and $\uptheta^1 = \frac{\ast\text{u}}{k}$. Under an infinitesimal Lorentz boost it transforms as
\begin{equation}
    \ast\text{A}\to \ast\text{A}-\text{d}\eta \, ,
\end{equation}
where $\ast\text{A}=\frac{1}{k^2}(\Theta^\ast  \text{u}-\Theta \ast\!\text{u})$.
Since we are in two dimensions, the curvatures of these connections are encoded in their dual scalars:
\begin{equation}\label{sdA}
    \ast\D \text{A}=F \, , \qquad 
    \ast \D \ast\! \text{A}=\frac{R}{2} \, .
\end{equation}
This sort of duality between the two sectors will regularly appear in our analysis, and it is ultimately rooted in their algebraic properties.

\paragraph{Recovering the Bondi gauge} The Bondi gauge is reached by choosing a specific Cartan frame for the boundary metric. This is performed by imposing the constraint $u_\phi=0$. 
The dictionary between $u_0$, $\ast u_0$, $\ast u_\phi$,  $\varepsilon$ and $\chi$ and the Bondi data is given in \cite{Ciambelli:2020eba, Ciambelli:2020ftk}.  
The previous constraint is not preserved by the residual symmetries \eqref{symmetry-group}, imposing thereof a reduction of the symmetry group. From \eqref{transfo-dyad_1} we see that $u_\phi$ is left unchanged if Lorentz boost and diffeomorphisms are related as
\begin{equation}
    \eta =-\frac{u_0 \, \partial_\phi {\xi}^0}{\ast u_\phi} \, .
\end{equation}
The symmetry group reduces to $\mathrm{Diff}\ltimes \mathrm{Weyl}$.

\subsection{Symplectic structure and boundary anomalies}\label{sec:symplectic_AdS}

\paragraph{Presymplectic potential} We now get into the core of this work, which is the analysis of the symplectic structure associated with the covariant Bondi gauge \eqref{DE}.
Calling the bulk metric $G_{MN}$, and starting from the bulk Einstein--Hilbert Lagrangian $L_{\text{EH}}[G]$, one can extract the presymplectic potential $\Theta_{\text{EH}}[G;\delta G]$ through the relation \cite{Lee:1990nz, Iyer:1994ys}
\beq
\delta L_{\text{EH}}[G] = \frac{\delta L_{\text{EH}}[G]}{\delta G_{MN}} \, \delta G_{MN} + \text{d}\Theta_{\text{EH}}[G;\delta G] \, .
\eeq
It reads explicitly 
\begin{equation}
\Theta_{\text{EH}}[G;\delta G] = \frac{1}{16 \pi \GN} \left[ \nabla^N \delta G_{PN} \, G^{PM} - \nabla^M \delta G_{PN} \, G^{PN} \right] \Sigma_M \, ,
\label{ThetaEH}
\end{equation}
where $\Sigma_M=\frac{\sqrt{-G}}{2}\,\epsilon_{MNP}\D x^N\wedge \D x^P$. The presymplectic potential is defined up to two types of ambiguities
\begin{equation}
    \Theta_{\text{EH}}[G;\delta G] \to \Theta_{\text{EH}}[G;\delta G] + \delta Z[G] - \D Y[G;\delta G] \, ,
    \label{ambiguities}
\end{equation}
where $Z[G]$ and $Y[G;\delta G]$ are two- and one-forms on spacetime \cite{Lee:1990nz,Iyer:1994ys}. The $Y$-ambiguity stems from the fact that the variation of the Lagrangian defines only the exterior derivative of the potential. 
The $Z$-ambiguity results from a different choice of boundary term in the action.\footnote{We should quote here that arguments have been proposed in the literature to extract the presymplectic potential from the Lagrangian, fixing thereby both ambiguities --- see e.g. \cite{Papadimitriou:2005ii, Compere:2008us, Freidel:2020xyx, Compere:2020lrt, Fiorucci:2020xto, Freidel:2021cjp, Ciambelli:2021nmv, Freidel:2021dxw, Speranza:2022lxr}.}
In this section we focus directly on the presymplectic potential, while in section~\ref{sec:first-order} we associate a variational principle with it working in the Chern--Simons formulation of three-dimensional gravity.

The conformal boundary being at $r\to\infty$, we focus on the $r$-component of the Einstein--Hilbert presymplectic potential \eqref{ThetaEH} in this limit:
\begin{equation}
\Theta^{(r)}_{\text{EH}} [G; \delta G] = r^2 \, \Theta_{(2)} + r \, \Theta_{(1)} + \Theta_{(0)} + \mathcal{O}\left(r^{-1}\right) \, ,
\end{equation}
where, in terms of boundary data, we have
\begin{subequations}
\begin{align}
\Theta_{(2)} & = - \frac{k}{8\pi \GN}\Big( \delta \ln \sqrt{-g}\Big) \, \text{Vol}_{\partial \mathcal{M}} \, ,\\[5pt]
\Theta_{(1)} & = \frac{1}{16\pi \GN k}  \left[  -2\, \frac{\delta ( \Theta \sqrt{- g} )}{\sqrt{- g}}  -  \nabla_{\!\mu} \delta u^\mu   \right] \text{Vol}_{\partial \mathcal{M}} \, , \\[5pt]
\Theta_{(0)}
& = \left( \frac{1}{2}\,
T^{\mu\nu}\delta g_{\mu\nu} + \frac{1}{2k\sqrt{-g}}\, \delta (\sqrt{-g}\, \ve) + \frac{1}{16\pi \GN k^3\sqrt{-g}}\, \delta \left[\sqrt{-g} \left(\Theta^2 - \Theta^{\ast 2} \right)\right] \right. \nn \\
& \phantom{=} \left. - \frac{1}{16 \pi \GN k\sqrt{-g}}\, \delta (\sqrt{-g}R)+ \frac{1}{8\pi \GN k^3}\, \nabla_{\!\mu} (\delta\Theta\, u^\mu) - \frac{1}{16\pi G k^3}\, \nabla_{\!\mu} \left[\delta (\Theta^\ast \ast\! u^\mu)\right] \right) \text{Vol}_{\partial \mathcal{M}} \, ,
\end{align}
\end{subequations}
with $\text{Vol}_{\partial \mathcal{M}}=\frac{\sqrt{-g}}{2}\,\epsilon_{r\mu\nu}\D x^\mu \wedge \D x^\nu$ being the boundary volume form. We also recall that
$\Theta$ and $\Theta^\ast$ denote the expansion and its dual defined in \eqref{def-expansions}.

This potential diverges, but both divergent terms are pure-ambiguity, as in \eqref{ambiguities}.
We can thus define the renormalized presymplectic potential of the theory as
\begin{equation}
\Theta_{\text{ren}}[G;\delta G] =
\Theta^{(r)}_{\text{EH}} [G;\delta G] + \delta Z[G] -\D Y[G;\delta G] \, ,
\end{equation}
where $Z$ and $Y$ have their own large-$r$ expansions
\begin{equation} \label{exp-amb}
Z[G]= r^2 \, Z_{(2)} + r \, Z_{(1)} + Z_{(0)} \, ,
 \qquad Y[G;\delta G] = r \, Y_{(1)} + Y_{(0)} \, ,
\end{equation}
whose divergent pieces are
\begin{equation}
Z_{(2)}= \frac{k}{8\pi \GN}\,\text{Vol}_{\partial \mathcal{M}} \, ,\qquad Z_{(1)}= \frac{1}{8\pi \GN k}\, \Theta\, \text{Vol}_{\partial \mathcal{M}} \, , \qquad
Y_{(1)} = \frac{1}{16\pi \GN k}\,  \cE_{\mu\alpha}\delta u^\alpha \D x^\mu \, .
\end{equation}
At this stage, a choice is expected for the zeroth order of the ambiguities, that is for $Z_{(0)}$ and $Y_{(0)}$ in \eqref{exp-amb}. The condition we wish to satisfy ---~which is natural in holography~--- is that the corresponding renormalized potential vanishes for Dirichlet boundary conditions on the boundary geometry. In other words, $\Theta_{\text{ren}}$ should be a linear combination of $\delta\text{u}$ and $\delta\!\ast\!\text{u}$, which ensures that its boundary value vanishes when the Cartan frame is fixed. The two choices for the zeroth order ambiguity that we propose are 
\begin{description}
\item[Weyl] 
\begin{equation}
Y_{(0)}=-\frac{\cE_{\mu\alpha}}{8\pi\mathcal{G}k^3} \left( u^\alpha \delta\Theta-\frac{\delta(\ast u^\alpha \Theta^\ast)}{2} \right)\D x^\mu, \quad
        Z_{(0)}=\left(-\frac{\varepsilon}{2k} -\frac{\Theta^2-\Theta^{\ast2}+k^2R}{16\pi \mathcal{G}k^3}\right)\text{Vol}_{\partial \mathcal{M}} \, , \label{Choice_1} 
\end{equation}
\item[Lorentz] 
\begin{equation}
Y_{(0)} =\frac{\cE_{\mu\alpha}}{16\pi \GN k^3} \left(\delta\ast\! u^\alpha \, \Theta^\ast-\ast u^\alpha \, \delta \Theta^\ast\right) \D x^\mu,\quad 
        Z_{(0)} = -\frac{\varepsilon}{2k}\, \text{Vol}_{\partial \mathcal{M}} \, ,
    \label{Choice_2}
\end{equation}
\end{description}
where we dubbed each choice according to which residual symmetry becomes an asymptotic symmetry. This will become clear when we study the associated surface charges in section~\ref{sec:conformal-gauge}. 

The key properties of the two choices in \eqref{Choice_1} and \eqref{Choice_2} are the following:
\begin{description}
\item[Weyl] 
This renormalized presymplectic potential  matches that obtained in the Fefferman--Graham gauge with the prescription for the boundary terms of \cite{deHaro:2000vlm, Skenderis:2002wp}:
\begin{equation}
    \left. \Theta_{\text{ren}}^{\mathrm{W}}[G;\delta G]\right\vert_{\partial \cM} =  \frac{1}{2}\,
T^{\mu\nu}\delta g_{\mu\nu} \text{Vol}_{\partial \mathcal{M}}=\frac{1}{k^2}\,
T^{\mu\nu}\big(-u_\mu \, \delta u_\nu+\ast u_\mu \, \delta \ast\! u_\nu\big) \text{Vol}_{\partial \mathcal{M}} \, .
\label{ThetaFG}
\end{equation}
As required, the presymplectic potential is proportional to $\delta \text{u}$ and  $\delta\ast\!\text{u}$ and therefore vanishes for Dirichlet boundary conditions on the Cartan frame. But it vanishes also for milder boundary conditions, since it is proportional to the variation of the boundary metric, meaning that one has to fix the Cartan frame only up to Lorentz transformations.

Since this presymplectic potential is the same as the usual one obtained in the Fefferman--Graham gauge, several results follow through.
In particular, one can compute the surface charges associated with each residual symmetry in \eqref{AKVGeneral} using the presymplectic current $\delta \Theta^{\text{W}}_{\text{ren}}$. The charges are the same as those obtained in \cite{Alessio:2020ioh, Fiorucci:2020xto} and the boundary Weyl anomaly is shown to lead to non-trivial but non-conserved Weyl charges. No charges are instead associated with Lorentz boosts, because $\delta_\eta T^{\mu\nu}=\delta_\eta g_{\mu\nu}=0$ as shown in \eqref{transfo metric} and \eqref{transfo stress}.

\item[Lorentz] The prescription, on which we focus in the following, leads to the renormalized presymplectic potential
\begin{equation}
    \left. \Theta_{\text{ren}}^{\mathrm{L}}[G;\delta G]\right\vert_{\partial \cM} = \big(J^\mu \, \delta u_\mu + J^\mu_\ast \, \delta \ast\! u_\mu\big) \text{Vol}_{\partial \mathcal{M}} \, ,
    \label{LorentzPotential}
\end{equation}
where we have introduced the currents
\begin{subequations}\label{currents}
\begin{align}
    J^\mu &= -\frac{1}{k^2}\, T^{\mu \nu} u_\nu  + \frac{1}{16 \pi \mathcal{G} k^5}\, u^\mu (\Theta^2 - {\Theta^\ast}^2) -\frac{1}{8 \pi \mathcal{G} k^3}\, \cE^{\mu\nu} \partial_\nu \Theta^\ast \, ,   \\[5pt]
    J^\mu_\ast &= \frac{1}{k^2}\, T^{\mu \nu} \!\ast\! u_\nu  - \frac{1}{16 \pi \mathcal{G} k^5} \ast\! u^\mu (\Theta^2 - {\Theta^\ast}^2) + \frac{1}{8 \pi \mathcal{G} k^3}\, \cE^{\mu\nu} \partial_\nu \Theta \, .
\end{align}
\end{subequations}
The presymplectic potential is again proportional to $\delta\text{u}$ and $\delta\ast\!\text{u}$ and therefore vanishes for Dirichlet boundary conditions on the Cartan frame. Similarly to the previous case, a milder condition can be imposed to get stationarity of the action: this time it is the conformal class of the Cartan frame that needs to be fixed, as will become clear in the remainder of this section. Most importantly, and this will be the subject of section~\ref{sec:Flat}, this potential admits a flat limit that will naturally endow the asymptotically flat covariant Bondi gauge with a symplectic structure.
\end{description}

\paragraph{Holographic interpretation of the Lorentz presymplectic potential}

The prescription \eqref{Choice_1} must be associated with suitable boundary terms to be added to the bulk action in order to reproduce the given presymplectic potential. We postpone the discussion of boundary terms in the action to section~\ref{sec:first-order}, where the issue will be approached within the Chern--Simons formulation. Here we simply assume that such an action (with specific boundary contributions) exists and denote it as $S_{\mathrm{L}}$. By definition, its on-shell variation corresponds to the boundary integral of the pull-back of $\Theta_{\text{ren}}^{\mathrm{L}}$ to the boundary defined in \eqref{LorentzPotential}, that is to
\begin{equation} \label{varS_Lorentz}
    \delta S_{\mathrm{L}}=\int \left(J^\mu \, \delta u_\mu + J^\mu_\ast \, \delta\! \ast\! u_\mu \right) \text{Vol}_{\partial \mathcal{M}} \, .
\end{equation} 
This variation has the familiar form (vev)$\,\times\,\delta$(source).
The two currents $(\text{J}, \text{J}_\ast)$ can be packaged into a energy--momentum tensor
\begin{equation} \label{stress-tensor_Lorentz}
    {\mathscr{T}^\mu}_\nu= J^\mu u_\nu + J^\mu_\ast \ast\! u_\nu \\
    = {T^\mu}_\nu  - \frac{1}{16 \pi \mathcal{G} k^3}\, \delta^\mu_\nu \Big(\Theta^2 - {\Theta^\ast}^2 \Big) -\frac{1}{8 \pi \mathcal{G} k^3}\, \cE^{\mu\rho} \Big( \partial_\rho \Theta^\ast u_\nu - \partial_\rho \Theta \ast \! u_\nu \Big) \, ,
\end{equation}
which should not be confused with the Brown--York energy--momentum tensor $T^{\mu\nu}$. When Einstein's equations are satisfied, this new energy--momentum tensor satisfies the identities
\begin{equation}
    \nabla_{\!\mu} \mathscr{T}^{\mu\nu} = - \frac{1}{8\pi \mathcal G k}\, F^{\mu \nu} A_\mu \, , \qquad \mathscr{T}^\mu{}_\mu = 0 \, , \qquad 
    \mathscr{T}_{[\mu\nu]} = \frac{1}{16\pi \mathcal{G}k}\, F_{\mu\nu} \, ,
    \label{Ward Identity}
\end{equation}
where $A_\m$ is the Weyl connection \eqref{WeylConnection} and $F_{\m\n}$ its curvature \eqref{WeylCurvature}. 
These relations, interpreted as holographic Ward identities, signal the presence of an anomaly in the Lorentz symmetry in the dual theory (see, e.g., section 12.5 of \cite{book:16319}).

When evaluated on a Weyl transformation of the dyad, the variation of the on-shell action \eqref{varS_Lorentz} is proportional to the trace of the energy--momentum tensor introduced in \eqref{stress-tensor_Lorentz}:
\begin{equation}
    \delta_{\sigma} S_{\mathrm{L}} = \int \sigma\, {\mathscr{T}^\mu}_\mu    \text{Vol}_{\partial \mathcal{M}}=0 \, .
\end{equation}
Equation \eqref{Ward Identity} thus implies that  it is sufficient to fix the conformal class of the boundary dyad in order
to impose Dirichlet boundary conditions, as anticipated. Considering the variation of the on-shell action under a general residual symmetry generated by the asymptotic Killing vector \eqref{AKVGeneral}, we obtain:
\begin{equation} \label{LorentzAnomaly}
\begin{split}
    \delta_{(\xi, \sigma, \eta)} S_{\mathrm{L}} &= \int \left[- \xi^\nu \left(\nabla_{\!\mu} {\mathscr{T}^\mu}_\nu + \frac{1}{8\pi \mathcal G k} F_{\mu \nu} A^\mu \right)      +\sigma {\mathscr{T}^\mu}_\mu+\eta\, \cE^{\mu\nu} \mathscr{T}_{[\mu\nu]}  \right]  \text{Vol}_{\partial \mathcal{M}} \\ & = \frac{1}{8\pi \mathcal{G}k} \int \eta \, F \,  \text{Vol}_{\partial \mathcal{M}} \, .
\end{split}
\end{equation}
Consistently with the holographic Ward identities \eqref{Ward Identity}, we conclude that  the only anomalous symmetry is Lorentz, with an anomaly weighted by the Weyl curvature, as opposed to the standard setup, in which the Weyl anomaly is weighted by the curvature of the boundary metric. As we shall discuss shortly, the displacement of the anomaly from the Weyl to the Lorentz symmetry can be seen as specifying two different representatives in the same cohomology class for the boundary theory. Indeed, the presence of an anomaly is a cohomological statement: the true information is encoded in the presence ---~and value~--- of central charges, which can appear in different places.

\paragraph{Holographic anomaly}

We now discuss the cohomological equivalence between the holographic anomalies induced by $\Theta^{\text{W}}$ defined in \eqref{ThetaFG} and $\Theta^{\text{L}}$ defined in \eqref{LorentzPotential}. As it will be shown in detail in a forthcoming publication \cite{preprint_Luca} and as resumed in appendix~\ref{app:BRST} for the sake of clarity, a classification of the anomaly structure of field theories possessing a Weyl--Lorentz frame bundle symmetry gives three inequivalent contributions:\footnote{The numerical factor in front of $c_3$ originates from the relationship $T^\mu{}_\mu=\frac{c_3}{24\pi}R$, see \cite{Henningson:1998gx}. Note also that we denote by $a^{(1,2)}$ the object involving the Hodge dual of the curvature two-forms appearing in \eqref{BRST-anomaly}.}
\begin{equation}
a^{(1,2)}= \frac{c_1}{2}\, \eta\, R + c_2 \, \sigma F + \frac{c_3}{24\pi} \, \sigma R \, . \label{LW}
\end{equation}
In this expression, $\{c_1,c_2,c_3\}$ are three central charges, and $\eta$ and $\sigma$ are the Lorentz and Weyl symmetry generators, see appendix~\ref{app:BRST}. 
We observe that $c_1$ is a pure Lorentz central charge (it corresponds to an anomaly involving the Lorentz symmetry generator and curvature), $c_2$ a pure Weyl central charge (Weyl symmetry generator and curvature), while $c_3$ is a mixed central charge (Weyl symmetry generator and Lorentz curvature).
The mixed term $c_3 \sigma R$ is proven in appendix~\ref{app:BRST} to be cohomologically equivalent to $2c_3 \eta F$ so that 
\begin{equation}
\frac{c_1}{2} \, \eta \, R + c_2 \, \sigma F + \frac{c_3}{24\pi} \, \sigma R \sim \frac{c_1}{2} \,  \eta \, R + c_2 \, \sigma F + \frac{c_3}{12\pi} \, \eta \, F \, . \label{coho}
\end{equation}
This means that  $\sigma R$ and $2\,\eta F$ source the same central charge, called here $c_3$.

We can now read off the central charge sourced in our holographic setup. From the choice \eqref{ThetaFG} for the presymplectic potential we get
\begin{equation}
    \delta_{(\xi,\sigma,\eta)}S_{\mathrm{W}}=\frac{1}{16\pi \mathcal{G}k}\int \sigma R\;\text{Vol}_{\partial \mathcal{M}} \, ,
\end{equation}
while the choice \eqref{LorentzPotential} gives
\begin{equation}\label{dSL}
    \delta_{(\xi,\sigma,\eta)}S_{\mathrm{L}}=\frac{1}{8\pi \mathcal{G}k} \int \eta\, F\;\text{Vol}_{\partial \mathcal{M}} \, ,
\end{equation}
where we stressed once again that our two prescriptions for the presymplectic potential must correspond to suitable choices of boundary terms to be added to the bulk action (see also section~\ref{sec:first-order}).
Comparing these results with \eqref{LW}, and taking into account \eqref{coho}, we obtain that the holographic central charges are, in both instances,
\begin{equation}
    c_1=0 \, , \qquad c_2=0 \, , \qquad c_3=\frac{3}{2 k \GN} \, .
\end{equation}
The central charge $c_3$ is the well-known Brown--Henneaux central charge \cite{Brown:1986nw}, related to the Weyl anomaly with holographic methods in \cite{Henningson:1998gx}. We conclude that our displacement of the anomaly from the Weyl to the Lorentz symmetry corresponds to moving from one representative to the other in the cohomological class. The two corresponding presymplectic potentials are related by a $(Z,Y)$-ambiguity \eqref{ambiguities} while their anomalies are related by a BRST  co-boundary term. It would be interesting to prove this interplay between bulk ambiguities and different representatives on the boundary in a wider class of theories. In section \ref{sec:Flat}, starting from the prescription displaying the central charge in the Lorentz sector, we shall perform the flat limit $k\to 0$, and reach a concrete holographic prediction for asymptotically flat gravity, where the boundary Carroll boost symmetry turns out to be anomalous.

\paragraph{Presymplectic form}

Considering the field variation of the presymplectic potential gives the renormalized presymplectic two-form on the boundary phase space, namely $\omega_{\text{ren}}=\delta \Theta_{\text{ren}}$. This yields
\begin{equation}\label{good presymplectic current}
\begin{split}
  \left. \omega_{\text{ren}}^{\mathrm{L}} \right\vert_{\partial {\cal \cM}} & =  \frac{1}{\sqrt{-g}}\Big(\delta( \sqrt{-g} \, J^\mu)\wedge \delta u_\mu+\delta( \sqrt{-g} \, J^\mu_\ast)\wedge \delta \!\ast\! u_\mu\Big)\text{Vol}_{\partial \mathcal{M}}\\
  & =  \left. \omega_{\text{ren}}^{\mathrm{W}} \right\vert_{\partial \cM} + \frac{1}{8\pi \GN k^3}\, \nabla_{\!\mu} \left[ \frac{\delta (\sqrt{-g}\, u^\mu)}{\sqrt{-g}}\wedge \delta \Theta - \frac{\delta (\sqrt{-g} \ast\! u^\mu )}{\sqrt{-g}}\wedge\delta  \Theta^\ast\right]   {\text{Vol}}_{\partial \mathcal{M}} \, ,
\end{split}
\end{equation}
where
\begin{equation}
    \left. \omega_{\text{ren}}^{\mathrm{W}}\right\vert_{\partial {\cal \cM}} =\frac{1}{2\sqrt{-g}}\, \delta \big(\sqrt{-g}\, T^{\mu\nu}\big)\wedge  \delta g_{\mu\nu}\text{Vol}_{\partial \mathcal{M}}
\end{equation}
is the presymplectic two-form associated with $\Theta_{\text{ren}}^{\mathrm{W}}$.
With the prescription for the ambiguities leading to $\omega_{\text{ren}}^{\mathrm{W}}$ the conjugate variables are thus the boundary metric and the Brown--York energy--momentum tensor, while with that leading to $\omega_{\text{ren}}^{\mathrm{L}}$ the conjugate variables are $(\text{u},\text{J})$ and $(\ast\text{u},\text{J}_\ast)$. We also observe that the presymplectic forms associated with our two prescriptions are related by an exact term. The latter corresponds to the $\delta$-variation of the difference between the two $Y_{(0)}$-ambiguities associated with the two choices in \eqref{Choice_1} and \eqref{Choice_2}. Furthermore, this exact term is anti-symmetric under the exchange $\text{u}\leftrightarrow\ast\text{u}$. As we shall see in section~\ref{sec:Flat}, it is this corner term that renders the presymplectic form finite in the $k\to 0$ limit, while in section~\ref{sec:first-order} we shall show that it arises naturally in the Chern--Simons formulation, which further justifies this heuristic prescription. With the presymplectic form at hand we can now compute the surface charges associated with the residual symmetries generated by the asymptotic Killing vectors \eqref{AKVGeneral}. For simplicity, we shall do this in the next section choosing the conformal gauge for the boundary metric. This will make the discussion on conservation/integrability more tractable and the interpretation of the holographic anomaly more transparent.

\subsection{Conformal gauge}
\label{sec:conformal-gauge}

The surface charges associated with the presymplectic current are generically non-integrable and non-conserved. While these features are common in the presence of gravitational radiation (see e.g.~\cite{Iyer:1994ys, Wald:1999wa, Barnich:2011mi}) they  might look awkward in a context without propagating degrees of freedom, as is three-dimensional gravity. Actually, the non-conservation can be interpreted holographically in three bulk dimensions as an anomalous Ward identity for the boundary Weyl symmetry \cite{Alessio:2020ioh}. In the present section, we make use of the new symplectic form $\omega_{\text{ren}}^{\mathrm{L}}$ to compute the charges after imposing the conformal gauge on the boundary metric, which makes their integrability manifest.

Since any two-dimensional metric is conformally flat, one can always choose a gauge in which the boundary metric reads
\begin{equation}
\D s^2=\text{e}^{2\varphi}\D x^+ \D x^- \, ,
\label{boundary gauge fixing}
\end{equation} 
where $\varphi (x^+, x^-)$ is an arbitrary function and $x^\pm = \phi \pm k \, u$ are light-cone coordinates. We choose the following parameterization for the corresponding Cartan frame $(\text{u},\ast \text{u})$:
\begin{equation} \label{velocity bgf}
\text{u}=-\frac{k}{2}\,\text{e}^{\varphi}\Big(\text{e}^{\zeta} \, \D 
x^+-\text{e}^{-\zeta} \, \D x^-\Big) \, , \qquad
\ast \text{u}=\frac{k}{2}\,\text{e}^{\varphi}\Big(\text{e}^{\zeta} \, \D x^++\text{e}^{-\zeta} \, \D x^-\Big) \, , 
\end{equation}
where $\zeta(x^+,x^-)$ is also an arbitrary boundary function. On the one hand, although this parameterization is always achievable locally, it could affect the global properties of the solution space. On the other hand, in this case it still allows to discuss all qualitative features of the asymptotic symmetries for the Lorentz presymplectic current \eqref{good presymplectic current}. 

According to eq.~\eqref{transfo metric}, demanding that the asymptotic Killing vectors \eqref{AKVGeneral} preserve the conformal gauge reduces the $\mathrm{Diff}$ symmetry to the set of conformal transformations. The latter is parameterized infinitesimally by a holomorphic and an anti-holomorphic function $Y^\pm(x^\pm)$. The symmetry parameters of the asymptotic Killing vector \eqref{AKVGeneral} become
\begin{subequations}\label{AKV conformal gauge}
\begin{align}
    &\xi^\mu\partial_\mu =Y^+\partial_+ + Y^-\partial_- , \\[4pt]
    &\eta = -\, h + \frac{1}{2} \left( \partial_+ Y^+ - \partial_- Y^- \right) +
   Y^+ \partial_+ \zeta + Y^- \partial_- \zeta  \, , \\
   &\sigma =  \varpi - \frac{1}{2} \left(\partial_+ Y^+ + \partial_- Y^-\right) - Y^+ \partial_+ \varphi - Y^- \partial_- \varphi \, ,
\end{align}
\end{subequations}
where we have purposefully implemented a field-dependent shift for the Lorentz and Weyl symmetry transformations and defined $h(x^+,x^-)$ and $\varpi(x^+,x^-)$ respectively.\footnote{Comparing with \cite{Ciambelli:2020eba, Ciambelli:2020ftk}, our new symmetry parameter $h$ is related to the old one through $h=\frac{2}{k}\text{e}^{-\varphi-\zeta}H_+-\varpi$.} This shift is such that the transformations of the Weyl factor $\varphi$ and Lorentz factor $\zeta$ in \eqref{velocity bgf} become very simple. Indeed, under \eqref{AKV conformal gauge} we obtain
\begin{equation}
       \delta_{(\xi, \varpi, h)} \varphi = \varpi \, , \qquad \delta_{(\xi, \varpi, h)} \zeta =h \, .
 \label{transfo zeta}
\end{equation}
Both $\varphi$ and $\zeta$ are shifted under the gauge transformation and, as we shall see, acting on global AdS$_3$ with the corresponding symmetry does not change its energy. For these reasons we refer to them as Goldstone modes for the Lorentz and Weyl symmetry. 
It turns out that this choice of parameters will render the charges integrable. This type of field-dependent redefinitions of the parameters have been exploited in the literature to find integrable slicings for the charges, see e.g.~\cite{Adami:2020ugu, Alessio:2020ioh, Adami:2020amw, Ruzziconi:2020wrb, Geiller:2021vpg, Adami:2021nnf}. 

In the conformal gauge, we can solve the conservation equations for the Brown--York energy--momentum tensor, whose components are given by
\begin{equation}
    T_{+-}=-\frac{1}{8\pi \mathcal{G}k}\, \partial_+\partial_-\varphi \, , \qquad T_{\pm\pm}=\frac{1}{8\pi \mathcal{G}k} \Big(\ell_\pm(x^\pm)+\partial_\pm^2\varphi-(\partial_\pm\varphi)^2\Big) \, ,
\end{equation}
where $\ell_\pm$ are defined from $\varepsilon$ and $\chi$ via
\begin{subequations} \label{ConformalStress}
\begin{align}
\varepsilon+\chi &= \frac{\text{e}^{-2(\varphi-\zeta)}}{2 \pi  \GN} \Big(\ell_-(x^-)
   -
   (\pa_-\zeta)^2-\pa_-^2 \zeta -\text{e}^{-2\zeta}\pa_-\pa_+\zeta
   \Big) \, , \\
   \varepsilon-\chi &= \frac{\text{e}^{-2( \varphi+\zeta)}}{2 \pi  \GN} \Big(\ell_+(x^+)
   -
   (\pa_+\zeta)^2+\pa_+^2 \zeta +\text{e}^{2\zeta}\pa_-\pa_+\zeta
   \Big) \, .
\end{align}
\end{subequations}
Using again the results in \cite{Ciambelli:2020eba, Ciambelli:2020ftk}, we compute the variation of the new fields $\ell_\pm$,
\begin{equation}
    \delta_{(\xi, \varpi, h)} \ell_\pm=Y^\pm\partial_\pm \ell_{\pm}+2 \, \partial_\pm Y^\pm \ell_\pm-\frac{1}{2} \, \partial_\pm^3Y^\pm \, .
    \label{transfo ell}
\end{equation} 
Starting from the modified Lie bracket \eqref{modified-Lie}, one can deduce that the algebra of residual symmetries is given by $(\text{Witt}\oplus\overline{\text{Witt}}) \oplus \text{Weyl} \oplus \mathfrak{so}(1,1)$, that is,
\begin{equation}
        \left[\left(Y^\pm_1, \varpi_1, h_1\right),\left(Y^\pm_2, \varpi_2, h_2\right)\right]_{\text{M}} =\left(Y^\pm_{12},\varpi_{12},h_{12}\right),
\end{equation}
with
\begin{equation} \label{conformal-gauge_algebra}
{Y^\pm_{12}} = {Y^\pm_2} \partial_\pm {Y^\pm_1} - {Y^\pm_1} \partial_\pm {Y^\pm_2} \, , \qquad
\varpi_{12} = 0 \, , \qquad
h_{12} = 0 \, .
\end{equation}

When choosing the conformal gauge for the boundary metric, the presymplectic potential and the presymplectic current \eqref{good presymplectic current} read
\begin{equation}
\Theta_{\text{ren}}^{\text{L}}\vert_{\pa {\cal M}}=\frac{\text{e}^{-2\varphi}}{8\pi \GN k}\, \zeta \, \delta \left(\text{e}^{2\varphi}\,F\right) \, \text{Vol}_{\partial \mathcal{M}} \, , \qquad
\omega_{\text{ren}}^{\text{L}}\vert_{\pa {\cal M}}=\frac{\text{e}^{-2\varphi}}{8\pi \GN k} \left[\delta \zeta \wedge \delta \left(\text{e}^{2\varphi}\,F\right) \right] \text{Vol}_{\partial \mathcal{M}} \, ,
\label{Presymplectic Current Conformal Gauge AdS}
\end{equation}
with
\begin{equation}
F=-4\,\text{e}^{-2\varphi}\partial_+\partial_-\zeta \, .
\end{equation}
We observe that the Lorentz Goldstone mode is the conjugate variable of the Weyl curvature \eqref{WeylF}. The charges can be computed using the codimension-two form $K_{(\xi, \varpi, h)}$ \cite{Iyer:1994ys, Wald:1999wa} satisfying 
\begin{equation}
\omega_{\text{ren}}[G ;\delta G, \delta_{(\xi, \varpi, h)} G]\vert_{\partial \cM} = \D K_{(\xi, \varpi, h)}[G; \delta G]\vert_{\partial \cM} \, .
\end{equation}
Assuming that the parameters $Y^\pm$, $\varpi$ and $h$ are field-independent, we find that the charges are integrable, that is, $K_{(\xi, \varpi, h)}=\delta H_{(\xi, \varpi, h)}$ and
\begin{equation}
    Q_{(\xi, \varpi, h)}=\int_0^{2\pi}H_{(\xi, \varpi, h)} \, .
\end{equation}
The Virasoro charges are given by the usual Brown--Henneaux expression
\begin{equation}
    Q_{Y^\pm}=\frac{1}{8\pi \mathcal{G}k}\int_0^{2\pi}\text{d}\phi \, \Big( Y^+\ell_+ - Y^-\ell_- \Big) \, ,
    \label{Virasoro charges}
\end{equation} 
while the charges associated with the Weyl and Lorentz symmetries read:
\begin{equation}
Q_{(\varpi,h)}= \frac{1}{8\pi \GN k}\int_0^{2\pi}\D \phi \, \Big( h \, (\partial_-  -\pa_+) \, \zeta - \zeta \, ( \partial_--\pa_+) \, h  \Big) \, .
\label{Weyl and Local Lorentz Charges}
\end{equation}
One thus finds that the charges associated with pure Weyl symmetries vanish identically. This implies that the Weyl symmetry is pure-gauge. 

Using the variations \eqref{transfo zeta} and \eqref{transfo ell}, one can compute the charge algebra. It is a representation of the asymptotic symmetry algebra, up to central extensions:
\begin{equation}
    \delta_{(\xi_1, h_1)} Q_{(\xi_2, h_2)} = Q_{[(\xi_1, h_1),(\xi_2, h_2)]_{\text{M}}} + \mathcal{K}_{(\xi_1, h_1),(\xi_2, h_2)} \, ,
\end{equation}
where we recall that $[\ ,\ ]_{\text{M}}$ stands for  the modified Lie bracket defined in \eqref{modified-Lie}.
The two copies of the $\text{Witt}$ algebra are promoted to a Virasoro algebra with central extension
\begin{equation}
    \mathcal{K}_{Y_1,Y_2}=\frac{1}{16\pi \mathcal{G}k}\int_0^{2\pi}\text{d}\phi \, \Big(Y_2^+\partial_+^3Y^+_1 - Y_2^-\partial_-^3Y^-_1 \Big) \, ,
\end{equation}
while the Abelian $\mathfrak{so}(1,1)$
algebra in \eqref{symmetry-group} is promoted to an affine algebra with central extension
\begin{equation}
    \mathcal{K}_{h_1,h_2}= \frac{1}{8\pi \GN k}\int_0^{2\pi}\D \phi \, \Big( h_2 \, (\partial_+  -\pa_-) \, h_1 - h_1 \, ( \partial_+-\pa_-) h_2 \, \Big) \, .
    \label{LorenzCentralCharge}
\end{equation} 
This concludes the study of the asymptotic charges associated with our new proposal of presymplectic potential \eqref{LorentzPotential} in conformal gauge. This prescription gives a physical meaning to the Lorentz factor $\zeta$, which has an associated non-vanishing, integrable and non-conserved charge. The Lorentz boosts form an affine $\mathfrak{so}(1,1)$ that is classically centrally extended with central extension given in \eqref{LorenzCentralCharge}.

\section{Flat limit and boundary Carroll frames} \label{sec:Flat}

Null gauges in AdS -- i.e.\ gauges whose constant-$r$ lines are null -- admit a proper flat-space limit \cite{Barnich:2012aw, Barnich:2013sxa, Ciambelli:2018wre, Poole:2018koa, Compere:2019bua,  Compere:2020lrt, Geiller:2022vto}. It is also the case for the covariant Bondi gauge as was shown in \cite{Campoleoni:2018ltl, Ciambelli:2020eba, Ciambelli:2020ftk}. This amounts to taking the large-$\ell$ (or small-$k$) limit, which turns the timelike AdS boundary into a null manifold, the null infinity of an asymptotically flat spacetime. The metric induced on the boundary becomes degenerate and therefore the geometry Carrollian. We will see that the flat-space symplectic structure mimics its  AdS counterpart, and in particular an ultra-relativistic version of the frame rotation will be charged. We also comment on anomalies for the putative holographic dual. 

\subsection{Asymptotically flat covariant Bondi gauge}

\paragraph{Solution space} Our goal is to take the flat limit of the two symplectic structures defined by \eqref{ThetaFG} and \eqref{LorentzPotential}. As we shall see only the second admits a proper flat limit. Following \cite{Campoleoni:2018ltl,Ciambelli:2020eba, Ciambelli:2020ftk}, we prescribe the following small-$k$ behavior for the various quantities appearing in the line element \eqref{DE}:
\begin{align}
\upmu = \lim_{k \to 0}  \frac{\text{u}}{k^2} \, ,\quad
\upmu^\ast = \lim_{k \to 0} \frac{\ast \text{u}}{k} \, , \quad 
\upsilon = \lim_{k \to 0} u \, , \quad
 \upsilon_\ast = \lim_{k \to 0} \frac{\ast u}{k} \, , \quad 
\alpha = \lim_{k\to 0} \frac{\chi}{k} \, ,\quad 
\epsilon = \lim_{k\to 0} \varepsilon \, , \label{falloffs in k}
\end{align}
where $(\text{u} = u_\m \D x^\m, \text{u} = \ast u_\m \D x^\m)$, while $(u = u^\m \partial_\m, \ast u = \ast u^\m \partial_\m)$. With this choice, the boundary metric becomes 
\begin{equation}
\D\ell^2 = \lim_{k\to 0} g_{\mu\nu} \, \D x^\mu \, \D x^\n = (\upmu^\ast)^2 \, ,
\end{equation}
which is manifestly degenerate. The degenerate direction corresponds to the span of the vector $\upsilon$, therefore the pair $(\upsilon,\D\ell^2 )$ forms a (weak) Carrollian structure \cite{Duval:2014lpa}.
The couple $(\upmu,\upmu^\ast)$ forms a dyad for this Carrollian structure while the couple of vectors $(\upsilon,\upsilon_\ast)$ corresponds to the dual basis:
\begin{equation}
    \upmu(\upsilon)=-1 \, , \quad 
    \upmu^\ast(\upsilon_\ast)=1 \, , \quad 
    \upmu(\upsilon_\ast)=0 \, , \quad\upmu^\ast(\upsilon)=0 \, .
\end{equation}
Notice that, due to the degeneracy of the metric, forms and vectors are not related to each other by lowering/raising indices, and this is why we have assigned different
symbols to them.
In analogy with the relativistic case, Carrollian expansions $\theta$ and $\theta^\ast$ can be defined via
\begin{equation} \label{Carrollian-expansions}
\text{d}\ast\!\upmu =\theta \ast \!  \upmu \wedge\upmu \, , 
\qquad
\text{d}\upmu = \theta^\ast \ast \! \upmu \wedge \upmu \, ,
\end{equation}
or, equivalently, via the Lie bracket of the Carrollian vectors $\upsilon$ and $\upsilon^\ast$,
\begin{equation}
\left[ \upsilon,\upsilon^\ast \right] = 
\theta^\ast 
\upsilon 
-\theta
\,\upsilon^\ast. \label{Carrollian-expansions_2}
\end{equation}
These are related to the relativistic expansions, defined in eq.~\eqref{def-expansions}, as
\begin{equation}
\theta = \lim_{k \to 0} \Theta \, , \qquad
\theta^\ast = \lim_{k \to 0} \frac{\Theta^\ast}{k} \, .
\end{equation}
With the scalings \eqref{falloffs in k}, the Weyl connection and its curvature also admit a flat limit
\begin{equation}
{\cal A} = \lim_{k \to 0} \text{A} = \upmu^\ast \, \theta^\ast-\upmu \, \theta \, , \qquad 
{\cal F}=\lim_{k\to 0} k \, F = \upsilon^\mu\partial_\mu\theta^\ast-\upsilon_\ast^\mu\partial_\mu\theta \, .
\label{FlatWeylConnection}
\end{equation}
The Carrollian connection $\mathcal{A}$ also transforms like a connection under Weyl rescalings of the Carrollian structure.

Using these ingredients, the flat limit of the line element \eqref{DE} is well-defined and reads
\begin{equation}
\D s^2_{\text{Flat}} = \lim_{k\to 0} \D s^2_{\text{AdS}} = 2\, \upmu \left(\D r+r{\cal A} \right) +r^2 \upmu^\ast\upmu^\ast +8\pi \GN \, \upmu \left(\epsilon \, \upmu + \alpha \, \upmu^\ast \right) .
\label{flat DE}
\end{equation}
For instance, the values of the various Carrollian quantities for the rotating cosmology of \cite{Barnich:2012aw} are
\begin{equation}
    \upmu=-\D u \, , \quad 
    \upmu^\ast=\D \phi \, , \quad \mathcal{A}=0 \, , \quad \epsilon=M,\quad \alpha=-J \, ,
\end{equation}
where $M$ and $J$ are the mass and the angular momentum of the cosmological solution. The one-form $\upmu$ is aligned with the retarded time while $\upmu^\ast$ is aligned with the angular direction. The metric $(\upmu^\ast)^2=\D \phi^2$ covers only spatial sections of null infinity.

Coming back to the general case, the zero-$k$ limit of the conservation equation $\nabla^\mu T_{\mu\nu} =0$, as written in \eqref{bella1}, is readily finite and gives the residual Ricci-flat Einstein equations for \eqref{flat DE}:
\begin{equation} \label{Ricciflat}
\upsilon^\mu(\pa_\mu +2 {\cal A}_\mu) \, \epsilon = \frac{1}{4\pi \GN}\, \upsilon_\ast^\mu (\pa_\mu +2 \, {\cal A}_\mu) \, {\cal F} \,, \qquad
\upsilon^\mu(\pa_\mu +2 \, {\cal A}_\mu) \, \alpha = -\upsilon_\ast^\mu(\pa_\mu +2 \, {\cal A}_\mu) \, \epsilon \,.
\end{equation}
These equations are Weyl-covariant because $\epsilon$ and $\alpha$ are scalars of Weyl-weight two. We refer to \cite{Campoleoni:2018ltl, Ciambelli:2020eba, Ciambelli:2020ftk} for more details on the Weyl--Carroll geometry induced on null infinity in the covariant Bondi gauge.

\paragraph{Symmetries} The asymptotic Killing vectors preserving the line element \eqref{flat DE} are obtained as the flat limit of the AdS Killing vectors \eqref{AKVGeneral}. They are parameterized by a boundary diffeomorphism and a boundary Weyl rescaling, $\xi^\mu(x)$ and $\sigma(x)$, which are the same as in AdS, and a boundary local Carroll boost
\begin{equation} \label{Carroll-boost}
\lambda(x) = \lim_{k\to 0} \frac{\eta(x)}{k} \, ,
\end{equation}
leading to
\begin{equation}
v=\left(\xi^\mu-\frac{1}{r}\, \lambda\, \upsilon_\ast^\mu \right) \partial_\mu + \left(r \, \sigma + \upsilon_\ast^\nu \partial_\nu \lambda +\theta^\ast \lambda + \frac{4\pi\mathcal{G}}{r}\, \alpha\,\lambda \right) \partial_r \, .
\end{equation}
Under the action of these asymptotic symmetries, the Carroll dyad $(\upmu, \upmu^\ast)$ transforms as
\begin{equation}
    \delta_{(\xi,\sigma, \lambda)} \upmu = \mathcal{L}_{\xi} \upmu + \sigma \, \upmu + \lambda \, \upmu^\ast, \qquad \delta_{(\xi, \sigma, \lambda)}  \upmu^\ast = \mathcal{L}_{\xi} \upmu^\ast + \sigma \, \upmu^\ast \, ,
    \label{transfo dyad flat}
\end{equation}
which justifies the names assigned to the symmetry parameters. We observe that $\upmu^\ast$ is not affected by the Carroll boost. This is the main difference with respect to the Lorentzian case, Carrollian frame rotations preserve the spatial metric $\D\ell^2=(\upmu^\ast)^2$ but  affect the temporal part $\upmu$ of the dyad. Finite Carroll boosts are coordinate transformations $t\to t+\vec{\lambda}\cdot \vec{x}$, where $\vec{\lambda}$ is a constant spatial vector. The infinitesimal version in two dimensions is $\delta t=\lambda\;\delta x$, which indeed implies $\delta_\lambda \upmu = \lambda  \upmu^\ast$.

We can also compute the variations of the Carrollian energy density and energy flow
\begin{subequations}
\begin{align} 
    \delta_{(\xi,\sigma, \lambda)}\epsilon &=\mathcal{L}_{\xi} \epsilon-2 \, \sigma \, \epsilon-\frac{1}{4\pi \mathcal{G}}\,\Big(\theta \, \upsilon_\ast^\mu \, \partial_\mu\lambda+\upsilon^\mu \, \partial_\mu(\upsilon_\ast^\nu \, \partial_\nu\lambda)-\mathcal{F} \, \lambda\Big)\,,\\
    \delta_{(\xi,\sigma, \lambda)}\alpha &=\mathcal{L}_{\xi} \alpha-2 \, \sigma \, \alpha-2 \, \lambda \, \epsilon+\frac{1}{4\pi \mathcal{G}}\, \Big(\theta^\ast \, \upsilon_\ast^\mu \,\partial_\mu\lambda+\upsilon_\ast^\mu \, \partial_\mu(\upsilon_\ast^\nu \,\partial_\nu\lambda)\Big) \, .
\end{align} 
\end{subequations}
These transformations contain a linear part, where we see that $\epsilon$ and $\alpha$ transform like scalars under diffeomorphisms and weight-two quantities under Weyl rescalings. The action of the Carroll boost is not symmetric and contains non-linear pieces.

\paragraph{Symplectic structure} We now perform the limit $k \to 0$ at the level of the presymplectic potential, using the same scalings in $k$ for the boundary data as in \eqref{falloffs in k}.
As stated earlier, the finiteness of this limit is a consequential  criterion for fixing the ambiguities in the definition of the presymplectic potential. 
The necessity to play with corner terms is reminiscent of the procedure applied in four-dimensional gravity to obtain a well-defined flat limit of the symplectic structure starting from the $\Lambda$-BMS phase space \cite{Compere:2020lrt} (see also \cite{Geiller:2022vto}). The method was later applied in three bulk dimensions to obtain a finite symplectic structure in Bondi gauge  \cite{Ruzziconi:2020wrb, Geiller:2021vpg}. Here, compared to the latter works, we must take care of the boundary Carroll boost symmetry, which is absent in the regular Bondi gauge, where the spatial part of $\upmu$ is required to vanish.

The flat limit of the renormalized presymplectic potential \eqref{LorentzPotential} reads
\begin{equation} \label{symplectic_flat}
        \Theta^{\mathrm{C}}_{\text{ren}} [G;\delta G]\vert_{\pa {\cal M}}=\lim_{k\to 0} \Theta_{\text{ren}}^{\mathrm{L}}[G;\delta G]\vert_{\pa {\cal M}} 
        = \big( j^\mu \, \delta \mu_\mu  +j^\mu_\ast \, \delta\mu_\mu^\ast \big)\text{vol}_{\partial \cM} \, ,
\end{equation}
where the superscript $\text{C}$ stands for ``Carrollian''. In this expression, we have introduced the components of the flat-space versions of the AdS holographic currents \eqref{currents}, which are the Carrollian vectors
\begin{subequations} \label{vol_flat}
\begin{align} 
        j &= \lim_{k \to 0} k^3 J = \frac{1}{2}\,   \epsilon\, \upsilon + \frac{1}{8\pi \GN}\, \upsilon_\ast \, \cF \, , \\
        j_\ast &= \lim_{k \to 0} k^2 J_\ast = \frac{1}{2}\,  \epsilon\, \upsilon_\ast + \frac{1}{2}\, \alpha\, \upsilon \, ,
    \end{align}
\end{subequations}
and defined
\begin{equation} \label{boundary-volume_flat}
\text{vol}_{\partial \cM}=
\lim_{k\to 0}\frac{\text{Vol}_{\partial \cM}}{k} \, ,
\end{equation}
as the volume form on null infinity.
Again, the potential vanishes when one imposes Dirichlet boundary conditions on the Carroll dyad $(\upmu, \upmu^\ast)$. We stress that the flat limit of the presymplectic potential renormalized as in \eqref{ThetaFG} diverges when employing the scalings \eqref{falloffs in k} and thus we work exclusively with the flat limit of \eqref{LorentzPotential} in the remaining of this section.

We can use the two currents to construct the Carrollian energy--momentum tensor 
\begin{equation}
    t^\mu{}_\nu =   j^\mu \, \mu_\nu + j^\mu_\ast \, \mu_\nu^\ast \, , 
\end{equation}
which can be equivalently defined as the limit of the relativistic energy--momentum tensor \eqref{stress-tensor_Lorentz} as
\begin{equation}
t^\mu{}_\nu = \lim_{k\to 0} k \mathscr{T}^\mu{}_\nu \, .
\end{equation}
The remaining Ricci-flat Einstein equations in the covariant Bondi gauge, i.e.\ eqs.~\eqref{Ricciflat}, imply that $t^{\mu}_{\hphantom{\mu}\nu}$ must satisfy the following Ward identities:
\begin{equation}
     D_\mu t^\mu{}_\nu = - \frac{1}{8\pi \mathcal G}\, \mathcal F_{\mu \nu} \, \mathcal A^\mu \, , \qquad 
     t^\mu{}_\mu = 0 \, , \qquad t^\mu{}_\nu \mu_\mu^\ast \, \upsilon^\nu = - \frac{\mathcal{F}}{8\pi \mathcal{G}} \, ,
\end{equation}
where
\begin{equation}
D_\mu t^\mu{}_\nu= \lim_{k\to 0} \nabla_{\!\mu} \mathscr{T}^\mu{}_\nu \, .
\end{equation}
These equations can also be obtained as the $k \to 0$ limit of the holographic Ward identities \eqref{Ward Identity}.

In terms of the energy--momentum tensor, the variation of the on-shell action (with appropriate boundary terms) becomes
\begin{equation} \label{flat-anomaly}
    \begin{split}
        \delta_{(v,\sigma, \lambda)} S_{\mathrm{C}} &= \int \Big[ -\xi^\nu (D_\mu \, {t^\mu}_\nu + \frac{1}{8\pi \mathcal G} \mathcal F_{\mu \nu} \, \mathcal A^\mu )  + \sigma \,  {t^\mu}_\mu - \lambda \, {t^\mu}_\nu \, \mu^\ast_\mu \, \upsilon^\nu \Big] \, \text{vol}_{\partial \cM}  = \int \Big( \lambda \, \frac{\mathcal F}{8\pi \mathcal{G}} \Big) \text{vol}_{\partial \cM} \, , 
    \end{split}
\end{equation} 
which reveals that the only anomalous symmetry is the Carroll boost, as was the Lorentz boost in~\eqref{dSL}.

The aftermath is engaging. If a holographic dictionary persists in the flat limit, i.e.\ a matching between the bulk action with prescribed boundary conditions and the partition function of a boundary theory, then the above result predicts an anomaly for the field theory dual.
The latter should be conformally coupled to a Carrollian dyad and is therefore an ultra-relativistic theory, naturally called Carrollian CFT. 
The flat background structure in this case is obtained by fixing the dyad to be $\upmu=-\D u$, $\upmu^\ast=\D \phi$.  
In this particular case, as we shall see in section~\ref{sec:flat-space-charges}, the asymptotic symmetry group includes the usual $\mathrm{BMS}_3$ group, which is in line with the fact that in three dimensions Carrollian Weyl-covariant theories  are $\mathrm{BMS}_3$-invariant. The Weyl anomaly was studied in \cite{Bagchi:2021gai} for such theories, while no results were provided for a possible Carroll boost anomaly. Our computation can be seen as a gravitational prediction for a boost anomaly in $\mathrm{BMS}_3$-invariant theories. 

Similarly, the presymplectic current \eqref{good presymplectic current} is finite in the limit $k \to 0$ and is given explicitly by
\begin{equation} \label{omega_ren_flat}
\omega^{\mathrm{C}}_{\text{ren}} \vert_{\pa {\cal M}}=\lim_{k\to 0} \omega^{\text{L}}_{\text{ren}}\vert_{\pa {\cal M}} 
    =\mathscr{D} ^{-1}\Big(\delta(\mathscr{D} \, j^\mu)\wedge \delta \mu^\ast_\mu+\delta(\mathscr{D} \, j^\mu_\ast)\wedge \delta \mu_\mu \Big)\text{vol}_{\partial \cM} \, ,
\end{equation}
where we have introduced the density
\begin{equation}
\mathscr{D} = \left|\varepsilon^{\mu\nu} \mu_\mu \mu^\star_\nu \right| = \lim_{k \to 0} \frac{\sqrt{-g}}{k} \, .
\end{equation}
The holographic currents are the conjugate variables of the dyad. This symplectic form will be used in the next section to compute charges in the conformal gauge.

\subsection{Flat-space charges in conformal gauge}\label{sec:flat-space-charges}

Our goal is here to go farther in the computation of charges in flat space. For that purpose we will extend the conformal gauge of section~\ref{sec:conformal-gauge} to the Carrollian framework. This  amounts to making a choice of coordinates $(u,\phi)$ such that the one-form $\upmu^\ast$ is aligned with the angular coordinate. This choice can also be recovered from the AdS conformal gauge imposing the scaling $\zeta=k \, \beta$ in \eqref{velocity bgf} and taking the $k$-to-zero limit.
We obtain the following expressions for the forms $\upmu$ and $\upmu^\ast$:
\begin{equation}
    \upmu=-\text{e}^{\varphi} \left( \text{d}u+ \beta \, \text{d}\phi \right) , \quad 
    \upmu^\ast=\text{e}^\varphi \, \text{d}\phi \, .
    \label{FlatMu}
\end{equation}
The asymptotic symmetry algebra is reduced to those vectors that preserve the conformal gauge. They are obtained by taking the flat limit of their AdS counterparts \eqref{AKV conformal gauge} with the following scalings for the symmetry parameters
\begin{equation}
    Y^\pm(x^\pm) = Y(\phi) \pm k \left( H(\phi) + u \, \partial_\phi Y(\phi) \right) , \quad
    h(x^+,x^-) = k \, \tilde{h}(u,\phi) \, ,
\end{equation}
while $\varpi$ is unaffected.
The variations of $\varphi$ and $\beta$ defined in \eqref{FlatMu} are
\begin{equation}
    \delta_{(H,Y,\varpi,\tilde{h})}\varphi=\varpi \, , \quad \delta_{(H,Y,\varpi,\tilde{h})}\beta=\tilde{h} \, ,
\end{equation}
where, in analogy with our discussion of the conformal gauge in AdS, we dub the conformal factor $\varphi$ and the field $\beta$ as Goldstone modes of the Weyl and Carroll boost symmetry, respectively. 
In this gauge one can completely solve the residual Ricci-flat equations \eqref{Ricciflat} for the energy density and the energy flow, obtaining
\begin{subequations} \label{FlatEqs}
\begin{align}
\epsilon &= \frac{\text{e}^{-2\varphi}}{8 \pi \GN} \Big(8 \pi \GN \, \varepsilon_0 - (\partial_u \beta)^2 + 2 \, \partial_u \partial_\phi \beta - 2 \, \beta \, \partial_u^2 \beta\Big) \, ,\label{FlatEpsilon}\\
\alpha &= \frac{\text{e}^{-2\varphi}}{4 \pi \GN} \Big(4 \pi \GN (\alpha_0 - u \, \partial_\phi \varepsilon_0 + 2 \, \beta \varepsilon_0) - \beta \left[(\partial_u \beta)^2 - 2 \, \partial_u \partial_\phi \beta + \beta \, \partial_u^2 \beta \right] - \partial_\phi^2 \beta + \partial_u \beta \, \partial_\phi \beta\Big) \, , \label{FlatAlpha}
\end{align}
\end{subequations}
where $\varepsilon_0 = \varepsilon_0(\phi)$ and $\alpha_0 = \alpha_0(\phi)$. These are linked to the $\ell_\pm$ fields of AdS by the following scalings:
\begin{equation}
\varepsilon_0(\phi) = \frac{1}{4 \pi \GN} \, \lim_{k\to0} \big( \ell_+ + \ell_- \big) \, , \qquad \alpha_0(\phi) - u \, \partial_\phi \varepsilon_0(\phi) = -\frac{1}{4 \pi \GN} \, \lim_{k\to0} \frac{\ell_+ - \ell_-}{k} \, .
\end{equation}
The infinitesimal variations of the on-shell boundary data read
\begin{subequations}
\label{FlatFieldVar}
\begin{align}
    \delta_{(H,Y,\varpi,\tilde{h})} \varepsilon_0 & = Y \, \partial_\phi \varepsilon_0 + 2 \, \varepsilon_0 \, \partial_\phi Y - \frac{1}{4 \pi \GN} \, \partial_\phi^3 \, Y \, , \\
    \delta_{(H,Y,\varpi,\tilde{h})} \alpha_0 & = Y \, \partial_\phi \alpha_0 + 2 \, \alpha_0 \, \partial_\phi Y - H \, \partial_\phi \varepsilon_0 - 2 \, \varepsilon_0 \, \partial_\phi H + \frac{1}{4 \pi \GN} \, \partial_\phi^3 \, H \, ,\\
    \delta_{(H,Y,\varpi,\tilde{h})} \varphi & = \varpi \, ,\\[4pt]
    \delta_{(H,Y,\varpi,\tilde{h})} \beta & = \tilde{h} \, .
\end{align}
\end{subequations}
Using these transformations, one shows that the residual-diffeomorphism algebra is a direct sum of three-dimensional BMS transformations, Weyl rescalings, and Carroll boosts, which are local $\mathbb{R}$-transformations:  $\mathrm{BMS}_3\oplus \mathrm{Weyl}\oplus \mathbb{R}$.\footnote{Akin to the AdS case, the field-dependent replacement $(\lambda,\sigma)\to(\tilde{h},\varpi)$ makes the sum be direct.}

We now have all ingredients for computing the asymptotic charges. The BMS charges read:
\begin{equation}
    Q_{(H,Y)}=\frac{1}{2} \int_0^{2\pi} \D \phi \, \Big(H \, \varepsilon_0 - Y \, \alpha_0\Big) \, ,
\end{equation}
as firstly obtained in \cite{Barnich:2006av}. The novelty comes from the new factors in our algebra of residual diffeomorphisms, i.e.\ the Carroll boosts. The charge associated with Weyl and boost transformations is integrable and reads
\begin{equation}
    Q_{(\varpi,\tilde{h})} = \frac{1}{8 \pi  G} \int_0^{2\pi} \text{d}\phi \, \Big(\partial_u \tilde{h} \, \beta - \tilde{h} \, \partial_u \beta\Big) \, .
\end{equation}
On the one hand, as in the anti-de Sitter framework, the Weyl parameter does not appear, meaning that Weyl rescalings are pure-gauge with this choice of symplectic structure. Therefore we can quotient them out. On the other hand, Carroll boosts are not pure-gauge and, since both the symmetry parameter $\tilde{h}$ and its Goldstone mode $\beta$ are unconstrained functions of $(u,\phi)$, the charge is manifestly not conserved. This is the avatar of the Carroll-boost  anomaly.

We can also compute the classical central charges of the resulting $\mathrm{BMS}_3\oplus \mathbb{R}$ asymptotic symmetry algebra. They come in couple, the first being the usual central extension of the $\mathrm{BMS}_3$ algebra
\begin{equation}
    \mathcal{K}_{(H_1,Y_1),(H_2,Y_2)} = \frac{1}{16
\pi \mathcal{G}} \int_0^{2\pi}  \text{d}\phi \, \Big(H_2 \, \partial_\phi^3 \, Y_1 + Y_2 \, \partial_\phi^3 \, H_1\Big) \, .
\end{equation}
The second is a central extension of the Abelian ultra-relativistic boosts at the boundary:
\begin{equation}
    \mathcal{K}_{\tilde{h}_1,\tilde{h}_2} = \frac{1}{8 \pi  \mathcal{G}} \int_0^{2\pi} \text{d}\phi \, \Big(\tilde{h}_2 \, \partial_u \tilde{h}_1 - \tilde{h}_1 \, \partial_u \tilde{h}_2\Big) \, .
\end{equation}
The latter has the same status as the Lorentz central extension in AdS, cf.~eq.~\eqref{LorenzCentralCharge}. The only difference is that the symmetry parameter $h$ is dimensionless in AdS, whereas the flat-space symmetry parameter $\tilde{h}$ is dimensionful. This is specific to flat space, where the only available scale is set by Newton's constant $\mathcal{G}$, which enters the central charges because it is combined with a dimensionful parameter.

\section{Conformal gauge in the Chern--Simons formulation}
\label{sec:first-order}

We now turn to the Chern--Simons formulation of three-dimensional Einstein's gravity \cite{Achucarro:1986uwr, Witten:1988hc}. After briefly reviewing it to fix conventions, we move to the computation of surface charges and their algebras for both asymptotically AdS and flat spacetimes, in the conformal gauge introduced in sections~\ref{sec:conformal-gauge} and \ref{sec:flat-space-charges}.  
An advantage of revisiting our previous discussion in this first-order formalism is that the  calculation of surface charges is more straightforward  than in the metric formulation: it does not require any sort of holographic renormalization because the dependence of the solutions on the null radial coordinate is encoded into a gauge transformation and it factors out from the charges at a very early stage. Moreover, there exists a simple choice for the Chern--Simons connection such that the computation of surface charges in AdS directly leads to those associated with the ``Lorentz'' presymplectic potential \eqref{LorentzPotential}, hence admitting a smooth flat limit. This gives further support to that choice for fixing the ambiguities appearing in the metric formulation. In the present section, we also clarify an issue that remained open up to now: any prescription for the $Z$-ambiguity in \eqref{ambiguities} must correspond to a specific boundary term to be added to the bulk action. The existence of such a boundary term was assumed in the previous sections, but no explicit realization was proposed. Here we provide a closed expression for the boundary term to be added to the bulk first-order action in order to obtain the presymplectic potential \eqref{LorentzPotential}.

\subsection{Asymptotically AdS spacetimes}
\label{subsec:first-order - AdS}
 
\paragraph{Chern--Simons formulation} The isometry algebra of AdS$_3$, i.e.\ the algebra $\mathfrak{so}(2,2)$, reads:
\begin{equation} \label{isometry}
[M_B,M_C] = \epsilon_{BCD} M^D , \quad
[M_B,P_C] = \epsilon_{BCD} P^D , \quad
[P_B,P_C] = \left( k \, \GN \right)^2 \epsilon_{BCD} M^D \, ,
\end{equation}
where $P_B$ and $M_B$ are the translation and  Lorentz generators (the latter are related to the customary Lorentz generators as $M_B = \tfrac{1}{2} \epsilon_{BCD} M^{CD}$). We recall that $\GN$ is Newton's constant and $k$ denotes the inverse of the AdS radius. In the vanishing-$k$ limit, which will be discussed in section~\ref{subsec:first-order - Flat}, one recovers the isometry algebra $\mathfrak{iso}(1,2)$ of three-dimensional Minkowski space.  We introduce a differential one-form, valued in the algebra \eqref{isometry}:
\begin{equation} \label{so(2,2)-connection}
\mathscr{A} = \frac{1}{\GN} \left( {E_N}^B P_B + {\omega_N}^B M_B \right) \D x^N ,
\end{equation}
where ${E_N}^B$ is the bulk triad and ${\omega_N}^B$ is its associated dualized spin connection (we label with $M,N,\ldots$ the bulk base-manifold indices as in previous sections). Up to boundary terms, one can rewrite the three-dimensional Einstein--Hilbert action as the following Chern--Simons action \cite{Achucarro:1986uwr,Witten:1988hc}:
\begin{equation} \label{CS_generic}
S_{\text{EH}} = \frac{1}{16\pi} \int \text{Tr} \left(\mathscr{A} \wedge \text{d}\mathscr{A} + \frac{2}{3} \, \mathscr{A} \wedge \mathscr{A} \wedge \mathscr{A} \right) ,
\end{equation}
where we introduced the Killing metric
\begin{equation}
\text{Tr} \left(M_B M_C\right) = \text{Tr} \left(P_B P_C\right) = 0 \, , \qquad 
\text{Tr} \left(M_B P_C\right) = \eta_{BC} \, ,
\end{equation}
with $\eta_{BC}$ the Minkowski metric whose signature is $(-, +, +)$ (we choose the convention $\epsilon^{012} = 1$ for the Levi--Civita symbol). 
For $k \neq 0$ we can take advantage of the isomorphism $\mathfrak{so}(2,2) \cong \mathfrak{sl}(2,\mathbb{R})\oplus \mathfrak{sl}(2,\mathbb{R})$ to rewrite the action \eqref{S_EH} as the difference of two $\mathfrak{sl}(2,\mathbb{R})$ Chern--Simons actions:
\begin{equation}\label{S_EH}
S_{\text{EH}} = S_{\text{CS}}\big[\,A\,\big] - S_{\text{CS}}\big[\,\widetilde{A}\,\big] \, ,
\end{equation}
with
\begin{equation}
S_{\text{CS}}[A] = \frac{1}{16\pi \GN k} \int \text{Tr} \left( A \wedge \text{d}A + \frac{2}{3} \, A \wedge A \wedge A \right) .
\label{actionCS}
\end{equation}
We have introduced the gauge connections\footnote{This connection should not be mistaken with the Weyl connection $\text{A} = k^{-2} \left(\Theta^\ast \!\ast\! \text{u}-\Theta\, \text{u} \right)$. The
meaning and distinction will always be clear from context. } $A = A^B J_B$ and  $\widetilde{A} = {\widetilde{A}}^B J_B$, which take values in the algebra $\mathfrak{sl}(2,\mathbb{R})$. For the latter, we use here the conventions
\begin{equation}
[{J_B},{J_C}] = {\epsilon_{BC}}^D \, J_D \, , \qquad 
\text{Tr} (J_B J_C) = \frac{1}{2}\,\eta_{BC} \, .
\end{equation}
The triad and the spin connection are related to the Chern--Simons forms $A$ and $\widetilde{A}$ as 
\begin{equation} \label{A-E}
A^B = \omega^B + k \, E^B \, , \qquad \widetilde{A}^B = \omega^B - k \, E^B \, .
\end{equation}

In the following, we compute the surface charges associated with the solution space determined by the line element \eqref{DE} (together with the differential conditions \eqref{holographic-Ward_1} or, equivalently, \eqref{bella1}) as well as their algebra using the methods of, e.g., \cite{Regge:1974zd, Banados:1994tn, Coussaert:1995zp, Banados:1998gg, Henneaux:1999ib, Bunster:2014mua}. For simplicity, we cover only the case in which the boundary metric is in the conformal gauge and we compare with previous results in the literature. 

\paragraph{Solution space in the first-order formalism} 

In the conformal gauge, the bulk metric \eqref{DE} reads 
\begin{equation}
\text{d}s^2_{\text{AdS}} = G_{MN} \, \text{d}x^M \, \text{d}x^N  = \frac{2}{k^2} \, \text{u} \left( \text{d}r + r \, \text{A} \right) + r^2\, \text{e}^{2\varphi}\D x^+ \D x^- + \frac{8\pi \GN}{k^4} \, \text{u} \left(\varepsilon \, \text{u} + \chi \ast\! \text{u} \right) ,
\label{ChernSimons - Relativistic Bulk Metric}
\end{equation}
where we have used the light-cone coordinates $x^\pm = \phi \pm k \, u$. In these coordinates, $\text{u}$ and $\ast\text{u}$ can be parameterized as in eq.~\eqref{velocity bgf},
\begin{equation}
\text{u}=-\frac{k}{2}\,\text{e}^{\varphi}\Big(\text{e}^{\zeta} \, \D 
x^+-\text{e}^{-\zeta} \, \D x^-\Big) \, , \qquad
\ast \text{u}=\frac{k}{2}\,\text{e}^{\varphi}\Big(\text{e}^{\zeta} \, \D x^++\text{e}^{-\zeta} \, \D x^-\Big) \, ,
\end{equation}
while  
$\varepsilon$ and $\chi$ are given by eq.~\eqref{ConformalStress},
\begin{equation}
\varepsilon\pm\chi = \frac{\text{e}^{-2 (\varphi\pm \zeta)}}{2 \pi  \GN} \,  \Big(\ell_\mp(x^\mp)
   -
   (\pa_\mp\zeta)^2\mp\pa_\mp^2 \zeta \mp \text{e}^{\mp2\zeta}\pa_-\pa_+\zeta
   \Big) \, .
\end{equation}

A natural, manifestly Weyl-invariant, choice for the triad associated with the metric \eqref{ChernSimons - Relativistic Bulk Metric} would be
\begin{equation}
 E^1 = \frac{r \, \text{u}}{k} \, , \quad
 E^{-1} = - \frac{1}{4 r k} \left(-r^2 \, \text{u} + 2 \, \text{d}r + 2 \, r \, \text{A} + \frac{8\pi \GN}{k^2} \, (\varepsilon \, \text{u} + \chi \ast\! \text{u}) \right) , \quad
 E^0 = \frac{r \ast\! \text{u}}{k} \, ,
\end{equation}
where we selected the components of the triad according to  the Minkowski metric
\begin{equation}
\eta_{BC} = \begin{pmatrix}0 & -2 & 0 \\ -2 & 0 & 0 \\ 0 & 0 & +1\end{pmatrix} ,
\label{MinkowskiMetricAdSChernSimons}
\end{equation}
so that $G_{MN} = {E_M}^B \eta_{BC} {E_N}^C$ with $B, C \in \{-1,0,1\}$. 
For the sake of computing the surface charges, it is however more convenient to perform a bulk Lorentz transformation and parameterize the connections appropriately. Let us introduce the $\mathfrak{sl}(2,\mathbb{R})$ basis
\begin{equation} \label{sl(2,R)}
[J_m,J_n] = (m-n) \, J_{m+n} \, , 
\qquad m,n \in \{ -1, 0, 1\} \, ,
\end{equation}
and implement the radial dependence of the connections $A$ and $\widetilde{A}$ as a gauge transformation:
\begin{subequations} \label{ChernSimons - Radial gauge A-field}
\begin{align}
A(x^+,x^-,r) & = b^{-1}(r)\, a_\mu(x^+,x^-)\, b(r) \, \text{d}x^\mu + b^{-1}(r) \partial_r \, b(r) \, \text{d}r \, ,  \\[5pt]
\widetilde{A}(x^+,x^-,r) & = \widetilde{b}^{-1}(r)\, \widetilde{a}_\mu(x^+,x^-)\, \widetilde{b}(r) \, \text{d}x^\mu + \widetilde{b}^{-1}(r) \, \partial_r \widetilde{b}(r) \, \text{d}r \, ,
\end{align}
\end{subequations}
with $b(r)$ and $\widetilde{b}(r)$ suitable $SL(2,\mathbb{R})$ group elements. Such a parameterization is always reachable, 
at least locally, see e.g.~\cite{Campoleoni:2010zq}. We also recall that the indices $\m, \n, \ldots$ label boundary coordinates. 
In the specific case of the metric \eqref{ChernSimons - Relativistic Bulk Metric}, one can choose
\begin{equation} \label{b}
b(r) = \text{exp}\left(r\, k \, J_{-1}\right)
\end{equation}
and
\begin{subequations}
\label{ChernSimons - Relativistic BC A-field}
\begin{align}
a_+ & = \text{e}^{\varphi+\zeta} \, J_1 - \text{e}^{-(\varphi+\zeta)} \left( \ell_+ - (\partial_+ \zeta)^2 + \partial_+^2  \zeta \right) J_{-1} + \partial_+ (\varphi - \zeta)\, J_0 \, , \\[5pt]
a_- & = - \text{e}^{-(\varphi+\zeta)} \, \partial_+ \partial_- \zeta \, J_{-1} + \partial_- (\varphi + \zeta)\, J_0
\end{align}
\end{subequations}
for the first gauge copy. The boundary connection thus depends on both the Weyl Goldstone mode $\varphi$ and the Lorentz Goldstone mode $\zeta$, and the integration constants that characterize the energy density and energy flow (see \eqref{ConformalStress}).

For the second gauge copy one can choose, correspondingly,
\begin{equation}
\widetilde{b} = \mathds{1} \quad \Rightarrow \quad {\widetilde{A}_r} = 0 \,,
\end{equation}
and
\begin{subequations}
\label{ChernSimons - Relativistic BC tildeA-field}
\begin{align}
&\widetilde{a}_+ = - \text{e}^{-(\varphi-\zeta)} \, \partial_+ \partial_- \zeta \, J_{-1} + \partial_+ (\varphi - \zeta) \, J_0 \, ,\\[1ex]
&\widetilde{a}_- = - \text{e}^{\varphi-\zeta} \, J_1 + \text{e}^{-(\varphi-\zeta)} \left(\ell_- - (\partial_- \zeta)^2 - \partial_-^2 \zeta \right) J_{-1} + \partial_- (\varphi + \zeta) \, J_0 \, .
\end{align}
\end{subequations}
Notice that the radial gauge fixing for the two connections introduced here differs from the standard literature on surface charges in the Chern--Simons formulation. Our unusual choice is locally legitimate  (starting from the previous definitions together with \eqref{A-E} one reproduces the line element \eqref{ChernSimons - Relativistic Bulk Metric}) and simplifies the technical analysis, while leading to the same charges as in section~\ref{sec:conformal-gauge}.  This matching excludes any possible global issue. 

\paragraph{Charges}

Once the boundary values of the gauge fields in a given radial gauge are set, one identifies the residual symmetries and  computes the associated surface charges. The computation for the two Chern--Simons gauge connections is similar. We will therefore display some details for $A$, while keeping the presentation minimal for $\widetilde{A}$. 

The first step in the agenda is to identify the gauge transformations, of the kind $\D A = \D \Lambda + [A, \Lambda]$, which preserve the form of the connection detailed above. To this end, we  consider gauge parameters of the type
\begin{equation} \label{expansion_parameter}
\Lambda(x^+,x^-,r) = b^{-1}(r) \left( \sum^{+1}_{m = -1} \epsilon^m(x^+,x^-) \, J_m \right) b(r) \, ,
\end{equation}
where $b(r)$ is the $SL(2,\mathbb{R})$ group element \eqref{b}, and the Lie-algebra-valued function of the boundary coordinates is decomposed in the basis \eqref{sl(2,R)}. This expression guarantees that the gauge transformations preserve the radial gauge fixing \eqref{ChernSimons - Relativistic BC tildeA-field}.
Preserving the form of the boundary connection \eqref{ChernSimons - Relativistic BC A-field} then fixes the components of the gauge parameter to
\begin{subequations}
\label{ChernSimons - Relativistic gauge parameters A-field}
\begin{align}
\epsilon^1 & = \text{e}^{\varphi+\zeta} \, Y^+ \, , \\[5pt]
\epsilon^0 & = h + \varpi - \partial_+ Y^+ - 2 \, Y^+ \, \partial_+ \zeta \, , \\
\epsilon^{-1} & = - \text{e}^{-(\varphi+\zeta)} \left( \ell_+ \, Y^+ - \frac{1}{2}\,\partial_+^2 Y^+ + \partial_+ h - \partial_+ Y^+ \, \partial_+ \zeta - Y^+  (\partial_+ \zeta)^2 \right) ,
\end{align}
\end{subequations}
where $\varpi$ and $h$ are arbitrary functions of the boundary coordinates, while $Y^+ = Y^+(x^+)$. These are precisely the symmetry parameters of the asymptotic Killing vectors \eqref{AKV conformal gauge}.
The variations of the boundary data ---~namely $\ell_+$, $\zeta$ and $\varphi$~--- under the above gauge transformations read
\begin{equation}
\label{ChernSimons - AdS Variation of Frame Function}
\delta_{\Lambda} \ell_+ = Y^+ \partial_+ \ell_+ + 2 \, \ell_+ \partial_+ Y^+ - \frac{1}{2} \, \partial_+^3 Y^+ \,, \qquad
\delta_{\Lambda} \varphi = \varpi \,, \qquad
\delta_{\Lambda} \zeta  = h \,,
\end{equation}
in agreement with eqs.~\eqref{transfo zeta} and \eqref{transfo ell}.\footnote{The algebra of residual gauge transformations
generated by \eqref{ChernSimons - Relativistic gauge parameters A-field} can  be computed in analogy with the metric formulation. 
Since the generators are field-dependent, one has to use a modified bracket, similarly to \eqref{modified-Lie}:
\[
[\Lambda_1,\Lambda_2]_{\text{M}} \equiv [\Lambda_1,\Lambda_2] - \delta_{\Lambda_1} \Lambda_2 + \delta_{\Lambda_2} \Lambda_1 \, ,
\]
where $\delta_{\Lambda_1} \Lambda_2$ stands for the variation of  $\Lambda_2$ under $\Lambda_1$. Under the residual gauge transformations \eqref{ChernSimons - Relativistic gauge parameters A-field}, the Lie algebra closes as $[\Lambda_1,\Lambda_2]_{\text{M}} = \Lambda_{12}$, where $\Lambda_{12}$ is built on $Y^+_{12} = {Y^+_2} \partial_+ {Y^+_1} - {Y^+_1} \partial_+ {Y^+_2}$ with $\varpi_{12} = h_{12} = 0$ as in eq.~\eqref{conformal-gauge_algebra}. \label{modified-commutator}} 

Next, the associated surface charges are obtained by integrating, if possible, the following variations calculated at fixed value of the coordinate $u$ \cite{Banados:1994tn}:
\begin{equation}
\delta Q[\Lambda] = - \frac{1}{8\pi \GN k} \int_0^{2\pi} \text{d}\phi \, \text{Tr} \left[ \Lambda \, \delta A_\phi \right] = - \frac{ 1}{8\pi \GN k} \int_0^{2\pi} \text{d}\phi \, \text{Tr} \left[ b \, \Lambda \, b^{-1} 
\left(\delta a_+ + \delta a_-\right) \right] .
\end{equation}
The last rewriting is manifestly independent of the null radial coordinate $r$ thanks to the cyclicity of the trace. Implementing the radial dependence as a gauge transformation (see  \eqref{ChernSimons - Radial gauge A-field}) thus makes the charges   finite in the  infinite-$r$ limit. Substituting the explicit parameterization of the generators of residual gauge transformations one obtains, up to integration by parts,
\begin{equation}
\begin{split}
\delta Q[\Lambda] = - \frac{1}{8\pi \GN k} \int_0^{2\pi} \text{d}\phi \, \Big[ & \delta \ell_+ Y^+ + \frac{\delta \zeta}{2} \, \Big( (3 \, \partial_+ - \partial_-) \, h + (\partial_+ - \partial_-) \, \varpi \Big) \\
 & + \frac{\delta \varphi}{2} \, \Big( (\partial_+ - \partial_-) \, h - (\partial_+ + \partial_-) \, \varpi \Big) \Big] \, .
\end{split}
\end{equation}
This expression is linear in the boundary data. Hence $\delta Q[\Lambda]$ is manifestly integrable, and gives the surface charges
\begin{equation}
\begin{split}
Q[\Lambda] = - \frac{1}{8\pi \GN k} \int_0^{2\pi} \text{d}\phi \, \Big[ & \ell_+ Y^+ + \frac{\zeta}{2} \, \Big( (3 \, \partial_+ - \partial_-) \, h + (\partial_+ - \partial_-) \, \varpi \Big)\\
&+ \frac{\varphi}{2} \, \Big( (\partial_+ - \partial_-) \, h - (\partial_+ + \partial_-) \, \varpi \Big) \Big] \, .
\end{split}
\end{equation}
This is the contribution to the total surface charges originating from the first gauge connection. 

We now turn to the second Chern--Simons copy.
The gauge field $\widetilde{A}$ depends on the same functions of the boundary coordinates $\varphi$ and $\zeta$ already appearing in $A$, together with the chiral function $\ell_-$ that was absent in \eqref{ChernSimons - Relativistic BC A-field}. Correspondingly, gauge parameters $\widetilde{\Lambda}$ generating variations preserving the conditions \eqref{ChernSimons - Radial gauge A-field} and \eqref{ChernSimons - Relativistic BC tildeA-field} depend on the functions $\varpi$ and $h$ already introduced in \eqref{ChernSimons - Relativistic gauge parameters A-field}, together with the chiral parameter $Y^- = Y^-(x^-)$. In addition to the variations of the boundary data summarized in \eqref{ChernSimons - AdS Variation of Frame Function}, we obtain
\begin{equation}
\delta_{\tilde{\Lambda}} \ell_- = Y^- \partial_- \ell_- + 2 \, \ell_- \partial_- Y^- - \frac{1}{2} \, \partial_-^3 Y^- \, ,
\end{equation}
and, altogether, the generators of residual gauge transformations in the Chern--Simons formulation are related to the asymptotic Killing vectors \eqref{AKV conformal gauge} as
\begin{equation}
\xi^M =\frac{1}{2\,k}\, {E_B}^M \left( \Lambda^B - \widetilde{\Lambda}^B \right),
\end{equation}
where ${E_B}^M$ is the inverse of the triad. 
The surface charges are finite and integrable also for the second gauge copy:
\begin{equation}
\begin{split}
\widetilde{Q}[\widetilde{\Lambda}] = - \frac{1}{8\pi \GN k} \int_0^{2\pi} \text{d}\phi \Big[ & \ell_- Y^- + \frac{\zeta}{2} \, \Big( (3 \, \partial_- - \partial_+) \, h + (\partial_+ - \partial_-) \, \varpi \Big)\\
& + \frac{\varphi}{2} \, \Big( (\partial_+ - \partial_-) \, h - (\partial_+ + \partial_-) \, \varpi \Big) \Big] \, .
\end{split}
\end{equation}
The total surface charges are then
\begin{equation}
Q_{\text{tot}}[\Lambda,\widetilde{\Lambda}] = Q[\Lambda] - \widetilde{Q}[\widetilde{\Lambda}] = - \frac{1}{8\pi \GN k} \int_0^{2\pi} \text{d}\phi \Big[ \ell_+ \, Y^+ - \ell_- \, Y^- + 2 \, \zeta \, \big(\partial_+ - \partial_- \big) \, h \Big] \, ,
\label{ChernSimons - Relativistic Total Charge}
\end{equation}
which are identical to those obtained in the metric formulation (cf.~\eqref{Virasoro charges} and \eqref{Weyl and Local Lorentz Charges}). 

It should be emphasized that the Weyl Goldstone mode and its associated shift parameter $\varpi$ are absent. Consequently the Weyl symmetry is pure-gauge and can be modded out. This is to compare with \cite{Ruzziconi:2020wrb}, where the charge associated with the Weyl sector is zero in Bondi gauge, contrary to the Fefferman--Graham gauge (see e.g. \cite{Troessaert:2013fma, Alessio:2020ioh}). We also observe that the charges \eqref{ChernSimons - Relativistic Total Charge} are not conserved in time, signaling the existence of a Lorentz anomaly associated with the shift of $\zeta$, thus confirming the analysis of section~\ref{sec:DE}. 

\paragraph{Asymptotic symmetry algebra}

The algebra of asymptotic symmetries is the canonical algebra of the charges \eqref{ChernSimons - Relativistic Total Charge}. We shall also show how to recover the two particular cases already analyzed in \cite{Campoleoni:2018ltl}, and discuss the relations with some results in the literature \cite{Troessaert:2013fma, Grumiller:2016pqb, Alessio:2020ioh}. In order to simplify the next steps, we introduce the following field redefinitions:
\begin{equation}
L_\pm(x^\pm) = \frac{\mp 1}{8\pi \GN k} \, \ell_\pm(x^\pm) \, , \qquad
Z(x^+,x^-) = \frac{1}{4 \pi \GN k} \, (\partial_- - \partial_+) \, \zeta(x^+,x^-) \, .
\end{equation}
The gauge variations of these functions are:
\begin{subequations}
\begin{align}
\delta_{(\Lambda,\tilde{\Lambda})} L_\pm & = \pm Y^\pm \, \partial_\pm  L_\pm \pm 2 \, L_\pm \, \partial_\pm Y^\pm \pm \frac{1}{16\pi \GN k} \, \partial_\pm^3 Y^\pm \, ,\\
\delta_{(\Lambda,\tilde{\Lambda})} Z & = \frac{1}{4\pi\GN k} \, (\partial_- - \partial_+) \, h \, ,
\end{align}
\end{subequations}
and the surface charges \eqref{ChernSimons - Relativistic Total Charge} can be recast as
\begin{equation}
Q_{\text{tot}}[\Lambda,\widetilde{\Lambda}] = \int_0^{2\pi} \text{d}\phi \, \left[ L_+ \, Y^+ + L_- \, Y^- + Z \, h \right] .
\label{ChernSimons - Relativistic Total Charge with Redefinitions}
\end{equation}
These charges generate global symmetries when acting on a generic functional $F$ of the phase space as $\delta_{(\Lambda,\tilde{\Lambda})} F = \{Q_{\text{tot}}[\Lambda,\widetilde{\Lambda}],F\}$. Their Poisson brackets are determined  on the solution space. 
Expanding the boundary data into Fourier modes as
\begin{equation}
L_\pm = - \frac{1}{2\pi} \sum_{p \in \mathbb{Z}} L_p^\pm \, \text{e}^{-\text{i} p x^\pm} \, , \qquad 
Z = - \frac{1}{2\pi}\, \sum_{p,q \in \mathbb{Z}} Z_{pq} \, \text{e}^{-\text{i} p x^+} \, \text{e}^{-\text{i} q x^-} \, ,
\end{equation}
the non-vanishing brackets are
\begin{subequations} \label{ChernSimons - Double Copy of Virasoro}
\begin{align}
\text{i} \, \left\{L_p^\pm,L_q^\pm\right\} & = (p-q) \, L_{p+q}^\pm + \frac{c}{12} \, p^3 \, \delta_{p+q,0} \, , \\
\text{i} \, \{Z_{pq},Z_{rs}\} & = - \frac{c}{3} \, (r-q) \, \text{e}^{2 \text{i} k (q+s) u} \, \delta_{p+r,q+s} \, ,
\end{align}
\end{subequations}
where all central charges are proportional to the Brown--Henneaux central charge
\begin{equation}
c = \frac{3}{2 k \GN} \, .
\end{equation}
This algebra has appeared in \cite{Alessio:2020ioh}, where the centrally extended local Abelian factor was due to the Weyl symmetry. In the present approach  Weyl symmetry is pure-gauge, and the asymptotic symmetry originates from Lorentz boosts --- a manifestation of the two-dimensional Lorentz--Weyl duality interplay. The central extension in the Lorentz sector depends explicitly on the time coordinate $u$, hence it depends on the point of the solution space one is considering. The above asymptotic symmetry algebra is rather a one-parameter family of algebras, the parameter being the value of $u$ at which the surface charges are determined.

If we further impose that the curvature of the Weyl connection vanishes, that is the extra boundary condition $F=0$, we obtain an (anti-)chiral splitting of the Lorentz Goldstone mode
\begin{equation}
Z_\pm(x^\pm) = \frac{\mp 1}{4\pi \GN k} \, \partial_\pm  \zeta^\pm(x^\pm) \,.
\end{equation}
The gauge parameter that preserves this condition also splits as
\begin{equation}
h = h^+(x^+) + h^-(x^-) \, ,
\end{equation}
so that
\begin{equation}
\delta_{(\Lambda,\tilde{\Lambda})} Z_\pm = \frac{\mp 1}{4\pi \GN k} \, \partial_\pm h^\pm \, .
\end{equation}
The surface charges \eqref{ChernSimons - Relativistic Total Charge with Redefinitions} then take the form
\begin{equation}
Q_{\text{tot}}[\Lambda,\widetilde{\Lambda}] = \int_0^{2\pi} \text{d}\phi \left[ L_+ \, Y^+ + L_- \, Y^- + Z_+ \, h^+ + Z_- \, h^- \right] ,
\end{equation}
and are now conserved. By decomposing these (anti-)chiral fields into Fourier modes,
\begin{equation}
Z_\pm = - \frac{1}{2\pi} \sum_{p \in \mathbb{Z}} Z_p^\pm \, \text{e}^{-\text{i} p x^\pm} \, ,
\end{equation}
the charge algebra then reads
\begin{subequations}\label{jj}
\begin{align}
&\text{i} \, \left\{L_p^\pm,L_q^\pm\right\} = (p-q) \, L_{p+q}^\pm + \frac{c}{12} \, p^3 \, \delta_{p+q,0} \, , \\
&\text{i} \, \left\{Z_p^\pm,Z_q^\pm\right\} = \frac{c}{3} \, p \, \delta_{p+q,0} \, .
\end{align}
\end{subequations}
In this case, the Poisson brackets do not depend on $u$ anymore. Henceforth, the algebra is the same at all values of time. The boundary condition imposed here is the equivalent in the Lorentz sector to what was done in \cite{Troessaert:2013fma} in the Weyl sector. There, the Weyl symmetry was charged and the condition $R=0$ allowed for the (anti-)chiral split of the modes. Our algebra \eqref{jj} is indeed isomorphic to the one found in \cite{Troessaert:2013fma}, but it is now associated with the Lorentz instead of Weyl symmetry. The same algebra was also obtained as a subcase of the analysis of the most general boundary conditions for asymptotically three-dimensional AdS spacetimes admitting a well-posed variational principle \cite{Grumiller:2016pqb}.

From this algebra, one can recover the two special cases anticipated in \cite{Campoleoni:2018ltl}, where the algebra of asymptotic symmetries was computed only for reduced phase spaces, in which the boundary conformal factor was turned off and only one of the pairs of functions $\ell_\pm$ or $\zeta^\pm$ appeared. This supports the claim made in \cite{Campoleoni:2018ltl} concerning the fluid/gravity correspondence: changing hydrodynamic frame influences the surface charges and thus the global properties of the boundary fluid.

\paragraph{Boundary Anomaly}

In the first order formulation we can go a step further and propose an expression for the action, including its boundary term, whose variation reproduces the presymplectic potential discussed in section \ref{sec:conformal-gauge}.
The on-shell variation of the Chern--Simons action \eqref{actionCS} reduces to the boundary term
\begin{equation}
\delta S_{\text{CS}}[A] = - \frac{1}{8\pi\GN k} \int \text{d}^2 x\, \text{Tr} \big( A_u \, \delta A_\phi \big) \, .
\end{equation}
To determine the variation of the Einstein--Hilbert action, we must subtract the contribution of the right sector from the left sector which yields
\begin{equation}
\delta S_{\text{EH}}=\delta S_{\text{CS}}[A] - \delta S_{\text{CS}}[\widetilde{A}] = - \frac{1}{8\pi \GN k} \int \text{d}^2 x \, \text{Tr} \left( A_u \, \delta A_\phi - \widetilde{A}_u \, \delta \widetilde{A}_\phi \right) .
\end{equation}
For the Brown--Henneaux boundary conditions one can obtain a well-posed variational problem by adding to the bulk action the Coussaert--Henneaux--Van Driel boundary term \cite{Coussaert:1995zp}:
\begin{equation}
S_{CHVD}=-\frac{1}{16 \pi \GN} \int \text{d}^2 x \, \text{Tr} \left(A_\phi^2 + \widetilde{A}_\phi^2 \right) .
\label{ChernSimons - CHVD Boundary Term}
\end{equation}
In our case, adding this boundary term the on-shell variation of the total action would not reproduce that in eq.~\eqref{dSL}, which is associated with the ``Lorentz'' presymplectic potential \eqref{LorentzPotential}. We observe that the surface charges computed in the Chern--Simons formulation are insensitive to the boundary terms added to the bulk action, needed to fully specify the variational principle. Nonetheless, the latter gives rise to the presymplectic structure of the theory, thus controlling the well-definiteness of the variation problem and consequently the appearance of anomalies. 

We can however choose a different boundary term, such that the total action reads
\begin{equation}
S_\text{tot}\big[A,\widetilde{A}\,\big] = S_{\text{EH}} + \frac{1}{16 \pi \GN k}\int \text{d}^2 x \, \text{Tr} \left(A_u \, A_\phi - \widetilde{A}_u \, \widetilde{A}_\phi \right) ,
\label{ChernSimons - AdS Boundary Term with Varphi non zero}
\end{equation}
and its on-shell variation gives
\begin{equation}
\delta S_\text{tot}\big[A,\widetilde{A}\,\big] = \frac{1}{2 \pi \GN k} \int \delta \zeta \, \text{e}^{-2\varphi} \, \partial_+  \partial_-  \zeta \, \text{Vol}_{\partial \cM} \, ,
\label{CS - delta S tot in AdS}
\end{equation}
where $\text{Vol}_{\partial \cM}$ is the boundary volume form. We recognize the variation of the action in eq.~\eqref{dSL} that corresponds to the ``Lorentz'' presymplectic potential \eqref{LorentzPotential} in conformal gauge. This remainder\footnote{Note that it has the same structure as the non-integrable term in equation (B.16) of \cite{Alessio:2020ioh}.} is not integrable, implying that, without further constraints on the solution space, no additional boundary term can make the variational problem well-defined, as in \cite{Alessio:2020ioh}. A sufficient boundary condition here is to fix the Lorentz Goldstone mode $\zeta$ by solving $F=0$. Doing so, one goes back to the (anti-)chiral split of the Lorentz sector, and retrieve the analogue of the results obtained in \cite{Troessaert:2013fma} for the Weyl sector, which were indeed derived by requiring a well-posed variational problem.

\subsection{Asymptotically flat spacetimes}
\label{subsec:first-order - Flat}

When the cosmological constant vanishes, i.e.\ in the flat limit $k \to 0$, it is still possible to write the Einstein--Hilbert action in the Chern--Simons form \eqref{CS_generic} with a $\mathfrak{iso}(1,2)$ gauge algebra \cite{Witten:1988hc}. The rewriting \eqref{S_EH} as the difference of two simpler Chern--Simons action is not available anymore, but one can apply the same techniques as in the previous section to compute asymptotic symmetries, surface charges and their algebra directly in the $\mathfrak{iso}(1,2)$ Chern--Simons theory.

\paragraph{Solution space in the first-order formalism} In the case of asymptotically flat spaces, in the conformal gauge the bulk metric \eqref{flat DE} reads 
\begin{equation}
\D s^2_{\text{Flat}} = 2\, \upmu \left(\D r+r{\cal A} \right) +r^2 \upmu^\ast\upmu^\ast +8\pi \GN \, \upmu \left(\epsilon \, \upmu + \alpha \, \upmu^\ast \right) ,
\label{flat DE - CS}
\end{equation}
where the values of the Carrollian Cartan frame are given in eq.~\eqref{FlatMu},
\begin{equation}
    \upmu=-\text{e}^{\varphi} \left( \text{d}u+ \beta \, \text{d}\phi \right) , \quad 
    \upmu^\ast=\text{e}^\varphi \, \text{d}\phi \, ,
    \label{FlatMu - CS}
\end{equation}
while the Carrollian energy density $\e$ and energy flow $\alpha$ can be parameterized as in eq.~\eqref{FlatEqs},
\begin{subequations} \label{FlatEqs - CS}
\begin{align}
\epsilon &= \frac{\text{e}^{-2\varphi}}{8 \pi \GN} \Big(8 \pi \GN \, \varepsilon_0 - (\partial_u \beta)^2 + 2 \, \partial_u \partial_\phi \beta - 2 \, \beta \, \partial_u^2 \beta\Big) \, , \\
\alpha &= \frac{\text{e}^{-2\varphi}}{4 \pi \GN} \Big(4 \pi \GN (\alpha_0 - u \, \partial_\phi \varepsilon_0 + 2 \, \beta \varepsilon_0) - \beta \left[(\partial_u \beta)^2 - 2 \, \partial_u \partial_\phi \beta + \beta \, \partial_u^2 \beta \right] - \partial_\phi^2 \beta + \partial_u \beta \, \partial_\phi \beta\Big) \, ,
\end{align}
\end{subequations}
with $\varepsilon_0 = \varepsilon_0(\phi)$ and $\alpha_0 = \alpha_0(\phi)$.

Also in this case, we can encode the dependence on the null radial coordinate $r$ in a gauge transformation:
\begin{equation}
\mathscr{A} = b^{-1} \left[ \text{a} + \text{d} \right] b \, ,
\end{equation}
where $\text{a} = a_\m(u,\phi) \D x^\m$ and $b$ is the $\text{ISO}(1,2)$ group element
\begin{equation}
b(r) = \text{exp}\left(\frac{r}{2}\, P_{-1} \right) .
\label{ChernSimons - Group Element of ISO(1,2)}
\end{equation}
We expressed the latter using the following convenient basis of the $\mathfrak{iso}(1,2)$ algebra:
\begin{equation} \label{iso(1,2)-cool-basis}
[M_m,M_n] = (m-n) \, M_{m+n} \, , \quad
[M_m,P_n] = (m-n) \, P_{m+n} \, , \quad
[P_m,P_n] = 0 \, ,
\end{equation}
with $m,n \in \{-1,0,1\}$.
In the same basis, the components of the boundary connection can be chosen as
\begin{subequations}
\label{ChernSimons - Carrollian BC}
\begin{align}
\begin{split}
a_\phi & = \frac{\text{e}^{-\varphi}}{\sqrt{2}} \, \Big( 4 \pi \GN \, \varepsilon_0 - \frac{1}{2}\, (\partial_u \beta)^2 + \partial_u \partial_\phi \beta \Big) \, M_1 - \Big(\partial_\phi \varphi - \partial_u \beta \Big) \, M_0 - \frac{\text{e}^{\varphi}}{\sqrt{2}} \, M_{-1} + \frac{\text{e}^{\varphi} \, \beta}{\sqrt{2}} \, P_{-1} \\
	  & + \frac{\text{e}^{-\varphi}}{\sqrt{2}} \, \Big( 4 \pi \GN\, ( \alpha_0 - u \, \partial_\phi \varepsilon_0 ) - \partial_\phi^2 \beta + \partial_u \beta \, \partial_\phi \beta  + \frac{\beta}{2} \, \big( 8 \pi \GN \, \varepsilon_0 - (\partial_u \beta)^2 + 2 \, \partial_u \partial_\phi \beta \big) \Big) \, P_1 \, ,
\end{split}\\[10pt]
\begin{split}
a_u & = \frac{\text{e}^{-\varphi} }{\sqrt{2}} \left[ \partial_u^2 \beta\, M_1 - \Big( 4 \pi \GN \, \varepsilon_0 - \frac{1}{2}\,(\partial_u \beta)^2 + \partial_u \partial_\phi \beta - \beta \, \partial_u^2 \beta \Big) \, P_1 \right] - \partial_u \varphi \, M_0  + \frac{\text{e}^{\varphi}}{\sqrt{2}} \, P_{-1} \, .
\end{split}
\end{align}
\end{subequations}

\paragraph{Charges}
\label{subsubsec:first-order - Flat Charges}

We now consider gauge parameters of the form 
\begin{equation}
\Lambda(r,u,\phi) = b^{-1}(r) \sum^{+1}_{m = -1} \Big[\epsilon^m(u,\phi) \, M_m + \sigma^m(u,\phi) \, P_m \Big] \, b(r) \, ,
\end{equation}
where $b(r)$ is the $\text{\text{ISO}}(1,2)$ group element \eqref{ChernSimons - Group Element of ISO(1,2)}.
Gauge transformations preserving the form \eqref{ChernSimons - Carrollian BC} of the boundary connection are generated by
\begin{subequations} \label{CS-parameters_flat}
\begin{align}
\epsilon^1 &= \frac{\text{e}^{-\varphi}}{2 \sqrt{2}} \, \Big( Y \big(8 \pi \GN \, \varepsilon_0 - (\partial_u \beta)^2 \big) - 2 \, \big( \partial_\phi Y \, \partial_u \beta - \partial_u \tilde{h} + \partial_\phi^2 Y \big)\Big) \, , \\
\epsilon^0 &= - \varpi + Y \, \partial_u \beta + \partial_\phi Y \, , \\
\epsilon^{-1} &= - \frac{Y \, \text{e}^{\varphi}}{\sqrt{2}} \, , \\
\begin{split}
\sigma^1 &= \frac{\text{e}^{-\varphi}}{2 \sqrt{2}} \Big( \beta \, \big(- Y \, \big((\partial_u \beta)^2-8 \pi \GN \, \varepsilon_0 \big)+ 2 \big(-\partial_\phi Y \, \partial_u \beta + \partial_u \tilde{h} - \partial_\phi^2 Y\big)\big)\\
		&\;\;\;\; + 2 \, Y \, \big(\partial_\phi \beta \, \partial_u \beta + 4 \pi \GN \big(\alpha_0 - u \, \partial_\phi \varepsilon_0 \big)\big) - 2 \, \sqrt{2}  \, \partial_\phi H \, \partial_u \beta - \sqrt{2} \, H \, (\partial_u \beta)^2\\
		&\;\;\;\; + 2 \, u \, \partial_\phi^2 Y \, \partial_u \beta + u \, \partial_\phi Y \, (\partial_u \beta)^2 + 2 \, \partial_\phi Y \, \partial_\phi \beta - 2 \, \partial_\phi \tilde{h} \\
		&\;\;\;\; + 8 \pi \GN \, \varepsilon_0 \, \big(\sqrt{2} \, H - u \, \partial_\phi Y\big) + 2 \, u \, \partial_\phi^3 Y - 2 \, \sqrt{2} \, \partial_\phi^2  H \Big) \, ,
\end{split}\\[3pt]
\sigma^0 &= \sqrt{2} \, H \, \partial_u \beta - u \, \partial_\phi Y \, \partial_u \beta - Y \, \partial_\phi \beta + \tilde{h} - u \, \partial_\phi^2 Y + \sqrt{2} \, \partial_\phi H \, , \\
\sigma^{-1} &= \frac{\sqrt{2}}{2} \, \text{e}^{\varphi} \, \left( Y \, \beta - \sqrt{2} \, H +  u \, \partial_\phi Y\right) ,
\end{align}
\end{subequations}
where $Y = Y(\phi)$, $H = H(\phi)$, $\varpi = \varpi(u,\phi)$ and $\tilde{h} = \tilde{h}(u,\phi)$ have the same properties as the corresponding functions featuring residual diffeomorphisms in the metric formulation, see eq.~\eqref{FlatFieldVar}. Gauge transformations generated by parameters of the form \eqref{CS-parameters_flat} also imply the same variations of the boundary data as in the latter equation.
As in the AdS case, these gauge parameters are field-dependent and thus define a Lie algebra under the modified bracket defined in footnote~\ref{modified-commutator} with
\begin{subequations}
\begin{align}
Y_{12} &= Y_2 \, \partial_\phi Y_1 - Y_1 \, \partial_\phi Y_2 \, , \\
H_{12} &= H_2 \, \partial_\phi Y_1 - H_1 \, \partial_\phi Y_2 + Y_2 \, \partial_\phi H_1 - Y_1 \, \partial_\phi H_2 \, ,\\
\varpi_{12} &= 0 \, ,\\
\tilde{h}_{12} &= 0 \, .
\end{align}
\end{subequations}
This gives the algebra 
$\mathrm{BMS}_3\oplus \mathbb{R} \oplus \mathbb{R}$ of residual gauge transformations, as anticipated in section~\ref{sec:flat-space-charges}.

The charges are finally obtained as in the previous section by integrating the variation
\begin{align}
\delta Q_{\text{tot}}[\Lambda] &= - \frac{ 1}{8\pi\GN k} \int_0^{2\pi} \text{d}\phi \, \text{Tr} \left[ b \, \Lambda \, b^{-1} \, \delta a_\phi\right] =
\int_0^{2\pi} \text{d}\phi \left[ \frac{1}{2} \left( H \,  \delta \varepsilon_0 - Y \, \delta \alpha_0 \right) + \frac{1}{4 \pi \GN} \, \partial_u \tilde{h} \, \delta \beta \right] .
\end{align}
The result is manifestly integrable and, in analogy with the AdS case, gives charges that are finite but non-conserved due to the arbitrary Carrollian boundary frame function $\beta(u,\phi)$.
We see that the charge associated with the Weyl symmetry, namely the variation of the arbitrary conformal factor $\varphi(u,\phi)$, vanishes. Again,  we can mod it out of the asymptotic symmetry group. One can check that these charges are the flat limit of \eqref{ChernSimons - Relativistic Total Charge}. We also stress that the non-conservation driven by the term involving the parameter $\tilde{h}$ signals the presence of an anomaly in the Carroll-boost sector.

\paragraph{Asymptotic symmetry algebra}

In order to derive the charge algebra, it is useful to introduce the function 
\begin{equation}
B(u,\phi) = \frac{1}{4 \pi \GN} \, \partial_u \beta(u,\phi) \, ,
\end{equation}
which enables one to rewrite the charges as
\begin{equation}
Q_{\text{tot}}[\Lambda] = \int_0^{2\pi} \text{d}\phi \, \Big(H \, \varepsilon_0 + Y \, \alpha_0 + \tilde{h} \, B \Big) \, .
\label{ChernSimons - Carrollian Total Charge with Redefinitions}
\end{equation}
These charges generate the following variations of the boundary data:
\begin{subequations}
\begin{align}
\delta_{(H,Y,\varpi,\tilde{h})} \varepsilon_0 & = Y \, \partial_\phi \varepsilon_0 + 2 \, \varepsilon_0 \, \partial_\phi Y + \frac{1}{8 \pi \GN} \, \partial_\phi^3 Y \, ,\\
\delta_{(H,Y,\varpi,\tilde{h})} \alpha_0 & = Y \, \partial_\phi \alpha_0 + 2 \, \alpha_0 \, \partial_\phi Y + H \, \partial_\phi \varepsilon_0 + 2 \, \varepsilon_0 \, \partial_\phi H + \frac{1}{8 \pi \GN} \, \partial_\phi^3 \, H \, ,\\[4pt]
\delta_{(H,Y,\varpi,\tilde{h})} \varphi & = \varpi \, , \\[4pt]
\delta_{(H,Y,\varpi,\tilde{h})} B & = \frac{1}{4 \pi \GN} \, \partial_u \tilde{h} \, .
\end{align}
\end{subequations}
Expanding the fields in Fourier modes,
\begin{subequations}
\begin{align}
\varepsilon_0(\phi) & = - \frac{1}{2\pi} \sum_{p \in \mathbb{Z}} T_p \, \text{e}^{-\text{i}p\phi} \, ,\\
\alpha_0(\phi) & = - \frac{1}{2\pi} \sum_{p \in \mathbb{Z}} Y_p \, \text{e}^{-\text{i}p\phi} \, ,\\
B(u,\phi) & = - \frac{1}{2\pi} \sum_{p,q \in \mathbb{Z}} B_{pq} \, \text{e}^{-\text{i} (p-q) \phi} \, \text{e}^{-\text{i} (p+q) u} \, ,
\end{align}
\end{subequations}
the non-vanishing Poisson brackets become
\begin{subequations}
\begin{align}
\text{i} \, \{Y_p,Y_q\} & = (p-q) \, Y_{p+q} \, ,\\
\text{i} \, \{Y_p,T_q\} & = (p-q) \, T_{p+q} + \frac{c_M}{12} \, p^3 \, \delta_{p+q,0} \, ,\\
\text{i} \, \{B_{pq},B_{rs}\} & = - \frac{c_M}{6} \, (r-q) \, \text{e}^{2 \text{i} (q+s) u} \, \delta_{p+r,q+s} \, .
\end{align}
\end{subequations}
The $Y_p$ and $T_p$ span a $\mathfrak{bms}_3$ algebra with central charge 
\begin{equation} \label{central-charge-flat}
c_M = \frac{3}{\GN} \, ,
\end{equation}
where we borrowed the notation $c_M$ from the papers \cite{Detournay:2016sfv, Grumiller:2017sjh} to which we shall compare our outcome shortly.
The central extension of the Carroll-boost sector depends explicitly on the time coordinate $u$, as for the Lorentz boosts in the anti-de Sitter  case, see eq.~\eqref{ChernSimons - Double Copy of Virasoro}. Again, as highlighted in the AdS analysis, the asymptotic symmetries of the case at hand realize a one-parameter family of algebras.
 
Imposing that the Weyl--Carroll curvature defined in \eqref{FlatWeylConnection} vanishes, i.e.\ $\mathcal{F}=0$, the Carroll-boost function becomes
\begin{equation}
\beta(u,\phi) = \beta_\phi(\phi) + u \, \beta_u(\phi) \, ,
\label{CS - split of Carroll boost}
\end{equation}
and its associated shift parameter also splits as
\begin{equation}
\tilde{h}(u,\phi) = \tilde{h}_\phi(\phi) + u \, \tilde{h}_u(\phi) \, .
\end{equation}
Upon rescaling of the latter fields as
\begin{equation}
B_\phi(\phi) = - \frac{1}{8\pi\GN} \, \beta_\phi(\phi) \, , \qquad B_u(\phi) = \frac{1}{8\pi\GN} \, \beta_u(\phi) \, ,
\end{equation}
we obtain the variations
\begin{equation}
\delta_{(H,Y,\varpi,\tilde{h})} B_\phi = - \frac{1}{8\pi\GN} \, \tilde{h}_\phi \, , \qquad \delta_{(H,Y,\varpi,\tilde{h})} B_u = \frac{1}{8\pi\GN} \, \tilde{h}_u \, .
\end{equation}
The total surface charges \eqref{ChernSimons - Carrollian Total Charge with Redefinitions} are then conserved and assume the form
\begin{equation}
Q_{\text{tot}}[\Lambda] = \int_0^{2\pi} \text{d}\phi \, \Big(H \, \varepsilon_0 + Y \, \alpha_0 + \tilde{h}_u \, B_\phi + \tilde{h}_\phi \, B_u \Big) \, .
\end{equation}
Decomposing the boundary data into Fourier series
\begin{equation}
\begin{aligned}
\varepsilon_0(\phi) & = - \frac{1}{2\pi} \sum_{p \in \mathbb{Z}} T_p \, \text{e}^{-\text{i}p\phi} \, , \qquad &\alpha_0(\phi) & = - \frac{1}{2\pi} \sum_{p \in \mathbb{Z}} Y_p \, \text{e}^{-\text{i}p\phi} \, ,\\
B_\phi(\phi) & = - \frac{1}{2\pi} \sum_{p \in \mathbb{Z}} P_{p} \, \text{e}^{-\text{i} p \phi} \, , \qquad & B_u(\phi) & = - \frac{1}{2\pi} \sum_{p \in \mathbb{Z}} U_{p} \, \text{e}^{-\text{i} p \phi} \, ,
\end{aligned}
\end{equation}
one obtains the following $\mathfrak{bms}_3$ $\oplus$ $\hat{\mathfrak{u}}(1)$ $\oplus$ $\hat{\mathfrak{u}}(1)$ algebra:
\begin{subequations}
\begin{align}
\text{i} \, \{Y_p,Y_q\} & = (p-q) \, Y_{p+q} \, ,\\
\text{i} \, \{Y_p,T_q\} & = (p-q) \, T_{p+q} + \frac{c_M}{12} \, p^3 \, \delta_{p+q,0} \, ,\\
\text{i} \, \{P_{p},U_{q}\} & = \frac{c_M}{12} \, p \, \delta_{p+q,0} \, , \label{CS - Poisson Bracket two u(1)}
\end{align}
\end{subequations}
with the same central charge as in eq.~\eqref{central-charge-flat}. The presence of two affine $\mathfrak{u}(1)$ current algebras can be made more explicit by redefining the modes
\begin{equation}
B_p^\pm = \frac{1}{\sqrt{2}} \left( U_p \pm P_p \right) ,
\end{equation}
which brings the last Poisson bracket \eqref{CS - Poisson Bracket two u(1)} in the form
\begin{equation}
\text{i} \, \left\{B_{p}^\pm,B_{q}^\pm\right\} = \frac{c_M}{12} \, p \, \delta_{p+q,0} \, .
\end{equation}
This leads to the asymptotic symmetry algebra of \cite{Detournay:2016sfv}, which is the generalization of the analysis of \cite{Troessaert:2013fma} to asymptotically flat spacetimes. We should however emphasize that, in analogy with the AdS case, this algebra is reached starting from different boundary conditions. The two affine $\mathfrak{u}(1)$s originate from the infinite-dimensional extensions of the boundary Carroll-boost symmetry, while in \cite{Detournay:2016sfv} they were associated with the boundary Weyl--Carroll symmetry. The same asymptotic symmetry algebra appears in \cite{Grumiller:2017sjh} --- obtained imposing a well-posed variational principle for gravity. One finally recovers the two special instances treated in \cite{Campoleoni:2018ltl} for a reduced phase space, where only one of the pairs of functions $\varepsilon_0$, $\alpha_0$ or $B_\phi$, $B_u$ was considered, and the boundary conformal factor was turned off.

\paragraph{Boundary anomalies} The first-order formulation is appropriate for designing an action that includes the necessary boundary term, reproducing the on-shell variation \eqref{flat-anomaly} in the conformal gauge. This was performed  in section~\ref{subsec:first-order - AdS} for anti-de Sitter. Here, the boundary term to be added to the bulk action \eqref{CS_generic} is
\begin{equation}
S_{\text{bdy}}[\mathscr{A}] = \frac{1}{8\pi\GN} \int \text{d}^2 x \, \text{Tr} \big( \mathscr{A}_\phi \, \mathscr{A}_u \big)
\end{equation}
and the on-shell variation of the total action is
\begin{equation}
\delta S_\text{tot}[\mathscr{A}] = \delta S_{\text{EH}}[\mathscr{A}] + \delta S_{\text{bdy}}[\mathscr{A}] = - \frac{1}{8\pi \GN} \int \delta \beta \, \text{e}^{-2 \varphi} \, \partial_u^2 \beta \, \text{vol}_{\partial \cM} \, ,
\end{equation}
where $\text{vol}_{\partial \cM}$ is the boundary volume form \eqref{boundary-volume_flat}. Unsurprisingly, the remaining term is not integrable and corresponds to the flat limit of \eqref{CS - delta S tot in AdS}. Hence, no further boundary term could help making the variational principle well-posed. Setting the Weyl--Carroll curvature to zero is instead a sufficient boundary condition for that purpose. In this case, the Carroll boost splits as in \eqref{CS - split of Carroll boost}, and we are back to the analogue of the analysis performed in \cite{Detournay:2016sfv} for the Weyl--Carroll sector.

In concluding the present analysis, let us emphasize that a non-trivial on-shell variation of the action should not be banned ipso facto. It may betray an anomalous physical contribution. This phenomenon is well-understood in anti-de Sitter  holography \cite{Henningson:1998gx, deHaro:2000vlm}, and is possibly of interest in the asymptotically flat context, where it emerges as a quantum feature of the assumed conformal Carrollian field theory defined on the null boundary. This is a notable prediction of our investigation.

\section{Conclusion}

The aim of the present article was to explore the charges of three-dimensional gravity in a manifestly covariant gauge with respect to boundary diffeomorphisms, Weyl rescalings and local frame boosts, whether Lorentz or Carroll --- in AdS or flat spacetimes --- all rooted in the bulk residual diffeomorphisms. This has been elegantly achieved using boundary Cartan dyads, which make the bulk gauge fixing incomplete with seemingly more boundary degrees of freedom, and open the Pandora box for addressing related questions: boundary conditions and boundary terms in the bulk action, variational principle, integrability and conservation of surface charges, asymptotic symmetry algebras, anomalies and their numerous manifestations, interplay between second-order metric and first-order Chern--Simons approaches.

Our first task has been the determination of the presymplectic potential in asymptotically anti-de Sitter spacetimes. This action raised immediately the question of treating ambiguities. Two distinct prescriptions emerged, leading to finite results. The first reproduces the same outcome of the Fefferman--Graham gauge, and Weyl rescalings become part of asymptotic symmetries while Lorentz boosts turn out to be pure gauge. 
The second promotes the hyperbolic rotations of the Cartan frame to genuine asymptotic symmetries, but Weyl rescalings become pure gauge. Differently from the previous one, this prescription admits a finite flat limit, in which the Carrollian boosts of the frame become part of the asymptotically flat symmetries.   

Before pursuing the summary of our achievements, one should emphasize that the latter result is rather intriguing. It seems indeed that the covariant, incomplete gauge we have set departing from Bondi is tailor-made to fix the ambiguities of the presymplectic potential, provided that one requires that the smoothness of the flat limit holds not only for the solution space but also for its symplectic structure. 
Additionally, our partial gauge fixing allows to understand the boundary conformal anomaly, appearing e.g.\ as an obstruction to the vanishing of on-shell variations of the action, either from Weyl or from Lorentz/Carroll perspectives, owing to the underlying cohomology properties. Notice however that these two viewpoints exhibit, once promoted in the bulk, a striking dissymmetry, and it is not clear whether a formulation --- i.e.\ a gauge --- exists, for which the presymplectic potential would have a finite flat limit respecting the Weyl instead of Lorentz symmetry, without eliminating the latter from the very beginning as in \cite{Detournay:2016sfv} or in its possible generalisation along the lines of \cite{Alessio:2020ioh}. 
Finally, it is suggested that a new anomaly should exist in Carrollian conformal field theories, out of reach for the moment within a genuine quantum computation. How all these issues should be understood in higher-dimensional theories, where boundary frames transform with the Lorentz or Carroll groups or, more generally, with the general linear group, deserves further investigation.

Computing the asymptotic charges was the next enterprise and possibly the original motivation of this work. This was performed in the first place in the metric approach. For a generic Cartan frame ---~be it Lorentz or Carroll, i.e.\ for asymptotically anti-de Sitter or flat spacetimes~--- the charges are neither integrable nor conserved. This is not a pathology, but rather a distinctive feature of the solution space, reflecting among others the gauge choice. When the boundary frame is tuned to the conformal gauge, the surface charges become however integrable. For the presymplectic potential  admitting a smooth flat limit, Weyl transformations are uncharged, as opposed to Lorentz/Carroll transformations, which provide integrable but non-conserved charges.
Their non-conservation is yet another avatar of the associated anomaly.

The first-order Chern--Simons formulation is the last side of our analysis. Its main advantages are the absence of holographic renormalization and the option of a Chern--Simons connection that fits naturally the Lorentz/Carroll presymplectic potential found in the metric approach. Its indisputable added value in this work is the concrete determination of the boundary term in the action required to obtain the specific presymplectic potential in use. 

The achievements we have presented here compose a wide picture that embraces several
previous works on asymptotic symmetries
in anti-de Sitter or flat three-dimensional spacetimes and completes the analysis initiated in \cite{Campoleoni:2018ltl, Ciambelli:2020eba, Ciambelli:2020ftk}. We have made a rewarding step beyond the standard complete gauge fixing; whether this is to be considered as the ``most general'' remains unclear.

We would like to terminate by highlighting some aspects of our results that we think deserve further study. As explained at length, relaxing the Bondi gauge and making a specific choice of presymplectic potential leads to a new boundary degree of freedom, which corresponds to rotations of the boundary dyad, at the expense of gauging the Weyl factor of the boundary metric. Therefore the total number of physical degrees of freedom remains unaltered. This phenomenon would have been hard to guess prior to computing the presymplectic potential explicitly. However it is in agreement with some counting arguments \cite{Grumiller:2020vvv, Adami:2020ugu} predicting that the maximal number of boundary degrees of freedom in three dimensions is equal to three, corresponding here to the number of generators of boundary diffeomorphisms and rotations of the dyad. 
It would be of general interest for the asymptotic-symmetry program to handle the conditions under which no extra degrees of freedom emerge from a gauge relaxation. These conditions might be the requirement of a finite presymplectic potential, vanishing under Dirichlet boundary conditions. This might also be related to recent developments in the corner proposal \cite{Donnelly:2016auv, Speranza:2017gxd,Geiller:2017whh,Freidel:2020xyx,Freidel:2020svx,Freidel:2020ayo,Donnelly:2020xgu,Ciambelli:2021vnn,Freidel:2021cjp,Ciambelli:2021nmv, Ciambelli:2022cfr}, where the so-called asymptotic corner symmetry group gives a maximal number of degrees of freedom that a theory of gravity can have. In this context, it is the Weyl symmetry that contributes to the asymptotic corner symmetry group, and understanding how this symmetry generator is intertwined with the local Lorentz generator can have important repercussions, especially in the coadjoint orbit method. It would be also interesting, following \cite{Donnay:2022aba,Bagchi:2022emh},  to connect the results on Carrollian holography discussed in this paper with the celestial holography proposal (see, e.g., \cite{Strominger:2017zoo, Pasterski:2021raf, McLoughlin:2022ljp} for reviews).

Another feature, illustrated here and expected to hold more generally, is the relation between 
presymplectic potentials descending from the same bulk action and cohomological classes of anomalies. In anti-de Sitter, an appropriate prescription for the ambiguous terms in the potential triggers a switch of the anomalous symmetry from Weyl to Lorentz. This is well-known to occur in field theory, see e.g.\ the two-dimensional analysis of \cite{Jackiw:1995qh}, where a change in the path-integral measure leads to a displacement of the anomaly from the Weyl symmetry to the diffeomorphisms. Our example points to the existence of general rules relating holographically this boundary displacement to a modification of the bulk symplectic structure.

Last, we would like to comment on the aforementioned extension to higher dimensions, where the Weyl and Lorentz groups are no longer isomorphic. In bulk dimension four, for instance, the former is still parameterized by one function of the boundary coordinates, while the latter contains three functions: one rotation and two boosts. The strict parallelism that was drawn in the present work between Weyl and Lorentz is disrupted. Although we expect a relaxation of the Bondi/Newmann--Unti gauge to persist giving rise to a physical Cartan frame on the boundary, the absence of Weyl anomaly blurs the three-dimensional pattern.

\paragraph{Acknowledgments} We would like to thank Francesco Alessio, Glenn Barnich, Adrien Fiorucci, Dario Francia, Marc Geiller, Daniel Grumiller, Carlo Heissenberg, Gerben Oling, Simon Pekar, Shahin Sheikh-Jabbari and C\'eline Zwikel for rich scientific exchanges. Luca Ciambelli would like to thank Rob Leigh and Weizhen Jia for useful discussions and for allowing him to share unpublished material. Andrea Campoleoni and Arnaud Delfante thank the Universit\'e Libre de Bruxelles and Marios Petropoulos thanks the Universit\'e de Mons for hospitality during various stages of this work. We all thank the organizers and participants of the Carroll workshop at TU Wien, where some preliminary results have been presented. The research of Andrea Campoleoni and Arnaud Delfante was partially supported by the Fonds de la Recherche Scientifique -- FNRS under Grants No.\ FC.41161, F.4503.20 (HighSpinSymm) and T.0022.19 (Fundamental issues in extended gravitational theories). Charles Marteau acknowledges support from NSERC. The research of Luca Ciambelli was partially supported by a Marina Solvay Fellowship, by the ERC Advanced Grant ``High-Spin-Grav'' and by the Fonds de la Recherche Scientifique -- FNRS (conventions FRFC PDR T.1025.14 and IISN 4.4503.15). Romain Ruzziconi  was supported by the Austrian Science Fund (FWF) START project Y 1447-N.

\appendix

\section{Classification of Weyl--Lorentz anomalies in two dimensions}\label{app:BRST}

In this appendix, we  explain how to obtain the anomaly structure \eqref{LW} applying BRST methods to field theories with Weyl--Lorentz symmetry. The first-order formulation of the problem is the core of a forthcoming paper, \cite{preprint_Luca}, from which these results are taken. 
Since the BRST tools employed in the classification are standard, we focus here on the novelty, which is their application to field theories with Weyl--Lorentz symmetries, referring the reader to \cite{Henneaux:1992ig} for more details on the techniques (see also \cite{Boulanger:2007ab, Boulanger:2007st} for examples of their usage in similar contexts).

In the BRST formulation, each field theory is associated with a BRST bicomplex, with exterior derivative
\beq
\hat \D=\D+s \, ,
\eeq
where $\D$ is the de Rham exterior derivative and $s$ the BRST operator. We use the notation $(p,q)$ for the bigrading, where $p$ is the (vertical) ghost number and $q$ the (horizontal) de Rham number. A Weyl--Lorentz structure can be seen as a $G$-structure in the frame bundle given by the total connection
\beq
\Upomega^a{}_b=\upomega^a{}_b+\text{A}\,\delta^a{}_b = \upomega\, \epsilon^a{}_b +\text{A} \, \delta^a{}_b \, .
\eeq
In two dimensions, the skew-symmetric Lorentz connection $\upomega^a{}_b$ has one independent component $\upomega$, and  $\text{A}$ denotes the Weyl connection. This is extended to the full bicomplex by adding the Lorentz ghost $\lambda$ and Weyl ghost $z$,
\beq \label{bicomplex}
\hat\Upomega^a{}_b=\hat\upomega\,\epsilon^a{}_b + \hat{\text{A}}\, \delta^a{}_b=(\upomega+\lambda)\,\epsilon^a{}_b + (\text{A}+z)\,\delta^a{}_b \, ,
\eeq
such that $\hat\upomega$ has degree $1$, $\upomega$  degree $(0,1)$, $\lambda$ degree $(1,0)$, and similarly for $\hat{\text{A}}$, $\text{A}$, and $z$. The total curvature is $\hat \D \hat \Upomega^a{}_b$. As a consequence of the Darboux--Maurer--Cartan--Ehresmann (DMCE) condition, it is required to be totally horizontal, that is, to contain only $(0,2)$ terms. This gives, as usual, the BRST transformations of the fields. 

Anomalies are  extracted from the cohomology of the space of local functionals that are top forms in spacetime with ghost number one, i.e., of degree $(1,2)$. For that purpose, one constructs the most general functional of total degree $3$, called $\hat{\text{a}}^{(3)}$, and solves the cohomology
\beq
\hat \D \, \hat{\text{a}}^{(3)}=0, \qquad \hat{\text{a}}^{(3)}\neq \hat \D \, \hat{\text{b}}^{(2)} \, ,
\eeq
where $\hat{\text{b}}^{(2)}$ is a co-boundary, i.e., a functional of total degree $2$. The $(1,2)$ term inside the bicomplex decomposition of $\hat{\text{a}}^{(3)}$ gives all possible anomalies and central charges of a theory with these symmetries.

In our specific case, setting $\hat{\text{F}}=\hat \D\hat{ \text{A}}$ and $\hat{\text{R}}=\hat \D\hat \upomega$, the most general $3$-form is
\beq
\hat{\text{a}}^{(3)}=c_1\, \hat \upomega \, \hat{\text{R}}+ c_2\, \hat{\text{A}} \, \hat{\text{F}} + \gamma_3\, \hat{\text{A}} \, \hat{\text{R}} + \gamma_4\, \hat \upomega \, \hat{\text{F}} \, ,
\eeq
where $\{c_1,c_2,\gamma_3,\gamma_4\}$ are the potential central charges. Solving its cohomology, we first note that
\beq
\hat \D\hat{\text{a}}^{(3)}=c_1 \,  \hat{\text{F}} \, \hat{\text{R}}+c_2 \, \hat{\text{F}} \, \hat{\text{F}}+\gamma_3 \, \hat{\text{F}} \, \hat{\text{R}}+\gamma_4 \, \hat{\text{R}} \, \hat{\text{F}}
=c_1 \, \text{F} \, \text{R}+c_2 \, \text{F} \, \text{F}+\gamma_3 \, \text{F} \, \text{R}+\gamma_4 \, \text{R} \, \text{F}=0 \, ,
\eeq
where we used $\hat{\D}^2=0$ and the DMCE condition. The functional is thus closed and it remains to remove $\hat \D$-exact terms (co-boundaries) in $\hat{\text{a}}^{(3)}$. This involves notably the mixed contributions 
\beq
\gamma_3 \, \hat{\text{A}} \, \hat{\text{R}}+\gamma_4 \, \hat \upomega \, \hat{\text{F}}=\gamma_3 \, \hat{\text{A}} \, \hat{\text{R}}-\gamma_4 \, \hat \D \,\big(\hat \upomega \, \hat{\text{A}}\big)+\gamma_4 \, \hat{\text{A}} \, \hat{\text{R}}=(\gamma_3+\gamma_4) \, \hat{\text{A}} \, \hat{\text{R}}-\gamma_4 \, \hat \D \, \big(\hat \upomega \, \hat{\text{A}}\big)\, .
\eeq
This instructs us that $\gamma_3$ and $\gamma_4$ are not independent central charges. Furthermore, it shows that for the horizontal quantities (using again the DMCE condition), $z \text{R}$ and $\lambda \text{F}$ are two representatives in the same cohomology class. Calling $\gamma_3+\gamma_4=\frac{c_3}{12\pi}$, and expanding in the horizontal and vertical fields, we can extract the final anomaly structure
\beq \label{BRST-anomaly}
\text{a}^{(1,2)}= c_1\,  \lambda\, \text{R} + c_2\, z\, \text{F} + \frac{c_3}{12\pi}\, z\, \text{R} \, ,
\eeq
where the last term is shown to be cohomologically equivalent to $\frac{c_3}{12\pi} \lambda \text{F}$. This equation,
derived in \cite{preprint_Luca}, is the starting point in section \ref{sec:DE} to match with the holographic result, upon performing the identification $\lambda=\eta$ and $z=\sigma$, and taking the Hodge dual of the curvatures, which gives
\beq
\ast \text{F}=F, \qquad \ast\text{R}=\frac{R}{2} \, ,
\eeq
where $F$ and $R$ (the Ricci scalar) are defined in \eqref{sdA}.


\bibliographystyle{uiuchept}
\bibliography{References}

\providecommand{\href}[2]{#2}\begingroup\raggedright\begin{thebibliography}{100}

\bibitem{Brown:1986nw}
J.~D. Brown and M.~Henneaux, ``{Central Charges in the Canonical Realization of
  Asymptotic Symmetries: An Example from Three-Dimensional Gravity},''
  \href{http://dx.doi.org/10.1007/BF01211590}{{\em Commun. Math. Phys.} {\bf
  104} (1986)  207--226}.

\bibitem{FG1}
C.~Fefferman and C.~R. Graham, ``Conformal invariants,'' in {\em \'Elie Cartan
  et les math\'ematiques d' aujourd'hui - Lyon, 25-29 juin 1984}, no.~S131 in
  Ast\'erisque, pp.~95--116.
\newblock Soci\'et\'e math\'ematique de France, 1985.

\bibitem{Fefferman:2007rka}
C.~Fefferman and C.~R. Graham, ``{The ambient metric},'' {\em Ann. Math. Stud.}
  {\bf 178} (2011)  1--128, \href{http://arxiv.org/abs/0710.0919}{{\tt
  arXiv:0710.0919 [math.DG]}}.

\bibitem{Troessaert:2013fma}
C.~Troessaert, ``{Enhanced asymptotic symmetry algebra of $AdS_{3}$},''
  \href{http://dx.doi.org/10.1007/JHEP08(2013)044}{{\em JHEP} {\bf 08} (2013)
  044}, \href{http://arxiv.org/abs/1303.3296}{{\tt arXiv:1303.3296 [hep-th]}}.

\bibitem{Alessio:2020ioh}
F.~Alessio, G.~Barnich, L.~Ciambelli, P.~Mao, and R.~Ruzziconi, ``{Weyl charges
  in asymptotically locally AdS$_3$ spacetimes},''
  \href{http://dx.doi.org/10.1103/PhysRevD.103.046003}{{\em Phys. Rev. D} {\bf
  103} (2021) no.~4, 046003}, \href{http://arxiv.org/abs/2010.15452}{{\tt
  arXiv:2010.15452 [hep-th]}}.

\bibitem{Compere:2013bya}
G.~Comp\`ere, W.~Song, and A.~Strominger, ``{New Boundary Conditions for
  AdS3},'' \href{http://dx.doi.org/10.1007/JHEP05(2013)152}{{\em JHEP} {\bf 05}
  (2013)  152}, \href{http://arxiv.org/abs/1303.2662}{{\tt arXiv:1303.2662
  [hep-th]}}.

\bibitem{Perez:2016vqo}
A.~P\'erez, D.~Tempo, and R.~Troncoso, ``{Boundary conditions for General
  Relativity on AdS$_{3}$ and the KdV hierarchy},''
  \href{http://dx.doi.org/10.1007/JHEP06(2016)103}{{\em JHEP} {\bf 06} (2016)
  103}, \href{http://arxiv.org/abs/1605.04490}{{\tt arXiv:1605.04490
  [hep-th]}}.

\bibitem{Afshar:2016kjj}
H.~Afshar, D.~Grumiller, W.~Merbis, A.~Perez, D.~Tempo, and R.~Troncoso,
  ``{Soft hairy horizons in three spacetime dimensions},''
  \href{http://dx.doi.org/10.1103/PhysRevD.95.106005}{{\em Phys. Rev. D} {\bf
  95} (2017) no.~10, 106005}, \href{http://arxiv.org/abs/1611.09783}{{\tt
  arXiv:1611.09783 [hep-th]}}.

\bibitem{Ojeda:2019xih}
E.~Ojeda and A.~P\'erez, ``{Boundary conditions for General Relativity in
  three-dimensional spacetimes, integrable systems and the KdV/mKdV
  hierarchies},'' \href{http://dx.doi.org/10.1007/JHEP08(2019)079}{{\em JHEP}
  {\bf 08} (2019)  079}, \href{http://arxiv.org/abs/1906.11226}{{\tt
  arXiv:1906.11226 [hep-th]}}.

\bibitem{Donnay:2015abr}
L.~Donnay, G.~Giribet, H.~A. Gonzalez, and M.~Pino, ``{Supertranslations and
  Superrotations at the Black Hole Horizon},''
  \href{http://dx.doi.org/10.1103/PhysRevLett.116.091101}{{\em Phys. Rev.
  Lett.} {\bf 116} (2016) no.~9, 091101},
\href{http://arxiv.org/abs/1511.08687}{{\tt arXiv:1511.08687 [hep-th]}}.

\bibitem{Grumiller:2019fmp}
D.~Grumiller, A.~P\'erez, M.~Sheikh-Jabbari, R.~Troncoso, and C.~Zwikel,
  ``{Spacetime structure near generic horizons and soft hair},''
  \href{http://dx.doi.org/10.1103/PhysRevLett.124.041601}{{\em Phys. Rev.
  Lett.} {\bf 124} (2020) no.~4, 041601},
  \href{http://arxiv.org/abs/1908.09833}{{\tt arXiv:1908.09833 [hep-th]}}.

\bibitem{Adami:2020ugu}
H.~Adami, M.~M. Sheikh-Jabbari, V.~Taghiloo, H.~Yavartanoo, and C.~Zwikel,
  ``{Symmetries at null boundaries: two and three dimensional gravity cases},''
  \href{http://dx.doi.org/10.1007/JHEP10(2020)107}{{\em JHEP} {\bf 10} (2020)
  107}, \href{http://arxiv.org/abs/2007.12759}{{\tt arXiv:2007.12759
  [hep-th]}}.

\bibitem{Adami:2020amw}
H.~Adami, D.~Grumiller, S.~Sadeghian, M.~M. Sheikh-Jabbari, and C.~Zwikel,
  ``{T-Witts from the horizon},''
  \href{http://dx.doi.org/10.1007/JHEP04(2020)128}{{\em JHEP} {\bf 04} (2020)
  128}, \href{http://arxiv.org/abs/2002.08346}{{\tt arXiv:2002.08346
  [hep-th]}}.

\bibitem{Adami:2021nnf}
H.~Adami, D.~Grumiller, M.~M. Sheikh-Jabbari, V.~Taghiloo, H.~Yavartanoo, and
  C.~Zwikel, ``{Null boundary phase space: slicings, news \& memory},''
  \href{http://dx.doi.org/10.1007/JHEP11(2021)155}{{\em JHEP} {\bf 11} (2021)
  155}, \href{http://arxiv.org/abs/2110.04218}{{\tt arXiv:2110.04218
  [hep-th]}}.

\bibitem{Grumiller:2016pqb}
D.~Grumiller and M.~Riegler, ``{Most general AdS$_{3}$ boundary conditions},''
  \href{http://dx.doi.org/10.1007/JHEP10(2016)023}{{\em JHEP} {\bf 10} (2016)
  023}, \href{http://arxiv.org/abs/1608.01308}{{\tt arXiv:1608.01308
  [hep-th]}}.

\bibitem{Papadimitriou:2005ii}
I.~Papadimitriou and K.~Skenderis, ``{Thermodynamics of asymptotically locally
  AdS spacetimes},''
  \href{http://dx.doi.org/10.1088/1126-6708/2005/08/004}{{\em JHEP} {\bf 08}
  (2005)  004}, \href{http://arxiv.org/abs/hep-th/0505190}{{\tt
  arXiv:hep-th/0505190}}.

\bibitem{Ciambelli:2019bzz}
L.~Ciambelli and R.~G. Leigh, ``{Weyl Connections and their Role in
  Holography},'' \href{http://dx.doi.org/10.1103/PhysRevD.101.086020}{{\em
  Phys. Rev. D} {\bf 101} (2020) no.~8, 086020},
  \href{http://arxiv.org/abs/1905.04339}{{\tt arXiv:1905.04339 [hep-th]}}.

\bibitem{Fiorucci:2020xto}
A.~Fiorucci and R.~Ruzziconi, ``{Charge algebra in Al(A)dS$_{n}$ spacetimes},''
  \href{http://dx.doi.org/10.1007/JHEP05(2021)210}{{\em JHEP} {\bf 05} (2021)
  210}, \href{http://arxiv.org/abs/2011.02002}{{\tt arXiv:2011.02002
  [hep-th]}}.

\bibitem{Ruzziconi:2020wrb}
R.~Ruzziconi and C.~Zwikel, ``{Conservation and Integrability in
  Lower-Dimensional Gravity},''
  \href{http://dx.doi.org/10.1007/JHEP04(2021)034}{{\em JHEP} {\bf 04} (2021)
  034}, \href{http://arxiv.org/abs/2012.03961}{{\tt arXiv:2012.03961
  [hep-th]}}.

\bibitem{Geiller:2021vpg}
M.~Geiller, C.~Goeller, and C.~Zwikel, ``{3d gravity in Bondi-Weyl gauge:
  charges, corners, and integrability},''
  \href{http://dx.doi.org/10.1007/JHEP09(2021)029}{{\em JHEP} {\bf 09} (2021)
  029}, \href{http://arxiv.org/abs/2107.01073}{{\tt arXiv:2107.01073
  [hep-th]}}.

\bibitem{Campoleoni:2018ltl}
A.~Campoleoni, L.~Ciambelli, C.~Marteau, P.~M. Petropoulos, and K.~Siampos,
  ``{Two-dimensional fluids and their holographic duals},''
  \href{http://dx.doi.org/10.1016/j.nuclphysb.2019.114692}{{\em Nucl. Phys. B}
  {\bf 946} (2019)  114692}, \href{http://arxiv.org/abs/1812.04019}{{\tt
  arXiv:1812.04019 [hep-th]}}.

\bibitem{Ciambelli:2020eba}
L.~Ciambelli, C.~Marteau, P.~M. Petropoulos, and R.~Ruzziconi, ``{Gauges in
  Three-Dimensional Gravity and Holographic Fluids},''
  \href{http://dx.doi.org/10.1007/JHEP11(2020)092}{{\em JHEP} {\bf 11} (2020)
  092}, \href{http://arxiv.org/abs/2006.10082}{{\tt arXiv:2006.10082
  [hep-th]}}.

\bibitem{Ciambelli:2020ftk}
L.~Ciambelli, C.~Marteau, P.~M. Petropoulos, and R.~Ruzziconi,
  ``{Fefferman-Graham and Bondi Gauges in the Fluid/Gravity Correspondence},''
  \href{http://dx.doi.org/10.22323/1.376.0154}{{\em PoS} {\bf CORFU2019} (2020)
   154}, \href{http://arxiv.org/abs/2006.10083}{{\tt arXiv:2006.10083
  [hep-th]}}.

\bibitem{Ciambelli:2018wre}
L.~Ciambelli, C.~Marteau, A.~C. Petkou, P.~M. Petropoulos, and K.~Siampos,
  ``{Flat holography and Carrollian fluids},''
  \href{http://dx.doi.org/10.1007/JHEP07(2018)165}{{\em JHEP} {\bf 07} (2018)
  165}, \href{http://arxiv.org/abs/1802.06809}{{\tt arXiv:1802.06809
  [hep-th]}}.

\bibitem{deHaro:2000vlm}
S.~de~Haro, S.~N. Solodukhin, and K.~Skenderis, ``{Holographic reconstruction
  of space-time and renormalization in the AdS / CFT correspondence},''
  \href{http://dx.doi.org/10.1007/s002200100381}{{\em Commun. Math. Phys.} {\bf
  217} (2001)  595--622},
\href{http://arxiv.org/abs/hep-th/0002230}{{\tt arXiv:hep-th/0002230
  [hep-th]}}.

\bibitem{Skenderis:2002wp}
K.~Skenderis, ``{Lecture notes on holographic renormalization},''
  \href{http://dx.doi.org/10.1088/0264-9381/19/22/306}{{\em Class. Quant.
  Grav.} {\bf 19} (2002)  5849--5876},
  \href{http://arxiv.org/abs/hep-th/0209067}{{\tt arXiv:hep-th/0209067}}.

\bibitem{Imbimbo:1999bj}
C.~Imbimbo, A.~Schwimmer, S.~Theisen, and S.~Yankielowicz, ``{Diffeomorphisms
  and holographic anomalies},''
  \href{http://dx.doi.org/10.1088/0264-9381/17/5/322}{{\em Class. Quant. Grav.}
  {\bf 17} (2000)  1129--1138}, \href{http://arxiv.org/abs/hep-th/9910267}{{\tt
  arXiv:hep-th/9910267}}.

\bibitem{Schwimmer:2000cu}
A.~Schwimmer and S.~Theisen, ``{Diffeomorphisms, anomalies and the
  Fefferman-Graham ambiguity},''
  \href{http://dx.doi.org/10.1088/1126-6708/2000/08/032}{{\em JHEP} {\bf 08}
  (2000)  032}, \href{http://arxiv.org/abs/hep-th/0008082}{{\tt
  arXiv:hep-th/0008082}}.

\bibitem{Rooman:2000ei}
M.~Rooman and P.~Spindel, ``{Uniqueness of the asymptotic AdS(3) geometry},''
  \href{http://dx.doi.org/10.1088/0264-9381/18/11/309}{{\em Class. Quant.
  Grav.} {\bf 18} (2001)  2117--2124},
  \href{http://arxiv.org/abs/gr-qc/0011005}{{\tt arXiv:gr-qc/0011005}}.

\bibitem{Rooman:2000zi}
M.~Rooman and P.~Spindel, ``{Holonomies, anomalies and the Fefferman-Graham
  ambiguity in AdS(3) gravity},''
  \href{http://dx.doi.org/10.1016/S0550-3213(00)00636-2}{{\em Nucl. Phys. B}
  {\bf 594} (2001)  329--353}, \href{http://arxiv.org/abs/hep-th/0008147}{{\tt
  arXiv:hep-th/0008147}}.

\bibitem{Anastasiou:2020zwc}
G.~Anastasiou, O.~Miskovic, R.~Olea, and I.~Papadimitriou, ``{Counterterms,
  Kounterterms, and the variational problem in AdS gravity},''
  \href{http://dx.doi.org/10.1007/JHEP08(2020)061}{{\em JHEP} {\bf 08} (2020)
  061}, \href{http://arxiv.org/abs/2003.06425}{{\tt arXiv:2003.06425
  [hep-th]}}.

\bibitem{Jia:2021hgy}
W.~Jia and M.~Karydas, ``{Obstruction tensors in Weyl geometry and holographic
  Weyl anomaly},'' \href{http://dx.doi.org/10.1103/PhysRevD.104.126031}{{\em
  Phys. Rev. D} {\bf 104} (2021) no.~12, 126031},
  \href{http://arxiv.org/abs/2109.14014}{{\tt arXiv:2109.14014 [hep-th]}}.

\bibitem{Henningson:1998gx}
M.~Henningson and K.~Skenderis, ``{The Holographic Weyl anomaly},''
  \href{http://dx.doi.org/10.1088/1126-6708/1998/07/023}{{\em JHEP} {\bf 07}
  (1998)  023}, \href{http://arxiv.org/abs/hep-th/9806087}{{\tt
  arXiv:hep-th/9806087}}.

\bibitem{Bondi:1962px}
H.~Bondi, M.~G.~J. van~der Burg, and A.~W.~K. Metzner, ``{Gravitational waves
  in general relativity. 7. Waves from axisymmetric isolated systems},''
\href{http://dx.doi.org/10.1098/rspa.1962.0161}{{\em Proc. Roy. Soc. Lond.}
  {\bf A269} (1962)  21}.

\bibitem{Sachs:1962wk}
R.~K. Sachs, ``{Gravitational waves in general relativity. 8. Waves in
  asymptotically flat space-times},''
\href{http://dx.doi.org/10.1098/rspa.1962.0206}{{\em Proc. Roy. Soc. Lond.}
  {\bf A270} (1962)  103--126}.

\bibitem{Sachs:1962zza}
R.~Sachs, ``{Asymptotic symmetries in gravitational theory},''
\href{http://dx.doi.org/10.1103/PhysRev.128.2851}{{\em Phys. Rev.} {\bf 128}
  (1962)  2851--2864}.

\bibitem{Barnich:2012aw}
G.~Barnich, A.~Gomberoff, and H.~A. Gonzalez, ``{The Flat limit of three
  dimensional asymptotically anti-de Sitter spacetimes},''
  \href{http://dx.doi.org/10.1103/PhysRevD.86.024020}{{\em Phys. Rev. D} {\bf
  86} (2012)  024020}, \href{http://arxiv.org/abs/1204.3288}{{\tt
  arXiv:1204.3288 [gr-qc]}}.

\bibitem{Barnich:2013sxa}
G.~Barnich and P.-H. Lambert, ``{Einstein-Yang-Mills theory: Asymptotic
  symmetries},'' \href{http://dx.doi.org/10.1103/PhysRevD.88.103006}{{\em Phys.
  Rev. D} {\bf 88} (2013)  103006}, \href{http://arxiv.org/abs/1310.2698}{{\tt
  arXiv:1310.2698 [hep-th]}}.

\bibitem{Poole:2018koa}
A.~Poole, K.~Skenderis, and M.~Taylor, ``{(A)dS$\mathbf{_4}$ in Bondi gauge},''
  \href{http://dx.doi.org/10.1088/1361-6382/ab117c}{{\em Class. Quant. Grav.}
  {\bf 36} (2019) no.~9, 095005}, \href{http://arxiv.org/abs/1812.05369}{{\tt
  arXiv:1812.05369 [hep-th]}}.

\bibitem{Compere:2019bua}
G.~Comp\`ere, A.~Fiorucci, and R.~Ruzziconi, ``{The $\Lambda$-BMS$_4$ group of
  dS$_4$ and new boundary conditions for AdS$_4$},''
  \href{http://dx.doi.org/10.1088/1361-6382/ab3d4b}{{\em Class. Quant. Grav.}
  {\bf 36} (2019) no.~19, 195017}, \href{http://arxiv.org/abs/1905.00971}{{\tt
  arXiv:1905.00971 [gr-qc]}}. [Erratum: Class.Quant.Grav. 38, 229501 (2021)].

\bibitem{Compere:2020lrt}
G.~Comp\`ere, A.~Fiorucci, and R.~Ruzziconi, ``{The $\Lambda$-BMS$_4$ charge
  algebra},'' \href{http://dx.doi.org/10.1007/JHEP10(2020)205}{{\em JHEP} {\bf
  10} (2020)  205}, \href{http://arxiv.org/abs/2004.10769}{{\tt
  arXiv:2004.10769 [hep-th]}}.

\bibitem{Geiller:2022vto}
M.~Geiller and C.~Zwikel, ``{The partial Bondi gauge: Further enlarging the
  asymptotic structure of gravity},''
  \href{http://arxiv.org/abs/2205.11401}{{\tt arXiv:2205.11401 [hep-th]}}.

\bibitem{Barnich:2012xq}
G.~Barnich, ``{Entropy of three-dimensional asymptotically flat cosmological
  solutions},'' \href{http://dx.doi.org/10.1007/JHEP10(2012)095}{{\em JHEP}
  {\bf 10} (2012)  095}, \href{http://arxiv.org/abs/1208.4371}{{\tt
  arXiv:1208.4371 [hep-th]}}.

\bibitem{Bagchi:2012xr}
A.~Bagchi, S.~Detournay, R.~Fareghbal, and J.~Sim\'on, ``{Holography of 3D Flat
  Cosmological Horizons},''
  \href{http://dx.doi.org/10.1103/PhysRevLett.110.141302}{{\em Phys. Rev.
  Lett.} {\bf 110} (2013) no.~14, 141302},
  \href{http://arxiv.org/abs/1208.4372}{{\tt arXiv:1208.4372 [hep-th]}}.

\bibitem{Bagchi:2012cy}
A.~Bagchi and R.~Fareghbal, ``{BMS/GCA Redux: Towards Flatspace Holography from
  Non-Relativistic Symmetries},''
  \href{http://dx.doi.org/10.1007/JHEP10(2012)092}{{\em JHEP} {\bf 10} (2012)
  092}, \href{http://arxiv.org/abs/1203.5795}{{\tt arXiv:1203.5795 [hep-th]}}.

\bibitem{Detournay:2014fva}
S.~Detournay, D.~Grumiller, F.~Sch\"oller, and J.~Sim\'on, ``{Variational
  principle and one-point functions in three-dimensional flat space Einstein
  gravity},'' \href{http://dx.doi.org/10.1103/PhysRevD.89.084061}{{\em Phys.
  Rev. D} {\bf 89} (2014) no.~8, 084061},
  \href{http://arxiv.org/abs/1402.3687}{{\tt arXiv:1402.3687 [hep-th]}}.

\bibitem{Bagchi:2014iea}
A.~Bagchi, R.~Basu, D.~Grumiller, and M.~Riegler, ``{Entanglement entropy in
  Galilean conformal field theories and flat holography},''
  \href{http://dx.doi.org/10.1103/PhysRevLett.114.111602}{{\em Phys. Rev.
  Lett.} {\bf 114} (2015) no.~11, 111602},
  \href{http://arxiv.org/abs/1410.4089}{{\tt arXiv:1410.4089 [hep-th]}}.

\bibitem{Bagchi:2015wna}
A.~Bagchi, D.~Grumiller, and W.~Merbis, ``{Stress tensor correlators in
  three-dimensional gravity},''
  \href{http://dx.doi.org/10.1103/PhysRevD.93.061502}{{\em Phys. Rev. D} {\bf
  93} (2016) no.~6, 061502}, \href{http://arxiv.org/abs/1507.05620}{{\tt
  arXiv:1507.05620 [hep-th]}}.

\bibitem{Hartong:2015usd}
J.~Hartong, ``{Holographic Reconstruction of 3D Flat Space-Time},''
  \href{http://dx.doi.org/10.1007/JHEP10(2016)104}{{\em JHEP} {\bf 10} (2016)
  104}, \href{http://arxiv.org/abs/1511.01387}{{\tt arXiv:1511.01387
  [hep-th]}}.

\bibitem{Basu:2017aqn}
R.~Basu, S.~Detournay, and M.~Riegler, ``{Spectral Flow in 3D Flat
  Spacetimes},'' \href{http://dx.doi.org/10.1007/JHEP12(2017)134}{{\em JHEP}
  {\bf 12} (2017)  134}, \href{http://arxiv.org/abs/1706.07438}{{\tt
  arXiv:1706.07438 [hep-th]}}.

\bibitem{Bhattacharyya:2007vjd}
S.~Bhattacharyya, V.~E. Hubeny, S.~Minwalla, and M.~Rangamani, ``{Nonlinear
  Fluid Dynamics from Gravity},''
  \href{http://dx.doi.org/10.1088/1126-6708/2008/02/045}{{\em JHEP} {\bf 02}
  (2008)  045}, \href{http://arxiv.org/abs/0712.2456}{{\tt arXiv:0712.2456
  [hep-th]}}.

\bibitem{Haack:2008cp}
M.~Haack and A.~Yarom, ``{Nonlinear viscous hydrodynamics in various dimensions
  using AdS/CFT},'' \href{http://dx.doi.org/10.1088/1126-6708/2008/10/063}{{\em
  JHEP} {\bf 10} (2008)  063}, \href{http://arxiv.org/abs/0806.4602}{{\tt
  arXiv:0806.4602 [hep-th]}}.

\bibitem{Hubeny:2011hd}
V.~E. Hubeny, S.~Minwalla, and M.~Rangamani, ``{The fluid/gravity
  correspondence},'' in {\em {Theoretical Advanced Study Institute in
  Elementary Particle Physics}: {String theory and its Applications: From meV
  to the Planck Scale}}, pp.~348--383.
\newblock 2012.
\newblock \href{http://arxiv.org/abs/1107.5780}{{\tt arXiv:1107.5780
  [hep-th]}}.

\bibitem{Ciambelli:2018xat}
L.~Ciambelli, C.~Marteau, A.~C. Petkou, P.~M. Petropoulos, and K.~Siampos,
  ``{Covariant Galilean versus Carrollian hydrodynamics from relativistic
  fluids},'' \href{http://dx.doi.org/10.1088/1361-6382/aacf1a}{{\em Class.
  Quant. Grav.} {\bf 35} (2018) no.~16, 165001},
  \href{http://arxiv.org/abs/1802.05286}{{\tt arXiv:1802.05286 [hep-th]}}.

\bibitem{Petkou:2022bmz}
A.~C. Petkou, P.~M. Petropoulos, D.~R. Betancour, and K.~Siampos,
  ``{Relativistic Fluids, Hydrodynamic Frames and their Galilean versus
  Carrollian Avatars},'' \href{http://arxiv.org/abs/2205.09142}{{\tt
  arXiv:2205.09142 [hep-th]}}.

\bibitem{Lee:1990nz}
J.~Lee and R.~M. Wald, ``{Local symmetries and constraints},''
\href{http://dx.doi.org/10.1063/1.528801}{{\em J. Math. Phys.} {\bf 31} (1990)
  725--743}.

\bibitem{Iyer:1994ys}
V.~Iyer and R.~M. Wald, ``{Some properties of Noether charge and a proposal for
  dynamical black hole entropy},''
  \href{http://dx.doi.org/10.1103/PhysRevD.50.846}{{\em Phys. Rev. D} {\bf 50}
  (1994)  846--864}, \href{http://arxiv.org/abs/gr-qc/9403028}{{\tt
  arXiv:gr-qc/9403028}}.

\bibitem{Wald:1999wa}
R.~M. Wald and A.~Zoupas, ``{A General definition of `conserved quantities' in
  general relativity and other theories of gravity},''
  \href{http://dx.doi.org/10.1103/PhysRevD.61.084027}{{\em Phys. Rev.} {\bf
  D61} (2000)  084027},
\href{http://arxiv.org/abs/gr-qc/9911095}{{\tt arXiv:gr-qc/9911095 [gr-qc]}}.

\bibitem{Compere:2008us}
G.~Compere and D.~Marolf, ``{Setting the boundary free in AdS/CFT},''
  \href{http://dx.doi.org/10.1088/0264-9381/25/19/195014}{{\em Class. Quant.
  Grav.} {\bf 25} (2008)  195014}, \href{http://arxiv.org/abs/0805.1902}{{\tt
  arXiv:0805.1902 [hep-th]}}.

\bibitem{Compere:2018ylh}
G.~Comp\`ere, A.~Fiorucci, and R.~Ruzziconi, ``{Superboost transitions,
  refraction memory and super-Lorentz charge algebra},''
  \href{http://dx.doi.org/10.1007/JHEP11(2018)200}{{\em JHEP} {\bf 11} (2018)
  200}, \href{http://arxiv.org/abs/1810.00377}{{\tt arXiv:1810.00377
  [hep-th]}}.

\bibitem{Freidel:2021fxf}
L.~Freidel, R.~Oliveri, D.~Pranzetti, and S.~Speziale, ``{The Weyl BMS group
  and Einstein\textquoteright{}s equations},''
  \href{http://dx.doi.org/10.1007/JHEP07(2021)170}{{\em JHEP} {\bf 07} (2021)
  170}, \href{http://arxiv.org/abs/2104.05793}{{\tt arXiv:2104.05793
  [hep-th]}}.

\bibitem{Chandrasekaran:2021vyu}
V.~Chandrasekaran, E.~E. Flanagan, I.~Shehzad, and A.~J. Speranza, ``{A general
  framework for gravitational charges and holographic renormalization},''
  \href{http://arxiv.org/abs/2111.11974}{{\tt arXiv:2111.11974 [gr-qc]}}.

\bibitem{Ciambelli:2021vnn}
L.~Ciambelli and R.~G. Leigh, ``{Isolated surfaces and symmetries of
  gravity},'' \href{http://dx.doi.org/10.1103/PhysRevD.104.046005}{{\em Phys.
  Rev. D} {\bf 104} (2021) no.~4, 046005},
  \href{http://arxiv.org/abs/2104.07643}{{\tt arXiv:2104.07643 [hep-th]}}.

\bibitem{Ciambelli:2021nmv}
L.~Ciambelli, R.~G. Leigh, and P.-C. Pai, ``{Embeddings and Integrable Charges
  for Extended Corner Symmetry},'' \href{http://arxiv.org/abs/2111.13181}{{\tt
  arXiv:2111.13181 [hep-th]}}.

\bibitem{Freidel:2021dxw}
L.~Freidel, ``{A canonical bracket for open gravitational system},''
  \href{http://arxiv.org/abs/2111.14747}{{\tt arXiv:2111.14747 [hep-th]}}.

\bibitem{Speranza:2022lxr}
A.~J. Speranza, ``{Ambiguity resolution for integrable gravitational
  charges},'' \href{http://arxiv.org/abs/2202.00133}{{\tt arXiv:2202.00133
  [hep-th]}}.

\bibitem{preprint_Luca}
L.~Ciambelli, R.~G. Leigh, and W.~Jia. {To appear}.

\bibitem{Henneaux:1992ig}
M.~Henneaux and C.~Teitelboim, {\em {Quantization of gauge systems}}.
\newblock Princeton University Press, 1992.

\bibitem{Boulanger:2007ab}
N.~Boulanger, ``{Algebraic Classification of Weyl Anomalies in Arbitrary
  Dimensions},'' \href{http://dx.doi.org/10.1103/PhysRevLett.98.261302}{{\em
  Phys. Rev. Lett.} {\bf 98} (2007)  261302},
  \href{http://arxiv.org/abs/0706.0340}{{\tt arXiv:0706.0340 [hep-th]}}.

\bibitem{Boulanger:2007st}
N.~Boulanger, ``{General solutions of the Wess-Zumino consistency condition for
  the Weyl anomalies},''
  \href{http://dx.doi.org/10.1088/1126-6708/2007/07/069}{{\em JHEP} {\bf 07}
  (2007)  069}, \href{http://arxiv.org/abs/0704.2472}{{\tt arXiv:0704.2472
  [hep-th]}}.

\bibitem{Duval:2014uva}
C.~Duval, G.~W. Gibbons, and P.~A. Horvathy, ``{Conformal Carroll groups and
  BMS symmetry},'' \href{http://dx.doi.org/10.1088/0264-9381/31/9/092001}{{\em
  Class. Quant. Grav.} {\bf 31} (2014)  092001},
  \href{http://arxiv.org/abs/1402.5894}{{\tt arXiv:1402.5894 [gr-qc]}}.

\bibitem{Duval:2014lpa}
C.~Duval, G.~W. Gibbons, and P.~A. Horvathy, ``{Conformal Carroll groups},''
  \href{http://dx.doi.org/10.1088/1751-8113/47/33/335204}{{\em J. Phys. A} {\bf
  47} (2014) no.~33, 335204}, \href{http://arxiv.org/abs/1403.4213}{{\tt
  arXiv:1403.4213 [hep-th]}}.

\bibitem{Ciambelli:2019lap}
L.~Ciambelli, R.~G. Leigh, C.~Marteau, and P.~M. Petropoulos, ``{Carroll
  Structures, Null Geometry and Conformal Isometries},''
  \href{http://dx.doi.org/10.1103/PhysRevD.100.046010}{{\em Phys. Rev. D} {\bf
  100} (2019) no.~4, 046010}, \href{http://arxiv.org/abs/1905.02221}{{\tt
  arXiv:1905.02221 [hep-th]}}.

\bibitem{Henneaux:1979vn}
M.~Henneaux, ``{Geometry of Zero Signature Space-times},'' {\em Bull. Soc.
  Math. Belg.} {\bf 31} (1979)  47--63.

\bibitem{Duval:1990hj}
C.~Duval, G.~W. Gibbons, and P.~Horvathy, ``{Celestial mechanics, conformal
  structures and gravitational waves},''
  \href{http://dx.doi.org/10.1103/PhysRevD.43.3907}{{\em Phys. Rev. D} {\bf 43}
  (1991)  3907--3922}, \href{http://arxiv.org/abs/hep-th/0512188}{{\tt
  arXiv:hep-th/0512188}}.

\bibitem{Duval:2014uoa}
C.~Duval, G.~W. Gibbons, P.~A. Horvathy, and P.~M. Zhang, ``{Carroll versus
  Newton and Galilei: two dual non-Einsteinian concepts of time},''
  \href{http://dx.doi.org/10.1088/0264-9381/31/8/085016}{{\em Class. Quant.
  Grav.} {\bf 31} (2014)  085016}, \href{http://arxiv.org/abs/1402.0657}{{\tt
  arXiv:1402.0657 [gr-qc]}}.

\bibitem{Ashtekar:2014zsa}
A.~Ashtekar, ``{Geometry and Physics of Null Infinity},''
  \href{http://arxiv.org/abs/1409.1800}{{\tt arXiv:1409.1800 [gr-qc]}}.

\bibitem{Herfray:2021qmp}
Y.~Herfray, ``{Carrollian manifolds and null infinity: A view from Cartan
  geometry},'' \href{http://arxiv.org/abs/2112.09048}{{\tt arXiv:2112.09048
  [gr-qc]}}.

\bibitem{Bergshoeff:2022eog}
E.~Bergshoeff, J.~Figueroa-O'Farrill, and J.~Gomis, ``{A non-lorentzian
  primer},'' \href{http://arxiv.org/abs/2206.12177}{{\tt arXiv:2206.12177
  [hep-th]}}.

\bibitem{Bagchi:2019xfx}
A.~Bagchi, A.~Mehra, and P.~Nandi, ``{Field Theories with Conformal Carrollian
  Symmetry},'' \href{http://dx.doi.org/10.1007/JHEP05(2019)108}{{\em JHEP} {\bf
  05} (2019)  108}, \href{http://arxiv.org/abs/1901.10147}{{\tt
  arXiv:1901.10147 [hep-th]}}.

\bibitem{Gupta:2020dtl}
N.~Gupta and N.~V. Suryanarayana, ``{Constructing Carrollian CFTs},''
  \href{http://dx.doi.org/10.1007/JHEP03(2021)194}{{\em JHEP} {\bf 03} (2021)
  194}, \href{http://arxiv.org/abs/2001.03056}{{\tt arXiv:2001.03056
  [hep-th]}}.

\bibitem{Bagchi:2022eav}
A.~Bagchi, A.~Banerjee, S.~Dutta, K.~S. Kolekar, and P.~Sharma, ``{Carroll
  covariant scalar fields in two dimensions},''
  \href{http://arxiv.org/abs/2203.13197}{{\tt arXiv:2203.13197 [hep-th]}}.

\bibitem{Rivera-Betancour:2022lkc}
D.~Rivera-Betancour and M.~Vilatte, ``{Revisiting the Carrollian Scalar
  Field},'' \href{http://arxiv.org/abs/2207.01647}{{\tt arXiv:2207.01647
  [hep-th]}}.

\bibitem{Baiguera:2022lsw}
S.~Baiguera, G.~Oling, W.~Sybesma, and B.~T. S\o{}gaard, ``{Conformal Carroll
  Scalars with Boosts},'' \href{http://arxiv.org/abs/2207.03468}{{\tt
  arXiv:2207.03468 [hep-th]}}.

\bibitem{Bagchi:2019clu}
A.~Bagchi, R.~Basu, A.~Mehra, and P.~Nandi, ``{Field Theories on Null
  Manifolds},'' \href{http://dx.doi.org/10.1007/JHEP02(2020)141}{{\em JHEP}
  {\bf 02} (2020)  141}, \href{http://arxiv.org/abs/1912.09388}{{\tt
  arXiv:1912.09388 [hep-th]}}.

\bibitem{Henneaux:2021yzg}
M.~Henneaux and P.~Salgado-Rebolledo, ``{Carroll contractions of
  Lorentz-invariant theories},''
  \href{http://dx.doi.org/10.1007/JHEP11(2021)180}{{\em JHEP} {\bf 11} (2021)
  180}, \href{http://arxiv.org/abs/2109.06708}{{\tt arXiv:2109.06708
  [hep-th]}}.

\bibitem{deBoer:2021jej}
J.~de~Boer, J.~Hartong, N.~A. Obers, W.~Sybesma, and S.~Vandoren, ``{Carroll
  Symmetry, Dark Energy and Inflation},''
  \href{http://dx.doi.org/10.3389/fphy.2022.810405}{{\em Front. in Phys.} {\bf
  10} (2022)  810405}, \href{http://arxiv.org/abs/2110.02319}{{\tt
  arXiv:2110.02319 [hep-th]}}.

\bibitem{Bagchi:2021gai}
A.~Bagchi, S.~Dutta, K.~S. Kolekar, and P.~Sharma, ``{BMS field theories and
  Weyl anomaly},'' \href{http://dx.doi.org/10.1007/JHEP07(2021)101}{{\em JHEP}
  {\bf 07} (2021)  101}, \href{http://arxiv.org/abs/2104.10405}{{\tt
  arXiv:2104.10405 [hep-th]}}.

\bibitem{Achucarro:1986uwr}
A.~Achucarro and P.~K. Townsend, ``{A Chern-Simons Action for Three-Dimensional
  anti-De Sitter Supergravity Theories},''
  \href{http://dx.doi.org/10.1016/0370-2693(86)90140-1}{{\em Phys. Lett. B}
  {\bf 180} (1986)  89}.

\bibitem{Witten:1988hc}
E.~Witten, ``{(2+1)-Dimensional Gravity as an Exactly Soluble System},''
  \href{http://dx.doi.org/10.1016/0550-3213(88)90143-5}{{\em Nucl. Phys. B}
  {\bf 311} (1988)  46}.

\bibitem{Loganayagam:2008is}
R.~Loganayagam, ``{Entropy Current in Conformal Hydrodynamics},''
  \href{http://dx.doi.org/10.1088/1126-6708/2008/05/087}{{\em JHEP} {\bf 05}
  (2008)  087}, \href{http://arxiv.org/abs/0801.3701}{{\tt arXiv:0801.3701
  [hep-th]}}.

\bibitem{Brown:1992br}
J.~D. Brown and J.~W. York, Jr., ``{Quasilocal energy and conserved charges
  derived from the gravitational action},''
  \href{http://dx.doi.org/10.1103/PhysRevD.47.1407}{{\em Phys. Rev. D} {\bf 47}
  (1993)  1407--1419}, \href{http://arxiv.org/abs/gr-qc/9209012}{{\tt
  arXiv:gr-qc/9209012}}.

\bibitem{Schwimmer:2008yh}
A.~Schwimmer and S.~Theisen, ``{Entanglement Entropy, Trace Anomalies and
  Holography},'' \href{http://dx.doi.org/10.1016/j.nuclphysb.2008.04.015}{{\em
  Nucl. Phys. B} {\bf 801} (2008)  1--24},
  \href{http://arxiv.org/abs/0802.1017}{{\tt arXiv:0802.1017 [hep-th]}}.

\bibitem{Barnich:2010eb}
G.~Barnich and C.~Troessaert, ``{Aspects of the BMS/CFT correspondence},''
  \href{http://dx.doi.org/10.1007/JHEP05(2010)062}{{\em JHEP} {\bf 05} (2010)
  062},
\href{http://arxiv.org/abs/1001.1541}{{\tt arXiv:1001.1541 [hep-th]}}.

\bibitem{Freidel:2020xyx}
L.~Freidel, M.~Geiller, and D.~Pranzetti, ``{Edge modes of gravity. Part I.
  Corner potentials and charges},''
  \href{http://dx.doi.org/10.1007/JHEP11(2020)026}{{\em JHEP} {\bf 11} (2020)
  026}, \href{http://arxiv.org/abs/2006.12527}{{\tt arXiv:2006.12527
  [hep-th]}}.

\bibitem{Freidel:2021cjp}
L.~Freidel, R.~Oliveri, D.~Pranzetti, and S.~Speziale, ``{Extended corner
  symmetry, charge bracket and Einstein\textquoteright{}s equations},''
  \href{http://dx.doi.org/10.1007/JHEP09(2021)083}{{\em JHEP} {\bf 09} (2021)
  083}, \href{http://arxiv.org/abs/2104.12881}{{\tt arXiv:2104.12881
  [hep-th]}}.

\bibitem{book:16319}
R.~A. Bertlmann, {\em Anomalies in quantum field theory}.
\newblock The International series of monographs on physics 91 Oxford science
  publications. Clarendon Press, 1996.

\bibitem{Barnich:2011mi}
G.~Barnich and C.~Troessaert, ``{BMS charge algebra},''
  \href{http://dx.doi.org/10.1007/JHEP12(2011)105}{{\em JHEP} {\bf 12} (2011)
  105}, \href{http://arxiv.org/abs/1106.0213}{{\tt arXiv:1106.0213 [hep-th]}}.

\bibitem{Barnich:2006av}
G.~Barnich and G.~Compere, ``{Classical central extension for asymptotic
  symmetries at null infinity in three spacetime dimensions},''
  \href{http://dx.doi.org/10.1088/0264-9381/24/5/F01}{{\em Class. Quant. Grav.}
  {\bf 24} (2007)  F15--F23}, \href{http://arxiv.org/abs/gr-qc/0610130}{{\tt
  arXiv:gr-qc/0610130}}.

\bibitem{Regge:1974zd}
T.~Regge and C.~Teitelboim, ``{Role of Surface Integrals in the Hamiltonian
  Formulation of General Relativity},''
  \href{http://dx.doi.org/10.1016/0003-4916(74)90404-7}{{\em Annals Phys.} {\bf
  88} (1974)  286}.

\bibitem{Banados:1994tn}
M.~Banados, ``{Global charges in Chern-Simons field theory and the (2+1) black
  hole},'' \href{http://dx.doi.org/10.1103/PhysRevD.52.5816}{{\em Phys. Rev. D}
  {\bf 52} (1996)  5816--5825}, \href{http://arxiv.org/abs/hep-th/9405171}{{\tt
  arXiv:hep-th/9405171}}.

\bibitem{Coussaert:1995zp}
O.~Coussaert, M.~Henneaux, and P.~van Driel, ``{The Asymptotic dynamics of
  three-dimensional Einstein gravity with a negative cosmological constant},''
  \href{http://dx.doi.org/10.1088/0264-9381/12/12/012}{{\em Class. Quant.
  Grav.} {\bf 12} (1995)  2961--2966},
  \href{http://arxiv.org/abs/gr-qc/9506019}{{\tt arXiv:gr-qc/9506019}}.

\bibitem{Banados:1998gg}
M.~Banados, ``{Three-dimensional quantum geometry and black holes},''
  \href{http://dx.doi.org/10.1063/1.59661}{{\em AIP Conf. Proc.} {\bf 484}
  (1999) no.~1, 147--169}, \href{http://arxiv.org/abs/hep-th/9901148}{{\tt
  arXiv:hep-th/9901148}}.

\bibitem{Henneaux:1999ib}
M.~Henneaux, L.~Maoz, and A.~Schwimmer, ``{Asymptotic dynamics and asymptotic
  symmetries of three-dimensional extended AdS supergravity},''
  \href{http://dx.doi.org/10.1006/aphy.2000.5994}{{\em Annals Phys.} {\bf 282}
  (2000)  31--66}, \href{http://arxiv.org/abs/hep-th/9910013}{{\tt
  arXiv:hep-th/9910013}}.

\bibitem{Bunster:2014mua}
C.~Bunster, M.~Henneaux, A.~Perez, D.~Tempo, and R.~Troncoso, ``{Generalized
  Black Holes in Three-dimensional Spacetime},''
  \href{http://dx.doi.org/10.1007/JHEP05(2014)031}{{\em JHEP} {\bf 05} (2014)
  031}, \href{http://arxiv.org/abs/1404.3305}{{\tt arXiv:1404.3305 [hep-th]}}.

\bibitem{Campoleoni:2010zq}
A.~Campoleoni, S.~Fredenhagen, S.~Pfenninger, and S.~Theisen, ``{Asymptotic
  symmetries of three-dimensional gravity coupled to higher-spin fields},''
  \href{http://dx.doi.org/10.1007/JHEP11(2010)007}{{\em JHEP} {\bf 11} (2010)
  007}, \href{http://arxiv.org/abs/1008.4744}{{\tt arXiv:1008.4744 [hep-th]}}.

\bibitem{Detournay:2016sfv}
S.~Detournay and M.~Riegler, ``{Enhanced Asymptotic Symmetry Algebra of 2+1
  Dimensional Flat Space},''
  \href{http://dx.doi.org/10.1103/PhysRevD.95.046008}{{\em Phys. Rev. D} {\bf
  95} (2017) no.~4, 046008}, \href{http://arxiv.org/abs/1612.00278}{{\tt
  arXiv:1612.00278 [hep-th]}}.

\bibitem{Grumiller:2017sjh}
D.~Grumiller, W.~Merbis, and M.~Riegler, ``{Most general flat space boundary
  conditions in three-dimensional Einstein gravity},''
  \href{http://dx.doi.org/10.1088/1361-6382/aa8004}{{\em Class. Quant. Grav.}
  {\bf 34} (2017) no.~18, 184001}, \href{http://arxiv.org/abs/1704.07419}{{\tt
  arXiv:1704.07419 [hep-th]}}.

\bibitem{Grumiller:2020vvv}
D.~Grumiller, M.~M. Sheikh-Jabbari, and C.~Zwikel, ``{Horizons 2020},''
  \href{http://dx.doi.org/10.1142/S0218271820430063}{{\em Int. J. Mod. Phys. D}
  {\bf 29} (2020) no.~14, 2043006}, \href{http://arxiv.org/abs/2005.06936}{{\tt
  arXiv:2005.06936 [hep-th]}}.

\bibitem{Donnelly:2016auv}
W.~Donnelly and L.~Freidel, ``{Local subsystems in gauge theory and gravity},''
  \href{http://dx.doi.org/10.1007/JHEP09(2016)102}{{\em JHEP} {\bf 09} (2016)
  102}, \href{http://arxiv.org/abs/1601.04744}{{\tt arXiv:1601.04744
  [hep-th]}}.

\bibitem{Speranza:2017gxd}
A.~J. Speranza, ``{Local phase space and edge modes for
  diffeomorphism-invariant theories},''
  \href{http://dx.doi.org/10.1007/JHEP02(2018)021}{{\em JHEP} {\bf 02} (2018)
  021}, \href{http://arxiv.org/abs/1706.05061}{{\tt arXiv:1706.05061
  [hep-th]}}.

\bibitem{Geiller:2017whh}
M.~Geiller, ``{Lorentz-diffeomorphism edge modes in 3d gravity},''
  \href{http://dx.doi.org/10.1007/JHEP02(2018)029}{{\em JHEP} {\bf 02} (2018)
  029}, \href{http://arxiv.org/abs/1712.05269}{{\tt arXiv:1712.05269 [gr-qc]}}.

\bibitem{Freidel:2020svx}
L.~Freidel, M.~Geiller, and D.~Pranzetti, ``{Edge modes of gravity. Part II.
  Corner metric and Lorentz charges},''
  \href{http://dx.doi.org/10.1007/JHEP11(2020)027}{{\em JHEP} {\bf 11} (2020)
  027}, \href{http://arxiv.org/abs/2007.03563}{{\tt arXiv:2007.03563
  [hep-th]}}.

\bibitem{Freidel:2020ayo}
L.~Freidel, M.~Geiller, and D.~Pranzetti, ``{Edge modes of gravity. Part III.
  Corner simplicity constraints},''
  \href{http://dx.doi.org/10.1007/JHEP01(2021)100}{{\em JHEP} {\bf 01} (2021)
  100}, \href{http://arxiv.org/abs/2007.12635}{{\tt arXiv:2007.12635
  [hep-th]}}.

\bibitem{Donnelly:2020xgu}
W.~Donnelly, L.~Freidel, S.~F. Moosavian, and A.~J. Speranza, ``{Gravitational
  edge modes, coadjoint orbits, and hydrodynamics},''
  \href{http://dx.doi.org/10.1007/JHEP09(2021)008}{{\em JHEP} {\bf 09} (2021)
  008}, \href{http://arxiv.org/abs/2012.10367}{{\tt arXiv:2012.10367
  [hep-th]}}.

\bibitem{Ciambelli:2022cfr}
L.~Ciambelli and R.~G. Leigh, ``{Universal Corner Symmetry and the Orbit Method
  for Gravity},'' \href{http://arxiv.org/abs/2207.06441}{{\tt arXiv:2207.06441
  [hep-th]}}.

\bibitem{Donnay:2022aba}
L.~Donnay, A.~Fiorucci, Y.~Herfray, and R.~Ruzziconi, ``{A Carrollian
  Perspective on Celestial Holography},''
  \href{http://arxiv.org/abs/2202.04702}{{\tt arXiv:2202.04702 [hep-th]}}.

\bibitem{Bagchi:2022emh}
A.~Bagchi, S.~Banerjee, R.~Basu, and S.~Dutta, ``{Scattering Amplitudes:
  Celestial and Carrollian},''
  \href{http://dx.doi.org/10.1103/PhysRevLett.128.241601}{{\em Phys. Rev.
  Lett.} {\bf 128} (2022) no.~24, 241601},
  \href{http://arxiv.org/abs/2202.08438}{{\tt arXiv:2202.08438 [hep-th]}}.

\bibitem{Strominger:2017zoo}
A.~Strominger, {\em {Lectures on the Infrared Structure of Gravity and Gauge
  Theory}}.
\newblock {Princeton University Press}, 2018.
\newblock
\href{http://arxiv.org/abs/1703.05448}{{\tt arXiv:1703.05448 [hep-th]}}.
\newblock

\bibitem{Pasterski:2021raf}
S.~Pasterski, M.~Pate, and A.-M. Raclariu, ``{Celestial Holography},'' in {\em
  {2022 Snowmass Summer Study}}.
\newblock 11, 2021.
\newblock \href{http://arxiv.org/abs/2111.11392}{{\tt arXiv:2111.11392
  [hep-th]}}.

\bibitem{McLoughlin:2022ljp}
T.~McLoughlin, A.~Puhm, and A.-M. Raclariu, ``{The SAGEX Review on Scattering
  Amplitudes, Chapter 11: Soft Theorems and Celestial Amplitudes},''
  \href{http://arxiv.org/abs/2203.13022}{{\tt arXiv:2203.13022 [hep-th]}}.

\bibitem{Jackiw:1995qh}
R.~Jackiw, ``{Another view on massless matter - gravity fields in
  two-dimensions},'' \href{http://arxiv.org/abs/hep-th/9501016}{{\tt
  arXiv:hep-th/9501016}}.

\end{thebibliography}\endgroup

\end{document}